\documentclass[a4paper,11pt]{article}
\pdfoutput=1
\usepackage{jheppub}

\usepackage{amssymb,amsmath,amsfonts}
\usepackage[normalem]{ulem}
\usepackage[utf8x]{inputenc}
\usepackage{slashed}
\usepackage{graphicx}
\usepackage{tabularx}
\usepackage{here}
\usepackage{color}
\usepackage{csquotes} 
\usepackage{comment}
\usepackage{mathrsfs}
\usepackage{float}
\usepackage{ascmac}
\usepackage{multirow}
\usepackage{longtable}
\usepackage{bm}
\usepackage{ulem}
\usepackage{tcolorbox}

\usepackage{booktabs}
\usepackage{array}

\usepackage[italicdiff]{physics}

\usepackage[hang,small,bf]{caption}
\usepackage[subrefformat=parens]{subcaption}
\makeatletter
\newcommand*\rel@kern[1]{\kern#1\dimexpr\macc@kerna}
\newcommand*\widebar[1]{%
  \begingroup
  \def\mathaccent##1##2{%
    \rel@kern{0.8}%
    \overline{\rel@kern{-0.8}\macc@nucleus\rel@kern{0.2}}%
    \rel@kern{-0.2}%
  }%
  \macc@depth\@ne
  \let\math@bgroup\@empty \let\math@egroup\macc@set@skewchar
  \mathsurround\z@ \frozen@everymath{\mathgroup\macc@group\relax}%
  \macc@set@skewchar\relax
  \let\mathaccentV\macc@nested@a
  \macc@nested@a\relax111{#1}%
  \endgroup
}
\makeatother

\numberwithin{equation}{section}

\preprint{
\begin{minipage}{5cm}
\small
\flushright
EPHOU-26-09\\
KYUSHU-HET-369
\end{minipage}}

\title{Textures of dimension-six operators in the SMEFT with non-invertible selection rules}

\author{Tatsuo Kobayashi$^{1}$,} 
\author{Hajime Otsuka$^{2,3}$,} 
\author{Morimitsu Tanimoto$^{4}$, and} 
\author{Kei Yamamoto$^{5}$}
\affiliation{
$^1$Department of Physics, Hokkaido University, Sapporo 060-0810, Japan}
\affiliation{
$^2$Department of Physics, Kyushu University, 744 Motooka, Nishi-ku, Fukuoka 819-0395, Japan}
\affiliation{
$^3$Quantum and Spacetime Research Institute (QuaSR), Kyushu University, 744 Motooka, Nishi-ku, Fukuoka 819-0395, Japan}
\affiliation{
$^4$Department of Physics, Niigata University, Ikarashi 2-8050, Niigata 950-2181, Japan}
\affiliation{
$^5$Faculty of Science and Engineering, Iwate University, Morioka, Iwate 020-8550, Japan}
\emailAdd{kobayashi@particle.sci.hokudai.ac.jp}
\emailAdd{otsuka.hajime@phys.kyushu-u.ac.jp}
\emailAdd{morimitsutanimoto@yahoo.co.jp}
\emailAdd{keiy@iwate-u.ac.jp}

\abstract{
We investigate the flavor structures of dimension-six operators in the Standard Model Effective Field Theory (SMEFT) subject to non-invertible selection rules. 
In particular, we classify the flavor textures of all baryon-number-conserving dimension-six SMEFT operators and determine the resulting constraints on their Wilson coefficients. The selection rules determine not only the texture zeros but also the allowed tensor structures of the Wilson coefficients, which can be expressed analytically in terms of a reduced number of independent parameters. 
We also find that the flavor structures of higher-dimensional operators are not necessarily aligned with those of the Yukawa couplings, 
in contrast to the Minimal Flavor Violation hypothesis, in which the Yukawa couplings govern the flavor structure of higher-dimensional operators. 
It turns out that the resulting flavor and chirality patterns differ from those typically obtained in SMEFT with conventional flavor symmetries, providing characteristic predictions for $B$-meson observables and charged-lepton-flavor-violating radiative decays.
}

\makeatletter
\gdef\@fpheader{}
\makeatother

\begin{document}

\maketitle

\section{Introduction}

The textures of Yukawa matrices play an important role in understanding the flavor structure of quarks and leptons with a limited number of free parameters \cite{Weinberg:1977hb,Fritzsch:1977vd}.
In conventional models based on group symmetries, nontrivial Yukawa textures are typically generated through vacuum expectation values of symmetry-breaking scalar fields, referred to as flavons. 
However, such constructions generally require a nontrivial scalar sector with additional fields and parameters in order to control spontaneous symmetry breaking.

Recently, it has been pointed out that Yukawa textures can instead be realized through {\it non-invertible selection rules}~\cite{Kobayashi:2024yqq,Kobayashi:2024cvp}. In this approach, the texture zeros and relations among Yukawa couplings follow directly from the underlying symmetry structure without introducing flavon fields. 
Furthermore, non-invertible selection rules can lead to Yukawa textures that cannot be reproduced by conventional group-based symmetries. Explicit constructions have been presented for the quark sector in Refs.~\cite{Kobayashi:2025znw,Liang:2025dkm} and for the lepton sector in Refs.~\cite{Kobayashi:2025ldi,Jiang:2025psz,Qu:2026omn,Kitagawa:2026eck} as well as for the quark-lepton sector consistent 
with the grand unified theory~\cite{Kobayashi:2025thd,Kobayashi:2025rpx,Chen:2026mvi}.

Flavor physics is relevant not only to Yukawa textures, but also to higher-dimensional operators, which can induce rare flavor processes. Such processes have been investigated with increasing experimental precision and may provide valuable probes of new physics (NP). In particular, the anomalies observed in $B$-meson decays~\cite{HeavyFlavorAveragingGroupHFLAV:2024ctg} may originate from contributions of higher-dimensional operators. 
In this work, we extend this framework to the Standard Model Effective Field Theory (SMEFT). For dimension-six operators, the baryon-number conserving SMEFT contains 59 independent operators when flavor structures and Hermitian conjugates are not distinguished~\cite{Grzadkowski:2010es}. Once the full flavor structure of three-generation fermions is taken into account, the number of independent parameters increases to 2499~\cite{Alonso:2013hga}. The operators involving the Higgs field and the four-fermion operators are respectively summarized in Tables~\ref{tab:dim6Higgs} and~\ref{tab:4fermi}. 
Their flavor structures are important not only for describing flavor-changing processes at low energies, but also for identifying possible imprints of an underlying ultraviolet (UV) theory.

\begin{table}[H]
\centering
\caption{Dimension-six operators including Higgs field $H$ and $\widetilde{H}=i\sigma^2 H^\ast$.}
\label{tab:dim6Higgs}
\scriptsize
\renewcommand{\arraystretch}{1.35}
\setlength{\tabcolsep}{3pt}
\resizebox{\textwidth}{!}{%
\begin{tabular}{||c|c||c|c||c|c||}
\hline\hline
\multicolumn{2}{||c||}{$X^3$} &
\multicolumn{2}{c||}{$H^6$ and $H^4D^2$} &
\multicolumn{2}{c||}{$\psi^2H^3$}\\
\hline
$Q_G$ &
$f^{ABC}G_{\mu}{}^{A\nu}G_{\nu}{}^{B\rho}G_{\rho}{}^{C\mu}$ &
$Q_H$ &
$(H^\dagger H)^3$ &
$Q_{eH}$ &
$(H^\dagger H)(\bar{l}_i e_j H)$\\
$Q_{\widetilde{G}}$ &
$f^{ABC}\widetilde{G}_{\mu}{}^{A\nu}G_{\nu}{}^{B\rho}G_{\rho}{}^{C\mu}$ &
$Q_{H\Box}$ &
$(H^\dagger H)\Box(H^\dagger H)$ &
$Q_{uH}$ &
$(H^\dagger H)(\bar{q}_i u_j\widetilde{H})$\\
$Q_W$ &
$\varepsilon^{IJK}W_{\mu}{}^{I\nu}W_{\nu}{}^{J\rho}W_{\rho}{}^{K\mu}$ &
$Q_{HD}$ &
$\left(H^\dagger D^\mu H\right)^*\left(H^\dagger D_\mu H\right)$ &
$Q_{dH}$ &
$(H^\dagger H)(\bar{q}_i d_j H)$\\
$Q_{\widetilde{W}}$ &
$\varepsilon^{IJK}\widetilde{W}_{\mu}{}^{I\nu}W_{\nu}{}^{J\rho}W_{\rho}{}^{K\mu}$ &
& &
&\\
\hline\hline
\multicolumn{2}{||c||}{$X^2H^2$} &
\multicolumn{2}{c||}{$\psi^2XH$} &
\multicolumn{2}{c||}{$\psi^2H^2D$}\\
\hline
$Q_{HG}$ &
$H^\dagger H\,G_{\mu\nu}^A G^{A\mu\nu}$ &
$Q_{eW}$ &
$(\bar{l}_i\sigma^{\mu\nu}e_j)\tau^I H W_{\mu\nu}^I$ &
$Q_{Hl}^{(1)}$ &
$\left(H^\dagger i\overleftrightarrow{D}_{\mu}H\right)(\bar{l}_i\gamma^\mu l_j)$\\
$Q_{H\widetilde{G}}$ &
$H^\dagger H\,\widetilde{G}_{\mu\nu}^A G^{A\mu\nu}$ &
$Q_{eB}$ &
$(\bar{l}_i\sigma^{\mu\nu}e_j)H B_{\mu\nu}$ &
$Q_{Hl}^{(3)}$ &
$\left(H^\dagger i\overleftrightarrow{D}_{\mu}^{\,I}H\right)(\bar{l}_i\tau^I\gamma^\mu l_j)$\\
$Q_{HW}$ &
$H^\dagger H\,W_{\mu\nu}^I W^{I\mu\nu}$ &
$Q_{uG}$ &
$(\bar{q}_i\sigma^{\mu\nu}T^A u_j)\widetilde{H}\,G_{\mu\nu}^A$ &
$Q_{He}$ &
$\left(H^\dagger i\overleftrightarrow{D}_{\mu}H\right)(\bar{e}_i\gamma^\mu e_j)$\\
$Q_{H\widetilde{W}}$ &
$H^\dagger H\,\widetilde{W}_{\mu\nu}^I W^{I\mu\nu}$ &
$Q_{uW}$ &
$(\bar{q}_i\sigma^{\mu\nu}u_j)\tau^I\widetilde{H}\,W_{\mu\nu}^I$ &
$Q_{Hq}^{(1)}$ &
$\left(H^\dagger i\overleftrightarrow{D}_{\mu}H\right)(\bar{q}_i\gamma^\mu q_j)$\\
$Q_{HB}$ &
$H^\dagger H\,B_{\mu\nu}B^{\mu\nu}$ &
$Q_{uB}$ &
$(\bar{q}_i\sigma^{\mu\nu}u_j)\widetilde{H}\,B_{\mu\nu}$ &
$Q_{Hq}^{(3)}$ &
$\left(H^\dagger i\overleftrightarrow{D}_{\mu}^{\,I}H\right)(\bar{q}_i\tau^I\gamma^\mu q_j)$\\
$Q_{H\widetilde{B}}$ &
$H^\dagger H\,\widetilde{B}_{\mu\nu}B^{\mu\nu}$ &
$Q_{dG}$ &
$(\bar{q}_i\sigma^{\mu\nu}T^A d_j)H\,G_{\mu\nu}^A$ &
$Q_{Hu}$ &
$\left(H^\dagger i\overleftrightarrow{D}_{\mu}H\right)(\bar{u}_i\gamma^\mu u_j)$\\
$Q_{HWB}$ &
$H^\dagger\tau^I H\,W_{\mu\nu}^I B^{\mu\nu}$ &
$Q_{dW}$ &
$(\bar{q}_i\sigma^{\mu\nu}d_j)\tau^I H\,W_{\mu\nu}^I$ &
$Q_{Hd}$ &
$\left(H^\dagger i\overleftrightarrow{D}_{\mu}H\right)(\bar{d}_i\gamma^\mu d_j)$\\
$Q_{H\widetilde{W}B}$ &
$H^\dagger\tau^I H\,\widetilde{W}_{\mu\nu}^I B^{\mu\nu}$ &
$Q_{dB}$ &
$(\bar{q}_i\sigma^{\mu\nu}d_j)H\,B_{\mu\nu}$ &
$Q_{Hud}$ &
$i\left(\widetilde{H}^{\dagger}D_\mu H\right)(\bar{u}_i\gamma^\mu d_j)$\\
\hline\hline
\end{tabular}%
}
\end{table}

\begin{table}[H]
\centering
\caption{Baryon-number-conserving four-fermion operators.}
\label{tab:4fermi}
\scriptsize
\renewcommand{\arraystretch}{1.35}
\setlength{\tabcolsep}{3pt}
\resizebox{\textwidth}{!}{%
\begin{tabular}{||c|c||c|c||c|c||}
\hline\hline
\multicolumn{2}{||c||}{$(\bar{L}L)(\bar{L}L)$} &
\multicolumn{2}{c||}{$(\bar{R}R)(\bar{R}R)$} &
\multicolumn{2}{c||}{$(\bar{L}L)(\bar{R}R)$}\\
\hline
$Q_{ll}$ &
$(\bar{l}_i\gamma_\mu l_j)(\bar{l}_k\gamma^\mu l_l)$ &
$Q_{ee}$ &
$(\bar{e}_i\gamma_\mu e_j)(\bar{e}_k\gamma^\mu e_l)$ &
$Q_{le}$ &
$(\bar{l}_i\gamma_\mu l_j)(\bar{e}_k\gamma^\mu e_l)$\\
$Q_{qq}^{(1)}$ &
$(\bar{q}_i\gamma_\mu q_j)(\bar{q}_k\gamma^\mu q_l)$ &
$Q_{uu}$ &
$(\bar{u}_i\gamma_\mu u_j)(\bar{u}_k\gamma^\mu u_l)$ &
$Q_{lu}$ &
$(\bar{l}_i\gamma_\mu l_j)(\bar{u}_k\gamma^\mu u_l)$\\
$Q_{qq}^{(3)}$ &
$(\bar{q}_i\gamma_\mu\tau^I q_j)(\bar{q}_k\gamma^\mu\tau^I q_l)$ &
$Q_{dd}$ &
$(\bar{d}_i\gamma_\mu d_j)(\bar{d}_k\gamma^\mu d_l)$ &
$Q_{ld}$ &
$(\bar{l}_i\gamma_\mu l_j)(\bar{d}_k\gamma^\mu d_l)$\\
$Q_{lq}^{(1)}$ &
$(\bar{l}_i\gamma_\mu l_j)(\bar{q}_k\gamma^\mu q_l)$ &
$Q_{eu}$ &
$(\bar{e}_i\gamma_\mu e_j)(\bar{u}_k\gamma^\mu u_l)$ &
$Q_{qe}$ &
$(\bar{q}_i\gamma_\mu q_j)(\bar{e}_k\gamma^\mu e_l)$\\
$Q_{lq}^{(3)}$ &
$(\bar{l}_i\gamma_\mu\tau^I l_j)(\bar{q}_k\gamma^\mu\tau^I q_l)$ &
$Q_{ed}$ &
$(\bar{e}_i\gamma_\mu e_j)(\bar{d}_k\gamma^\mu d_l)$ &
$Q_{qu}^{(1)}$ &
$(\bar{q}_i\gamma_\mu q_j)(\bar{u}_k\gamma^\mu u_l)$\\
& &
$Q_{ud}^{(1)}$ &
$(\bar{u}_i\gamma_\mu u_j)(\bar{d}_k\gamma^\mu d_l)$ &
$Q_{qu}^{(8)}$ &
$(\bar{q}_i\gamma_\mu T^A q_j)(\bar{u}_k\gamma^\mu T^A u_l)$\\
& &
$Q_{ud}^{(8)}$ &
$(\bar{u}_i\gamma_\mu T^A u_j)(\bar{d}_k\gamma^\mu T^A d_l)$ &
$Q_{qd}^{(1)}$ &
$(\bar{q}_i\gamma_\mu q_j)(\bar{d}_k\gamma^\mu d_l)$\\
& &
& &
$Q_{qd}^{(8)}$ &
$(\bar{q}_i\gamma_\mu T^A q_j)(\bar{d}_k\gamma^\mu T^A d_l)$\\
\hline\hline
\multicolumn{6}{||c||}{$(\bar{L}R)(\bar{R}L)$ and $(\bar{L}R)(\bar{L}R)$}\\
\hline
$Q_{ledq}$ &
$(\bar{l}_i^{\,p}e_j)(\bar{d}_k q_l^{\,p})$ &
$Q_{quqd}^{(1)}$ &
$(\bar{q}_i^{\,p}u_j)\varepsilon_{pq}(\bar{q}_k^{\,q}d_l)$ &
$Q_{quqd}^{(8)}$ &
$(\bar{q}_i^{\,p}T^A u_j)\varepsilon_{pq}(\bar{q}_k^{\,q}T^A d_l)$\\
$Q_{lequ}^{(1)}$ &
$(\bar{l}_i^{\,p}e_j)\varepsilon_{pq}(\bar{q}_k^{\,q}u_l)$ &
$Q_{lequ}^{(3)}$ &
$(\bar{l}_i^{\,p}\sigma_{\mu\nu}e_j)\varepsilon_{pq}
(\bar{q}_k^{\,q}\sigma^{\mu\nu}u_l)$ &
&\\
\hline\hline
\end{tabular}%
}
\end{table}

A famous approach to addressing the flavor structure of SMEFT is based on the global flavor symmetry of the Standard Model, i.e., $U(3)^5$. 
Under the Minimal Flavor Violation (MFV) hypotheses~\cite{Chivukula:1987py,DAmbrosio:2002vsn}, 
the Yukawa couplings are the only sources of flavor symmetry breaking, which naturally suppresses flavor-violating observables in the SMEFT.
When we turn on the third generation of quarks and leptons, the symmetry is reduced to $U(2)^5$, whose minimal breaking 
provides a natural framework for understanding the hierarchical structure of the Yukawa couplings
~\cite{Barbieri:2011ci,Barbieri:2012uh,Faroughy:2020ina}.\footnote{For more details, see Refs.~\cite{Brivio:2017vri,Isidori:2023pyp} for reviews.} 
Flavor structures in SMEFT have also been investigated using $U(1)$ flavor symmetry~\cite{Greljo:2022cah,Loisa:2024xuk}, modular flavor symmetries~\cite{Kobayashi:2021pav,Kobayashi:2022jvy,Moreno-Sanchez:2025bzz,Kang:2026qgi}, and string-theoretic selection rules~\cite{Kobayashi:2021uam}.

In this paper, we investigate the textures of Wilson coefficients of dimension-six SMEFT operators arising from non-invertible selection rules. 
So far, the flavor structure of the SMEFT operators has been understood via MFV hypotheses which regard Yukawa couplings as spurion fields under the flavor symmetries. 
It indicates that the flavor structure of Yukawa couplings and higher-dimensional operators are correlated with each other. 
By contrast, the non-invertible selection rules can directly address the flavor structure of the SMEFT operators without relying on the MFV hypotheses. 
Since matter fields are assigned to basis elements of a fusion algebra characterized by non-invertible fusion rules, it is expected that one can realize the different flavor structure of higher-dimensional operators from that of Yukawa couplings. 

In the framework of non-invertible selection rules, one can realize Yukawa textures reproducing realistic fermion masses and mixing angles after renormalization-group evolution. 
In this respect, it provides a natural framework in which the flavor structures of Yukawa couplings and higher-dimensional operators can be studied simultaneously. 
By choosing appropriate assignments for the Standard Model fields, we show that realistic Yukawa textures and nontrivial textures of flavor-dependent Wilson coefficients can be realized within a unified symmetry framework. 
This construction reduces the number of independent flavor parameters as in the MFV hypotheses and leads to testable correlations among Wilson coefficients. These correlations would be probed by current and future flavor experiments. As a phenomenological application, in particular, we focus on semileptonic processes 
and lepton flavor-violating processes, and investigate the consequences of the resulting Wilson-coefficient textures.

In Sec.~\ref{sec:setup}, we propose a four-dimensional effective field theory governed by fusion rules arising from the $\mathbb{Z}_2$ gauging of $\mathbb{Z}_N$ symmetries. 
These fusion rules lead to the textures of Yukawa matrices that cannot be obtained by group-based symmetries. 
In Sec.~\ref{sec:classification}, we classify the textures of all dimension-six operators in the SMEFT. 
The corresponding Wilson coefficients are constrained by the fusion rules. 
Sec.~\ref{sec:pheno} is devoted to the phenomenology of four-fermion operators and leptonic dipole operators. 
Finally, we summarize our results in Sec.~\ref{sec:con}.

\section{Setup}
\label{sec:setup}
In this section, we introduce the fusion rules used to constrain interactions in the effective field theory. We then apply them to a type-II two Higgs doublet model and identify matter-field assignments that reproduce the Yukawa textures.

\subsection{Fusion rules}

Let us consider a four-dimensional effective field theory in which each field is assigned to a basis element of a certain fusion algebra. 
In particular, quarks, leptons and Higgs fields are assigned to conjugacy classes of a finite group $G$, rather than to irreducible representations of $G$, as motivated by effective field theories of string theory~\cite{Kobayashi:2024yqq,Dong:2025pah,Kobayashi:2025ocp}. 
When we denote a conjugacy class of $G$ by $[g_i]$, 
a matter field $\phi_i$ is assigned to $[g_i]$, and its complex conjugate $\phi_i^\ast$ is naturally associated with the inverse class $[g_i^{-1}]$. For the moment, we omit indices of fields unless they are explicitly required.

The selection rules for interactions among matter fields are constrained by the algebra of conjugacy classes, which defines a commutative fusion algebra, or equivalently, a finite hypergroup. The corresponding multiplication law is written as
\begin{align}
\label{eq:rules}
[g_i]\cdot[g_j]
=
\sum_k N_{ij}^{\phantom{ij}k}[g_k],
\end{align}
where the structure constants $N_{ij}^{\phantom{ij}k}$ are non-negative integers. Obviously, there is the identity $e$ in the group $G$. Its conjugacy class $[e]$ includes only $e$ and satisfies $[g_i]\cdot[e]=[g_i]$ for any $[g_i]$. 
In particular, the product of a conjugacy class and its inverse class necessarily contains the identity class:
\begin{align}
[g_i]\cdot[g_i^{-1}]
=
[e]+\cdots.
\end{align}
Here, we choose the normalization of the conjugacy-class basis such that
\begin{align}
N_{[g_i],[g_i^{-1}]}^{[e]}=1.
\end{align}

This fusion structure determines whether a local interaction is allowed. An operator of the form
\begin{align}
\phi_1\phi_2\cdots\phi_n
\end{align}
is permitted if and only if one can choose representatives
\begin{align}
\widetilde{g}_i\in[g_i]
\end{align}
such that their ordered product is equal to the identity element $e\in G$:
\begin{align}
\widetilde{g}_1\widetilde{g}_2\cdots\widetilde{g}_n=e.
\end{align}
Equivalently, the interaction is allowed when the repeated fusion product
\begin{align}
[g_1]\cdot[g_2]\cdots[g_n]
\end{align}
contains the identity class $[e]$.

To understand these selection rules in more detail, 
let us focus on the $\mathbb{Z}_2$ gauging of a $\mathbb{Z}_5$ symmetry as proposed in Ref.~\cite{Kobayashi:2024cvp}, referred to as $\tilde{\mathbb{Z}}_5$.\footnote{Other gaugings were classified in Ref.~\cite{Dong:2025jra}.} 
When the generator of $\mathbb{Z}_5$ is represented by $g$ with $g^5=e$, the outer automorphism of $\mathbb{Z}_5$, i.e., $\mathbb{Z}_2$ is defined such that 
\begin{align}
r^2 = e, \quad r g^k r^{-1} = g^{-k},
\end{align}
where $r$ denotes the $\mathbb{Z}_2$ generator. 
Then, one can introduce the $\mathbb{Z}_2$-invariant conjugacy classes of the dihedral group $D_5 \cong \mathbb{Z}_5 \rtimes \mathbb{Z}_2$: 
\begin{align}
[g^{(k)}] = \{ r^n g^k r^{-n} \mid n=0,1 \} = \{ g^k, g^{-k} \}.
\end{align}
with $k=0,1,2$. 
Thanks to the $\mathbb{Z}_2$ identification, we have three distinct classes: $\{[g^0], [g^1], [g^2]\}$. 
Specifically, the elements of these classes are given by $[g^0]=\{e\}$, $[g^1]=\{g^1,g^4\}$ and $[g^2]=\{g^2,g^3\}$. 
They obey the following commutative fusion rules:
\begin{align}
    [g^0]\cdot[g^0] &= [g^0]\,,
    \nonumber\\
    [g^0]\cdot [g^i] &= [g^i]\,,
    \nonumber\\
    [g^1]\cdot [g^1] &= [g^0] + [g^2]\,,
    \nonumber\\
    [g^1]\cdot [g^2] &= [g^1] + [g^2]\,,
    \nonumber\\
    [g^2]\cdot [g^2] &= [g^0] + [g^1]\,,
\end{align}
with $i=1,2$. 
These are the fusion rules for $\tilde{\mathbb{Z}}_5$, but a generic fusion rule under $\tilde{\mathbb{Z}}_N$ is described by
\begin{align}
\label{eq:non-invertible-SR}
[g^{(k_1)}] \cdot [g^{(k_2)}] = [g^{(k_1+k_2)}] + [g^{(k_1-k_2)}],
\end{align}
with $k_1,k_2 = 0, 1, ..., \lfloor \frac{N}{2} \rfloor$, 
where $\lfloor \cdot \rfloor$ represents the floor function.

\subsection{Textures based on $\tilde{\mathbb{Z}}_5\times \tilde{\mathbb{Z}}_5$}

Let us apply this fusion rule to the Standard Model. 
To realize realistic Yukawa texture of quarks and leptons, 
we consider a type-II two Higgs doublet model in which matter fields are assigned to basis elements of fusion algebras given by the $\mathbb{Z}_2$ gauging of two $\mathbb{Z}_5$ symmetries, i.e.,  $\tilde{\mathbb{Z}}_5^{(1)}\times \tilde{\mathbb{Z}}_5^{(2)}$. 
By introducing two Higgs doublets $\Phi_u$ and $\Phi_d$ each with a hypercharge of $Y = 1/2$, the Yukawa interactions are given by
\begin{align}
-{\cal L}_{\mathrm{Yukawa}}= Y_u \bar{q}_i \widetilde{\Phi}_u u_j + Y_d \bar{q}_i \Phi_d d_j + Y_e \bar{\ell}_i \Phi_d e_j + Y_{\cal W}\,\frac{\ell\widetilde{\Phi}_u \ell\widetilde{\Phi}_u}{\Lambda} \, ,
\label{eq:WMSSM}
\end{align}
where $\widetilde{\Phi}_u=i\sigma^2 \Phi_u$. 
The neutrino masses are generated by the Weinberg operator.

For illustrative purposes, we focus on the following Yukawa textures:
\begin{align}
Y_u = \begin{pmatrix}
* & 0 & 0 \\
0 & * & 0 \\
0 & 0 & *
\end{pmatrix} \, , ~ 
Y_d = \begin{pmatrix}
0 & * & 0 \\
* & * & * \\
0 & * & *
\end{pmatrix} \, , ~ 
Y_e = \begin{pmatrix}
0 & * & 0 \\
* & * & * \\
0 & * & *
\end{pmatrix} \, , ~ 
Y_{\cal W} = \begin{pmatrix}
* & 0 & 0 \\
0 & * & 0 \\
0 & 0 & *
\end{pmatrix} \, .
\label{eq:Yftexture}
\end{align} 
We systematically search for charge assignments of matter fields under $\tilde{\mathbb{Z}}_5^{(1)}\times \tilde{\mathbb{Z}}_5^{(2)}$. 
It turns out that there are 952 possible charge assignments (including permutations of fields) leading to the above textures. 
These assignments satisfy the following relations: 
\begin{align}
\label{eq:charge}
    [q_i]=[\ell_i], \quad
    [d_i]=[e_i].
\end{align}
Furthermore, we find that the textures of $(\bar{L}L)(\bar{L}L)$ four-fermion operators can be classified into five types as shown in the next section. Hence, we analyze the flavor structure of dimension-six operators by focusing on five representative charge assignments of matter fields in Table~\ref{tab:Z5Z5_tex2_1}. These assignments lead to five distinct textures of $(\bar{L}L)(\bar{L}L)$ four-fermion operators.

\begin{table}[H]
\centering
\caption{Representative charge assignments of the matter fields.}
\label{tab:Z5Z5_tex2_1}
\small
\setlength{\tabcolsep}{4pt}
\renewcommand{\arraystretch}{1.1}
\resizebox{\textwidth}{!}{%
\begin{tabular}{|c|c|c|c|c|c|}
\hline
& $q=\ell$ & $u$ & $d=e$ & $\widetilde{\Phi}_u$ & $\Phi_d$\\
\hline
(a), 1 &
$\{[0]_1[0]_2,\,[1]_1[1]_2,\,[0]_1[1]_2\}$ &
$\{[0]_1[0]_2,\,[1]_1[1]_2,\,[0]_1[1]_2\}$ &
$\{[1]_1[1]_2,\,[2]_1[2]_2,\,[2]_1[1]_2\}$ &
$[0]_1[0]_2$ &
$[2]_1[2]_2$\\
\hline
(b), 300 &
$\{[0]_1[1]_2,\,[1]_1[2]_2,\,[1]_1[1]_2\}$ &
$\{[1]_1[1]_2,\,[0]_1[2]_2,\,[2]_1[1]_2\}$ &
$\{[2]_1[0]_2,\,[2]_1[1]_2,\,[1]_1[1]_2\}$ &
$[1]_1[0]_2$ &
$[2]_1[2]_2$\\
\hline
(c), 351 &
$\{[0]_1[1]_2,\,[2]_1[2]_2,\,[0]_1[2]_2\}$ &
$\{[2]_1[1]_2,\,[1]_1[2]_2,\,[2]_1[2]_2\}$ &
$\{[2]_1[1]_2,\,[1]_1[2]_2,\,[1]_1[1]_2\}$ &
$[2]_1[0]_2$ &
$[1]_1[1]_2$\\
\hline
(d), 651 &
$\{[1]_1[1]_2,\,[1]_1[2]_2,\,[0]_1[2]_2\}$ &
$\{[2]_1[1]_2,\,[0]_1[2]_2,\,[1]_1[2]_2\}$ &
$\{[1]_1[1]_2,\,[2]_1[2]_2,\,[2]_1[1]_2\}$ &
$[1]_1[0]_2$ &
$[2]_1[1]_2$\\
\hline
(e), 951 &
$\{[2]_1[2]_2,\,[2]_1[1]_2,\,[1]_1[1]_2\}$ &
$\{[2]_1[2]_2,\,[2]_1[1]_2,\,[1]_1[1]_2\}$ &
$\{[1]_1[0]_2,\,[2]_1[2]_2,\,[2]_1[0]_2\}$ &
$[0]_1[0]_2$ &
$[1]_1[1]_2$\\
\hline
\end{tabular}%
}
\end{table}

\section{Classification of flavor-dependent dimension-six operators}
\label{sec:classification}

In this section, we count the number of 
independent Wilson coefficients in the so-called Warsaw basis~\cite{Grzadkowski:2010es}. 
In particular, we analyze the four-fermion operators in Sec.~\ref{sec:4fermi} and dimension-six operators including Higgs and fermions in Sec.~\ref{sec:2fermi}.

\subsection{Four-fermion operators}
\label{sec:4fermi}

In this section, we classify the textures of four-fermion operators under the charge assignments in Table~\ref{tab:Z5Z5_tex2_1}.

\subsubsection{$(\bar{L}L)(\bar{L}L)$}

First, let us consider the four-fermion operators in the so-called Warsaw basis~\cite{Grzadkowski:2010es}:
\begin{align}
    C_{ijkl}Q_{\ell\ell} &= C_{ijkl}(\bar{\ell}_L^i\gamma_\mu \ell^j_L)(\bar{\ell}^k_L \gamma^\mu \ell^l_L),
    \nonumber\\
    C_{ijkl}Q_{qq}^{(1)} &= C_{ijkl}(\bar{q}_L^i\gamma_\mu q^j_L)(\bar{q}^k_L \gamma^\mu q^l_L),
    \nonumber\\
    C_{ijkl}Q_{qq}^{(3)} &= C_{ijkl}(\bar{q}_L^i\gamma_\mu \tau^I q^j_L)(\bar{q}^k_L \gamma^\mu \tau^I q^l_L),
    \nonumber\\
    C_{ijkl}Q_{\ell q}^{(1)} &= C_{ijkl}(\bar{\ell}_L^i\gamma_\mu \ell^j_L)(\bar{q}^k_L \gamma^\mu q^l_L),
    \nonumber\\
    C_{ijkl}Q_{\ell q}^{(3)} &= C_{ijkl}(\bar{\ell}_L^i\gamma_\mu \tau^I \ell^j_L)(\bar{q}^k_L \gamma^\mu \tau^I q^l_L).
\end{align}

Among the original 952 possible charge assignments, we find that there are five physically distinct textures realized by five representative assignments of matter fields in Table~\ref{tab:Z5Z5_tex2_1}. Furthermore, all these operators exhibit the same flavor support because of the identical charge assignments of $[q]$ and $[\ell]$. For notational simplicity, we use the common symbol $C_{ijkl}$ to represent their textures, while the nonzero Wilson coefficients of different operators
are understood to be independent unless otherwise specified. 
In the following, we present an explicit texture for the five cases in Table~\ref{tab:Z5Z5_tex2_1}. 
\begin{enumerate}
    \item Assignment (a) yields 21 nonzero entries.

{\small
\setlength{\arraycolsep}{3pt}
\renewcommand{\arraystretch}{0.95}
\begin{align}
C_{ijkl}^{(a)}&= 
\begin{pmatrix}
[0]_1^2 [0]_2^2 & [0]_1 [1]_1 [0]_2 [1]_2 & [0]_1^2 [0]_2 [1]_2 \\
[0]_1 [1]_1 [0]_2 [1]_2 & [1]_1^2 [1]_2^2 & [0]_1 [1]_1 [1]_2^2 \\
[0]_1^2 [0]_2 [1]_2 & [0]_1 [1]_1 [1]_2^2 & [0]_1^2 [1]_2^2
\end{pmatrix}_{\ell\ell}
\begin{pmatrix}
[0]_1^2 [0]_2^2 & [0]_1 [1]_1 [0]_2 [1]_2 & [0]_1^2 [0]_2 [1]_2 \\
[0]_1 [1]_1 [0]_2 [1]_2 & [1]_1^2 [1]_2^2 & [0]_1 [1]_1 [1]_2^2 \\
[0]_1^2 [0]_2 [1]_2 & [0]_1 [1]_1 [1]_2^2 & [0]_1^2 [1]_2^2
\end{pmatrix}_{qq}
\nonumber\\
&=
\begin{array}{cc}
& \begin{array}{ccc} \hspace{0.5em} l=1 \hspace{0.5em} & \hspace{0.5em} l=2 \hspace{0.5em} & \hspace{0.5em} l=3 \hspace{0.5em} \end{array} \\
\begin{array}{c} k=1 \\ \\ \\ k=2 \\ \\ \\ k=3 \end{array} & 
\left( \begin{array}{ccc|ccc|ccc}
\checkmark & 0 & 0 & 0 & \checkmark & 0 & 0 & 0 & \checkmark \\
0 & \checkmark & 0 & \checkmark & 0 & 0 & 0 & 0 & 0 \\
0 & 0 & \checkmark & 0 & 0 & 0 & \checkmark & 0 & 0 \\
\hline
0 & \checkmark & 0 & \checkmark & 0 & 0 & 0 & 0 & 0 \\
\checkmark & 0 & 0 & 0 & \checkmark & 0 & 0 & 0 & \checkmark \\
0 & 0 & 0 & 0 & 0 & \checkmark & 0 & \checkmark & 0 \\
\hline
0 & 0 & \checkmark & 0 & 0 & 0 & \checkmark & 0 & 0 \\
0 & 0 & 0 & 0 & 0 & \checkmark & 0 & \checkmark & 0 \\
\checkmark & 0 & 0 & 0 & \checkmark & 0 & 0 & 0 & \checkmark
\end{array} \right).
\end{array}
\label{eq:4LLLL(a)}
\end{align}
}
Remarkably, we find that $C_{ijkl}^{(a)}$ has the following property:
\begin{align}
    C_{ijkl}^{(a)}= f_{ijkl}\left(\delta_{ij}\delta_{kl} +\delta_{ik}\delta_{jl} + \delta_{il}\delta_{jk}\right),
    \label{CLLL(a)}
\end{align}
where $f_{ijkl}$ is an arbitrary tensor parametrizing the non-vanishing entries.

This texture is simple. 
The reason is as follows.
Every class is self-conjugate. 
That is, $[g^k][g^k]$ always includes $[g^0]$.
By use of this, we contract the index $i$ of $\bar{\ell}_L^i$
with one of $\ell^j_L$, $\bar{\ell}^k_L$, and $\ell^l_L$, and contract the other two indices.
That leads to this texture.
On top of that, this assignment includes only two classes, $[g^0]$ and $[g^1]$.
The multiplication $[g^1][g^1]$ includes $[g^2]$ as well as $[g^0]$.
However, the allowed four-point coupling must correspond to $[g^1][g^1][g^1][g^1]$, because this assignment does not include $[g^2]$. This result is consistent with the above discussion. 
Thus, this assignment leads to the simple texture. 
Similarly, other textures of four-fermion operators can be understood by non-invertible selection rules.

Also, the non-vanishing structure can be understood in terms of a $\mathbb{Z}_2\times\mathbb{Z}_2$ selection rule effectively. Assigning the charges for each generation of fermions under $\mathbb{Z}_2\times\mathbb{Z}_2$: 
\begin{align}
q_1=(1,0),\qquad
q_2=(0,1),\qquad
q_3=(1,1)
\pmod 2,
\end{align}
a component of $C_{ijkl}$ can be non-vanishing only when
\begin{align}
\label{eq:LLLLcasea_cond}
q_i+q_j+q_k+q_l=(0,0)\pmod 2.
\end{align}
This condition is equivalent to requiring each generation index to appear an even number of times.
Defining
\begin{align}
\eta_i^{(a)}
\equiv
(-1)^{q_i^{(a)}},
\qquad
a=1,2,
\end{align}
we have
\begin{align}
\eta^{(1)}
&=
(-1,+1,-1),
&
\eta^{(2)}
&=
(+1,-1,-1).
\end{align}
The Wilson coefficients can then be written as
\begin{align}
C_{ijkl}^{(a)}=
\frac{f_{ijkl}}{4}
\left(
1+
\eta_i^{(1)}
\eta_j^{(1)}
\eta_k^{(1)}
\eta_l^{(1)}
\right)
\left(
1+
\eta_i^{(2)}
\eta_j^{(2)}
\eta_k^{(2)}
\eta_l^{(2)}
\right),
\end{align}
where $f_{ijkl}$ parametrizes the non-vanishing entries. The zero structure is therefore governed by a $\mathbb{Z}_2\times\mathbb{Z}_2$ symmetry generated by
\begin{align}
S_1=\operatorname{diag}(-1,1,-1),
\qquad
S_2=\operatorname{diag}(1,-1,-1).
\end{align}

The zero structure is also invariant under simultaneous permutations of the three generation labels. Indeed, for any $\sigma\in S_3$, the condition~\eqref{eq:LLLLcasea_cond} is preserved under
\begin{align}
(i,j,k,l)
\longrightarrow
\bigl(
\sigma(i),\sigma(j),\sigma(k),\sigma(l)
\bigr).
\end{align}
This can also be seen directly from the charge assignment. The three charges $q_1$, $q_2$, and $q_3$ are precisely the three non-trivial elements of $\mathbb{Z}_2\times\mathbb{Z}_2$, which are permuted by its automorphism group:
\begin{align}
\operatorname{Aut}
\left(
\mathbb{Z}_2\times\mathbb{Z}_2
\right)
\simeq
S_3.
\end{align}
Thus, the $S_3$ symmetry of the zero pattern is the automorphism symmetry of the underlying $\mathbb{Z}_2\times\mathbb{Z}_2$ selection rule.

For a generic $f_{ijkl}$, however, this $S_3$ acts only on the zero structure and is not a symmetry of the full tensor. The $S_3$ symmetry is promoted to a symmetry of $C_{ijkl}$ if
\begin{align}
f_{\sigma(i)\sigma(j)\sigma(k)\sigma(l)}
=f_{ijkl},
\qquad
\sigma\in S_3.
\end{align}
Under this condition, $C_{\sigma(i)\sigma(j)\sigma(k)\sigma(l)}^{(a)}
= C_{ijkl}^{(a)}$, so that the full coefficient tensor is invariant under simultaneous permutations of the three generations.

The resulting symmetry is then larger than $\mathbb{Z}_2\times\mathbb{Z}_2$. Since $S_3$ acts by automorphisms on the three non-trivial elements of $\mathbb{Z}_2\times\mathbb{Z}_2$, the full symmetry is given by the semidirect product
\begin{align}
\left(
\mathbb{Z}_2\times\mathbb{Z}_2
\right)
\rtimes S_3
\simeq
S_4.
\end{align}
Therefore, the zero structure always exhibits a $\mathbb{Z}_2\times\mathbb{Z}_2$ selection rule together with an $S_3$ automorphism, while the symmetry of the full tensor is enhanced to $S_4$ when the non-vanishing coefficients are constant along the corresponding $S_3$ orbits.

    \item Assignment (b) yields 41 nonzero entries.

{\small
\setlength{\arraycolsep}{3pt}
\renewcommand{\arraystretch}{0.95}
\begin{align}
C_{ijkl}^{(b)}&= 
\begin{array}{cc}
& \begin{array}{ccc} \hspace{0.5em} l=1 \hspace{0.5em} & \hspace{0.5em} l=2 \hspace{0.5em} & \hspace{0.5em} l=3 \hspace{0.5em} \end{array} \\
\begin{array}{c} k=1 \\ \\ \\ k=2 \\ \\ \\ k=3 \end{array} & 
\left( \begin{array}{ccc|ccc|ccc}
\checkmark & 0 & 0 & 0 & \checkmark & \checkmark & 0 & \checkmark & \checkmark \\
0 & \checkmark & \checkmark & \checkmark & 0 & 0 & \checkmark & 0 & 0 \\
0 & \checkmark & \checkmark & \checkmark & 0 & 0 & \checkmark & 0 & 0 \\
\hline
0 & \checkmark & \checkmark & \checkmark & 0 & 0 & \checkmark & 0 & 0 \\
\checkmark & 0 & 0 & 0 & \checkmark & \checkmark & 0 & \checkmark & \checkmark \\
\checkmark & 0 & 0 & 0 & \checkmark & \checkmark & 0 & \checkmark & \checkmark \\
\hline
0 & \checkmark & \checkmark & \checkmark & 0 & 0 & \checkmark & 0 & 0 \\
\checkmark & 0 & 0 & 0 & \checkmark & \checkmark & 0 & \checkmark & \checkmark \\
\checkmark & 0 & 0 & 0 & \checkmark & \checkmark & 0 & \checkmark & \checkmark
\end{array} \right)
\end{array}
.
\end{align}
}
This structure can be interpreted as a $\mathbb{Z}_2$ selection rule. 
Assigning the charges for each generation of fermions under $\mathbb{Z}_2$:
\begin{align}
q_1=0,\qquad q_2=q_3=1 \pmod 2,
\end{align}
and defining
\begin{align}
\eta_i\equiv (-1)^{q_i},
\end{align}
i.e., $\eta_1=+1$, $\eta_2=\eta_3=-1$, 
the Wilson coefficients can be written as
\begin{align}
C_{ijkl}^{(b)}=f_{ijkl}\left(\frac{1+\eta_i\eta_j\eta_k\eta_l}{2}\right),    
\end{align}
where $f_{ijkl}$ is an arbitrary tensor parametrizing the non-vanishing entries. 
Hence, $C_{ijkl}^{(b)}$ can be non-vanishing only when
\begin{align}
\eta_i\eta_j\eta_k\eta_l=1,
\end{align}
or equivalently,
\begin{align}
q_i+q_j+q_k+q_l=0\pmod 2.
\end{align}
Thus, the zero structure of $C_{ijkl}^{(b)}$ obeys a $\mathbb{Z}_2$ selection rule generated by
\begin{align}
S=\operatorname{diag}(1,-1,-1).
\end{align}
Since the second and third generations carry the same $\mathbb{Z}_2$ charge, the zero structure is also invariant under the interchange $2\leftrightarrow3$. This does not, however, imply a $2\leftrightarrow3$ symmetry of the full tensor, since the non-vanishing entries parametrized by $f_{ijkl}$ are arbitrary.

Since the second and third generations carry the same $\mathbb{Z}_2$ charge, the zero structure is invariant under their interchange. Denoting the corresponding transformation by
\begin{align}
R_{23}
=\begin{pmatrix}
1&0&0\\
0&0&1\\
0&1&0
\end{pmatrix},
\end{align}
we arrive at the following algebra:
\begin{align}
S^2=R_{23}^2=\mathbf{1},
\qquad
SR_{23}=R_{23}S.
\end{align}
For a generic $f_{ijkl}$, however, $R_{23}$ is a symmetry only of the zero structure. It is promoted to a symmetry of the full coefficient tensor if
\begin{align}
f_{\sigma_{23}(i)\sigma_{23}(j)\sigma_{23}(k)\sigma_{23}(l)}
=f_{ijkl},
\end{align}
where $\sigma_{23}$ denotes the interchange of the second and third generations. Under this condition, $C_{\sigma_{23}(i)\sigma_{23}(j)\sigma_{23}(k)\sigma_{23}(l)}^{(b)}
=C_{ijkl}^{(b)}$, and the symmetry of the full tensor is enhanced to
\begin{align}
\langle S,R_{23}\rangle
\simeq
\mathbb{Z}_2\times\mathbb{Z}_2.
\end{align}
Unlike the previous $\mathbb{Z}_2\times\mathbb{Z}_2$ example, no larger non-Abelian enhancement follows from the zero structure alone, since the first generation is distinguished from the degenerate pair of the second and third generations.

    \item Assignment (c) yields 41 nonzero entries.

{\small
\setlength{\arraycolsep}{3pt}
\renewcommand{\arraystretch}{0.95}
\begin{align}
C_{ijkl}^{(c)}&= 
\begin{array}{cc}
& \begin{array}{ccc} \hspace{0.5em} l=1 \hspace{0.5em} & \hspace{0.5em} l=2 \hspace{0.5em} & \hspace{0.5em} l=3 \hspace{0.5em} \end{array} \\
\begin{array}{c} k=1 \\ \\ \\ k=2 \\ \\ \\ k=3 \end{array} & 
\left( \begin{array}{ccc|ccc|ccc}
\checkmark & 0 & \checkmark & 0 & \checkmark & 0 & \checkmark & 0 & \checkmark \\
0 & \checkmark & 0 & \checkmark & 0 & \checkmark & 0 & \checkmark & 0 \\
\checkmark & 0 & \checkmark & 0 & \checkmark & 0 & \checkmark & 0 & \checkmark \\
\hline
0 & \checkmark & 0 & \checkmark & 0 & \checkmark & 0 & \checkmark & 0 \\
\checkmark & 0 & \checkmark & 0 & \checkmark & 0 & \checkmark & 0 & \checkmark \\
0 & \checkmark & 0 & \checkmark & 0 & \checkmark & 0 & \checkmark & 0 \\
\hline
\checkmark & 0 & \checkmark & 0 & \checkmark & 0 & \checkmark & 0 & \checkmark \\
0 & \checkmark & 0 & \checkmark & 0 & \checkmark & 0 & \checkmark & 0 \\
\checkmark & 0 & \checkmark & 0 & \checkmark & 0 & \checkmark & 0 & \checkmark
\end{array} \right),
\end{array}
\end{align}
}
suggesting the following property. 
Denoting
\[
P_{ij}=
\begin{cases}
1 & (i+j\ \mathrm{even}),\\
0 & (i+j\ \mathrm{odd}),
\end{cases}
\qquad
\mathrm{i.e.,}
\qquad
P=
\begin{pmatrix}
1&0&1\\
0&1&0\\
1&0&1
\end{pmatrix},
\]
the non-vanishing pattern is determined by the parity of \(i+j\). In particular, 
we find
\[
C_{ijkl}^{(c)}\neq 0 \quad \Longleftrightarrow \quad P_{ij}=P_{kl},
\]
i.e.,
\[
C_{ijkl}^{(c)}
=
f_{ijkl}\left[
P_{ij}P_{kl}+(1-P_{ij})(1-P_{kl})
\right],
\]
where $f_{ijkl}$ is an arbitrary tensor parametrizing the non-vanishing entries. Since
\begin{align}
P_{ij}=\frac{1+(-1)^{i+j}}{2},
\end{align}
this expression can equivalently be written as
\begin{align}
C_{ijkl}^{(c)}= f_{ijkl}
\left(\frac{1+(-1)^{i+j+k+l}}{2}\right).
\end{align}
Thus, a component of $C_{ijkl}^{(c)}$ can be non-vanishing only when
\begin{align}
i+j+k+l=0\pmod 2.
\end{align}
This structure can be understood as a $\mathbb{Z}_2$ selection rule. Defining
\begin{align}
\eta_i\equiv (-1)^{i+1},
\end{align}
i.e., $\eta_1=+1$, $\eta_2=-1$ and $\eta_3=+1$, the coefficient tensor can then be written as
\begin{align}
C_{ijkl}^{(c)}=f_{ijkl}
\left(\frac{1+\eta_i\eta_j\eta_k\eta_l}{2}\right).
\end{align}
The zero structure is therefore governed by the $\mathbb{Z}_2$ transformation generated by
\begin{align}
S=\operatorname{diag}(1,-1,1).
\end{align}

Since the first and third generations carry the same $\mathbb{Z}_2$ charge, the zero structure is invariant under their interchange. Denoting the corresponding transformation by
\begin{align}
R_{13}=
\begin{pmatrix}
0&0&1\\
0&1&0\\
1&0&0
\end{pmatrix},
\end{align}
we have
\begin{align}
S^2=R_{13}^2=\mathbf{1},
\qquad
SR_{13}=R_{13}S.
\end{align}
For a generic $f_{ijkl}$, however, the $1\leftrightarrow3$ interchange is a symmetry only of the zero structure and not necessarily of the full coefficient tensor. It is promoted to a symmetry of $C_{ijkl}^{(c)}$ if
\begin{align}
f_{\sigma_{13}(i)\sigma_{13}(j)\sigma_{13}(k)\sigma_{13}(l)}
=f_{ijkl},
\end{align}
where $\sigma_{13}$ denotes the interchange of the first and third generations. Under this condition, we obtain $C_{\sigma_{13}(i)\sigma_{13}(j)\sigma_{13}(k)\sigma_{13}(l)}^{(c)}=C_{ijkl}^{(c)}$, and the symmetry of the full tensor is enhanced to
\begin{align}
\langle S,R_{13}\rangle
\simeq
\mathbb{Z}_2\times\mathbb{Z}_2.
\end{align}
Thus, for a generic $f_{ijkl}$, the zero structure exhibits the original $\mathbb{Z}_2$ selection rule together with a $1\leftrightarrow3$ exchange symmetry. When the non-vanishing coefficients are also invariant under this interchange, the symmetry of the full tensor is enhanced from $\mathbb{Z}_2$ to $\mathbb{Z}_2\times\mathbb{Z}_2$.

    \item Assignment (d) yields 41 nonzero entries.

{\small
\setlength{\arraycolsep}{3pt}
\renewcommand{\arraystretch}{0.95}
\begin{align}
C_{ijkl}^{(d)}&= 
\begin{array}{cc}
& \begin{array}{ccc} \hspace{0.5em} l=1 \hspace{0.5em} & \hspace{0.5em} l=2 \hspace{0.5em} & \hspace{0.5em} l=3 \hspace{0.5em} \end{array} \\
\begin{array}{c} k=1 \\ \\ \\ k=2 \\ \\ \\ k=3 \end{array} & 
\left( \begin{array}{ccc|ccc|ccc}
\checkmark & \checkmark & 0 & \checkmark & \checkmark & 0 & 0 & 0 & \checkmark \\
\checkmark & \checkmark & 0 & \checkmark & \checkmark & 0 & 0 & 0 & \checkmark \\
0 & 0 & \checkmark & 0 & 0 & \checkmark & \checkmark & \checkmark & 0 \\
\hline
\checkmark & \checkmark & 0 & \checkmark & \checkmark & 0 & 0 & 0 & \checkmark \\
\checkmark & \checkmark & 0 & \checkmark & \checkmark & 0 & 0 & 0 & \checkmark \\
0 & 0 & \checkmark & 0 & 0 & \checkmark & \checkmark & \checkmark & 0 \\
\hline
0 & 0 & \checkmark & 0 & 0 & \checkmark & \checkmark & \checkmark & 0 \\
0 & 0 & \checkmark & 0 & 0 & \checkmark & \checkmark & \checkmark & 0 \\
\checkmark & \checkmark & 0 & \checkmark & \checkmark & 0 & 0 & 0 & \checkmark
\end{array} \right)
\end{array}
\end{align}
}
The non-vanishing entries of $C_{ijkl}^{(d)}$ exhibit the following structure. 
Defining
\begin{equation}
P=
\begin{pmatrix}
1 & 1 & 0 \\
1 & 1 & 0 \\
0 & 0 & 1
\end{pmatrix},
\end{equation}
the selection rule can be expressed as
\begin{equation}
C_{ijkl}^{(d)} = f_{ijkl}\left\{ P_{ij}P_{kl} + (1 - P_{ij})(1 - P_{kl}) \right\},
\end{equation}
where $f_{ijkl}$ is an arbitrary tensor parametrizing the values of the non-vanishing entries. Equivalently, a component of $C_{ijkl}$ can be non-vanishing only when
\begin{align}
P_{ij}=P_{kl}.
\end{align}

This structure can be understood as a $\mathbb{Z}_2$ selection rule. 
Assigning the charges for each generation of fermions:
\begin{align}
q_1=q_2=0,\qquad q_3=1\pmod 2,
\end{align}
and defining
\begin{align}
\eta_i\equiv (-1)^{q_i},
\end{align}
we have
\begin{align}
\eta_1=\eta_2=+1,\qquad \eta_3=-1.
\end{align}
The binary matrix $P$ can then be written as
\begin{align}
P_{ij}=\frac{1+\eta_i\eta_j}{2}.
\end{align}
Substituting this expression into the Wilson coefficients, we obtain
\begin{align}
C_{ijkl}^{(d)}=f_{ijkl}
\left(\frac{1+\eta_i\eta_j\eta_k\eta_l}{2}\right).
\end{align}
Thus, $C_{ijkl}^{(d)}$ can be non-vanishing only when
\begin{align}
\eta_i\eta_j\eta_k\eta_l=1,
\end{align}
or equivalently,
\begin{align}
q_i+q_j+q_k+q_l=0\pmod 2.
\end{align}
The zero structure of $C_{ijkl}^{(d)}$ therefore obeys the $\mathbb{Z}_2$ selection rule generated by
\begin{align}
S=\operatorname{diag}(1,1,-1).
\end{align}

Since the first and second generations carry the same $\mathbb{Z}_2$ charge, the zero structure is invariant under their interchange. Denoting the corresponding transformation by
\begin{align}
R_{12}=
\begin{pmatrix}
0&1&0\\
1&0&0\\
0&0&1
\end{pmatrix},
\end{align}
we obtain
\begin{align}
S^2=R_{12}^2=\mathbf{1},
\qquad
SR_{12}=R_{12}S.
\end{align}
For a generic $f_{ijkl}$, however, the $1\leftrightarrow2$ interchange is a symmetry only of the zero structure. It becomes a symmetry of the full coefficient tensor if
\begin{align}
f_{\sigma_{12}(i)\sigma_{12}(j)\sigma_{12}(k)\sigma_{12}(l)}
=f_{ijkl},
\end{align}
where $\sigma_{12}$ denotes the interchange of the first and second generations. Under this condition, we obtain 
$C_{\sigma_{12}(i)\sigma_{12}(j)\sigma_{12}(k)\sigma_{12}(l)}^{(d)}
=C_{ijkl}^{(d)}$, and the symmetry of the full tensor is enhanced to
\begin{align}
\langle S,R_{12}\rangle
\simeq
\mathbb{Z}_2\times\mathbb{Z}_2.
\end{align}
Thus, the zero structure exhibits the original $\mathbb{Z}_2$ selection rule together with a $1\leftrightarrow2$ exchange symmetry. When the non-vanishing coefficients are also invariant under this interchange, the symmetry of the full tensor is enhanced from $\mathbb{Z}_2$ to $\mathbb{Z}_2\times\mathbb{Z}_2$. 

Note that the Wilson coefficients in cases (b)--(d) exhibit the following structure:
\begin{align}
C^{(b)}
\sim
C^{(c)}
\sim
C^{(d)},
\end{align}
where $\sim$ denotes equivalence under the same permutation acting on all four indices,
\begin{align}
(i,j,k,l)
\longrightarrow
\bigl(
\sigma(i),\sigma(j),\sigma(k),\sigma(l)
\bigr),
\qquad
\sigma\in S_3.
\end{align}
More explicitly, $C^{(c)}$ is obtained from $C^{(b)}$ by the interchange $1\leftrightarrow2$, while $C^{(d)}$ is obtained from $C^{(b)}$ by the interchange $1\leftrightarrow3$. Equivalently, $C^{(c)}$ and $C^{(d)}$ are related by $2\leftrightarrow3$.

    \item All entries are allowed for assignment (e).

\end{enumerate}

\subsubsection{$(\bar{L}L)(\bar{R}R)$}

In this section, we deal with the four-fermion operators:
\begin{align}
    D_{ijkl}Q_{\ell e} &= D_{ijkl}(\bar{\ell}_L^i\gamma_\mu \ell^j_L)(\bar{e}^k_R \gamma^\mu e^l_R),
    \nonumber\\
    D_{ijkl}Q_{\ell d} &= D_{ijkl}(\bar{\ell}_L^i\gamma_\mu \ell^j_L)(\bar{d}^k_R \gamma^\mu d^l_R),
    \nonumber\\
   D_{ijkl} Q_{q e} &= D_{ijkl}(\bar{q}_L^i\gamma_\mu q^j_L)(\bar{e}^k_R \gamma^\mu e^l_R),
    \nonumber\\
    E_{ijkl}Q_{\ell u} &= E_{ijkl}(\bar{\ell}_L^i\gamma_\mu \ell^j_L)(\bar{u}^k_R \gamma^\mu u^l_R),
    \nonumber\\
    E_{ijkl}Q_{qu}^{(1)} &= E_{ijkl}(\bar{q}_L^i\gamma_\mu q^j_L)(\bar{u}^k_R \gamma^\mu u^l_R),
    \nonumber\\
    E_{ijkl}Q_{qu}^{(8)} &= E_{ijkl}(\bar{q}_L^i\gamma_\mu T^A q^j_L)(\bar{u}^k_R \gamma^\mu T^A u^l_R),
    \nonumber\\
    D_{ijkl}Q_{qd}^{(1)} &= D_{ijkl}(\bar{q}_L^i\gamma_\mu q^j_L)(\bar{d}^k_R \gamma^\mu d^l_R),
    \nonumber\\
    D_{ijkl}Q_{qd}^{(8)} &= D_{ijkl}(\bar{q}_L^i\gamma_\mu T^A q^j_L)(\bar{d}^k_R \gamma^\mu T^A d^l_R).
\end{align}
Note that some operators exhibit the same textures due to the same charge assignments of $q=\ell$ and $d=e$. 
We therefore denote the Wilson coefficients by the same symbols $D_{ijkl}$ and $E_{ijkl}$. 

Among the 952 possibilities, we find 8 and 34 physically distinct textures 
for $D_{ijkl}$ and $E_{ijkl}$, respectively. 
Let us focus on five representative assignments of matter fields listed in Table~\ref{tab:Z5Z5_tex2_1}.

\begin{enumerate}
    \item Assignment (a):

{\small
\setlength{\arraycolsep}{3pt}
\renewcommand{\arraystretch}{0.95}    
\begin{align}
D_{ijkl}^{(a)}&= 
\left( \begin{array}{ccc|ccc|ccc}
\checkmark & 0 & 0 & 0 & \checkmark & 0 & 0 & 0 & 0 \\
0 & \checkmark & 0 & \checkmark & \checkmark & \checkmark & 0 & \checkmark & \checkmark \\
0 & 0 & \checkmark & 0 & \checkmark & 0 & 0 & \checkmark & 0 \\
\hline
0 & \checkmark & 0 & \checkmark & \checkmark & \checkmark & 0 & \checkmark & \checkmark \\
\checkmark & \checkmark & \checkmark & \checkmark & \checkmark & \checkmark & \checkmark & \checkmark & \checkmark \\
0 & \checkmark & 0 & \checkmark & \checkmark & \checkmark & \checkmark & \checkmark & \checkmark \\
\hline
0 & 0 & 0 & 0 & \checkmark & \checkmark & \checkmark & 0 & 0 \\
0 & \checkmark & \checkmark & \checkmark & \checkmark & \checkmark & 0 & \checkmark & \checkmark \\
0 & \checkmark & 0 & \checkmark & \checkmark & \checkmark & 0 & \checkmark & \checkmark
\end{array} \right)
,\quad
\operatorname{supp}(E_{ijkl}^{(a)})
=
\operatorname{supp}(C_{ijkl}^{(a)}).
\label{Dijkl}
\end{align}
}
Throughout this paper, the rows and columns of the $9\times 9$ matrix are ordered as in the $(\bar{L}L)(\bar{L}L)$ case unless otherwise specified. 
The coefficient $E_{ijkl}^{(a)}$ has the same texture as $C_{ijkl}^{(a)}$, as shown in Eq.~\eqref{eq:LLLLcasea_cond} and therefore obeys the same selection rule. The texture of $D_{ijkl}^{(a)}$, on the other hand, is more involved.

The texture of $D_{ijkl}^{(a)}$ can be organized as a $3\times3$ block matrix:
\begin{align}
D_{ijkl}^{(a)}
=f_{ijkl}^{(a)}P_{ijkl}^{D(a)},
\end{align}
where $f_{ijkl}^{(a)}$ denotes an independent coefficient for each allowed component. The support tensor $P_{ijkl}^{D(a)}$ is represented by
\begin{align}
P^{D(a)}
=
\begin{pmatrix}
A_D^{(1)} & A_D^{(2)} & A_D^{(3)} \\
A_D^{(2)} & A_D^{(4)} & A_D^{(5)} \\
A_D^{(3)} & A_D^{(5)} & A_D^{(6)}
\end{pmatrix}_{(k,l)},
\end{align}
where the block rows and columns are labeled by $k$ and $l$, respectively, while each block is a $3\times3$ matrix in the indices $i$ and $j$, i.e.,
\begin{align}
A_D^{(1)}
&=
\begin{pmatrix}
1 & 0 & 0 \\
0 & 1 & 0 \\
0 & 0 & 1
\end{pmatrix},
\quad
A_D^{(2)}
=
\begin{pmatrix}
0 & 1 & 0 \\
1 & 1 & 1 \\
0 & 1 & 0
\end{pmatrix},
\quad
A_D^{(3)}
=
\begin{pmatrix}
0 & 0 & 0 \\
0 & 1 & 1 \\
0 & 1 & 0
\end{pmatrix},
\nonumber\\
A_D^{(4)}
&=
\begin{pmatrix}
1 & 1 & 1 \\
1 & 1 & 1 \\
1 & 1 & 1
\end{pmatrix},
\quad
A_D^{(5)}
=
\begin{pmatrix}
0 & 1 & 1 \\
1 & 1 & 1 \\
1 & 1 & 1
\end{pmatrix},
\quad
A_D^{(6)}
=
\begin{pmatrix}
1 & 0 & 0 \\
0 & 1 & 1 \\
0 & 1 & 1
\end{pmatrix}.
\end{align}
Note that when the coefficients satisfy $f_{ijkl}^{(a)}=f_{ijlk}^{(a)}$, the full tensor is symmetric under the exchange $k\leftrightarrow l$.

    \item Assignment (b):

{\small
\setlength{\arraycolsep}{3pt}
\renewcommand{\arraystretch}{0.95}    
\begin{align}
D_{ijkl}^{(b)}&= 
\left( \begin{array}{ccc|ccc|ccc}
\checkmark & 0 & \checkmark & 0 & \checkmark & 0 & 0 & \checkmark & 0 \\
0 & \checkmark & 0 & \checkmark & \checkmark & \checkmark & \checkmark & \checkmark & \checkmark \\
\checkmark & 0 & \checkmark & 0 & \checkmark & 0 & 0 & \checkmark & 0 \\
\hline
0 & \checkmark & 0 & \checkmark & \checkmark & \checkmark & 0 & \checkmark & \checkmark \\
\checkmark & \checkmark & \checkmark & \checkmark & \checkmark & \checkmark & \checkmark & \checkmark & \checkmark \\
0 & \checkmark & 0 & \checkmark & \checkmark & \checkmark & \checkmark & \checkmark & \checkmark \\
\hline
0 & \checkmark & 0 & 0 & \checkmark & \checkmark & \checkmark & 0 & 0 \\
\checkmark & \checkmark & \checkmark & \checkmark & \checkmark & \checkmark & 0 & \checkmark & \checkmark \\
0 & \checkmark & 0 & \checkmark & \checkmark & \checkmark & 0 & \checkmark & \checkmark
\end{array} \right)
,\quad
E_{ijkl}^{(b)}= 
\left( \begin{array}{ccc|ccc|ccc}
\checkmark & 0 & 0
& 0 & \checkmark & \checkmark
& 0 & \checkmark & \checkmark
\\
0 & \checkmark & \checkmark
& \checkmark & 0 & 0
& \checkmark & \checkmark & \checkmark
\\
0 & \checkmark & \checkmark
& \checkmark & 0 & 0
& \checkmark & \checkmark & \checkmark
\\
\hline
0 & \checkmark & \checkmark
& \checkmark & 0 & 0
& 0 & 0 & 0
\\
\checkmark & 0 & 0
& 0 & \checkmark & \checkmark
& 0 & \checkmark & \checkmark
\\
\checkmark & 0 & 0
& 0 & \checkmark & \checkmark
& 0 & \checkmark & \checkmark
\\
\hline
0 & \checkmark & \checkmark
& 0 & 0 & 0
& \checkmark & \checkmark & \checkmark
\\
\checkmark & \checkmark & \checkmark
& 0 & \checkmark & \checkmark
& \checkmark & \checkmark & \checkmark
\\
\checkmark & \checkmark & \checkmark
& 0 & \checkmark & \checkmark
& \checkmark & \checkmark & \checkmark
\end{array} \right).
\end{align}
}

The textures of $D_{ijkl}^{(b)}$ and $E_{ijkl}^{(b)}$ can be organized as $3\times3$ block matrices. Let us introduce the support tensors $P_{ijkl}^{D(b)}$ and $P_{ijkl}^{E(b)}$ through
\begin{align}
D_{ijkl}^{(b)}
&=
d_{ijkl}^{(b)}P_{ijkl}^{D(b)},
&
E_{ijkl}^{(b)}
&=
e_{ijkl}^{(b)}P_{ijkl}^{E(b)},
\end{align}
where $d_{ijkl}^{(b)}$ and $e_{ijkl}^{(b)}$ denote independent coefficients for the allowed components.
For $D_{ijkl}^{(b)}$, the support tensor takes the form
\begin{align}
P^{D(b)}
=
\begin{pmatrix}
A_D^{(1)} & A_D^{(2)} & A_D^{(2)} \\
A_D^{(2)} & \mathbf{J} & A_D^{(3)} \\
A_D^{(2)} & A_D^{(3)} & A_D^{(4)}
\end{pmatrix}_{(k,l)},
\end{align}
where the block rows and columns are labeled by $k$ and $l$, respectively, while each block is a $3\times3$ matrix in the indices $i$ and $j$. 
Here and in what follows, $\mathbf{J}$ denotes the $3\times3$ matrix whose entries are all equal to one,
i.e.,
\begin{align}
    \mathbf{J}=
    \begin{pmatrix}
        1 & 1 & 1\\
         1 & 1 & 1\\
          1 & 1 & 1\\
    \end{pmatrix},
\end{align}
and
\begin{align}
A_D^{(1)}
&=
\begin{pmatrix}
1 & 0 & 1 \\
0 & 1 & 0 \\
1 & 0 & 1
\end{pmatrix},
\quad
A_D^{(2)}
=
\begin{pmatrix}
0 & 1 & 0 \\
1 & 1 & 1 \\
0 & 1 & 0
\end{pmatrix},
\quad
A_D^{(3)}
=
\begin{pmatrix}
0 & 1 & 1 \\
1 & 1 & 1 \\
1 & 1 & 1
\end{pmatrix},
\quad
A_D^{(4)}
=
\begin{pmatrix}
1 & 0 & 0 \\
0 & 1 & 1 \\
0 & 1 & 1
\end{pmatrix}.
\end{align}

A unit entry in $P^{D(b)}$ specifies an allowed component of $D_{ijkl}^{(b)}$, whereas a vanishing entry denotes a forbidden component. The equality of the blocks at $(k,l)$ and $(l,k)$ shows that the support is invariant under $k\leftrightarrow l$. Each block is also symmetric in $i$ and $j$, and hence
\begin{align}
P_{ijkl}^{D(b)}
=P_{jikl}^{D(b)}
=P_{ijlk}^{D(b)}.
\end{align}

For $E_{ijkl}^{(b)}$, the support tensor is given by
\begin{align}
P^{E(b)}
=
\begin{pmatrix}
A_E^{(1)} & A_E^{(2)} & A_E^{(3)} \\
A_E^{(2)} & A_E^{(1)} & A_E^{(4)} \\
A_E^{(3)} & A_E^{(4)} & \mathbf{J}
\end{pmatrix}_{(k,l)},
\end{align}
with
\begin{align}
A_E^{(1)}
&=
\begin{pmatrix}
1 & 0 & 0 \\
0 & 1 & 1 \\
0 & 1 & 1
\end{pmatrix},
\quad
A_E^{(2)}
=
\begin{pmatrix}
0 & 1 & 1 \\
1 & 0 & 0 \\
1 & 0 & 0
\end{pmatrix},
\quad
A_E^{(3)}
=
\begin{pmatrix}
0 & 1 & 1 \\
1 & 1 & 1 \\
1 & 1 & 1
\end{pmatrix},
\quad
A_E^{(4)}
=
\begin{pmatrix}
0 & 0 & 0 \\
0 & 1 & 1 \\
0 & 1 & 1
\end{pmatrix}.
\end{align}

As in the case of $D_{ijkl}^{(b)}$, the block structure of $P^{E(b)}$ is symmetric under $k\leftrightarrow l$, and each block is symmetric under $i\leftrightarrow j$. Therefore,
\begin{align}
P_{ijkl}^{E(b)}
=P_{jikl}^{E(b)}
=P_{ijlk}^{E(b)}.
\end{align}

These relations refer to the support of the tensors. The full coefficients possess the same exchange symmetries only when
\begin{align}
d_{ijkl}^{(b)}
&=
d_{jikl}^{(b)}
=d_{ijlk}^{(b)},
\quad
e_{ijkl}^{(b)}
=
e_{jikl}^{(b)}
=
e_{ijlk}^{(b)}.
\end{align}

    \item Assignment (c):

{\small
\setlength{\arraycolsep}{3pt}
\renewcommand{\arraystretch}{0.95}    
\begin{align}
D_{ijkl}^{(c)}&= 
\left( \begin{array}{ccc|ccc|ccc}
\checkmark & 0 & \checkmark & 0 & \checkmark & 0 & 0 & \checkmark & 0 \\
0 & \checkmark & 0 & \checkmark & \checkmark & \checkmark & \checkmark & \checkmark & \checkmark \\
\checkmark & 0 & \checkmark & 0 & \checkmark & 0 & 0 & \checkmark & 0 \\
\hline
0 & \checkmark & 0 & \checkmark & \checkmark & \checkmark & \checkmark & \checkmark & \checkmark \\
\checkmark & \checkmark & \checkmark & \checkmark & \checkmark & \checkmark & \checkmark & \checkmark & \checkmark \\
0 & \checkmark & 0 & \checkmark & \checkmark & \checkmark & \checkmark & \checkmark & \checkmark \\
\hline
0 & \checkmark & 0 & \checkmark & \checkmark & \checkmark & \checkmark & \checkmark & \checkmark \\
\checkmark & \checkmark & \checkmark & \checkmark & \checkmark & \checkmark & \checkmark & \checkmark & \checkmark \\
0 & \checkmark & 0 & \checkmark & \checkmark & \checkmark & \checkmark & \checkmark & \checkmark
\end{array} \right)
,\quad
E_{ijkl}^{(c)}= 
\left( \begin{array}{ccc|ccc|ccc}
\checkmark & 0 & \checkmark
& 0 & \checkmark & 0
& \checkmark & 0 & \checkmark
\\
0 & \checkmark & 0
& \checkmark & \checkmark & \checkmark
& 0 & \checkmark & 0
\\
\checkmark & 0 & \checkmark
& 0 & \checkmark & 0
& \checkmark & 0 & \checkmark
\\
\hline
0 & \checkmark & 0
& \checkmark & \checkmark & \checkmark
& 0 & \checkmark & 0
\\
\checkmark & \checkmark & \checkmark
& \checkmark & \checkmark & \checkmark
& \checkmark & \checkmark & \checkmark
\\
0 & \checkmark & 0
& \checkmark & \checkmark & \checkmark
& 0 & \checkmark & 0
\\
\hline
\checkmark & 0 & \checkmark
& 0 & \checkmark & 0
& \checkmark & 0 & \checkmark
\\
0 & \checkmark & 0
& \checkmark & \checkmark & \checkmark
& 0 & \checkmark & 0
\\
\checkmark & 0 & \checkmark
& 0 & \checkmark & 0
& \checkmark & 0 & \checkmark
\end{array} \right).
\end{align}
}

The textures of $D_{ijkl}^{(c)}$ and $E_{ijkl}^{(c)}$ can be organized as $3\times3$ block matrices. We introduce the support tensors $P_{ijkl}^{D(c)}$ and $P_{ijkl}^{E(c)}$ through
\begin{align}
D_{ijkl}^{(c)}
&=
d_{ijkl}^{(c)}P_{ijkl}^{D(c)},
&
E_{ijkl}^{(c)}
&=
e_{ijkl}^{(c)}P_{ijkl}^{E(c)},
\end{align}
where $d_{ijkl}^{(c)}$ and $e_{ijkl}^{(c)}$ denote independent coefficients for the allowed components.

For $D_{ijkl}^{(c)}$, the support tensor takes the form
\begin{align}
P^{D(c)}
=
\begin{pmatrix}
A_D^{(1)} & A_D^{(2)} & A_D^{(2)} \\
A_D^{(2)} & \mathbf{J} & \mathbf{J} \\
A_D^{(2)} & \mathbf{J} & \mathbf{J}
\end{pmatrix}_{(k,l)},
\end{align}
where the block rows and columns are labeled by $k$ and $l$, respectively, while each block is a $3\times3$ matrix in the indices $i$ and $j$. Here, we define
\begin{align}
A_D^{(1)}
&=
\begin{pmatrix}
1 & 0 & 1 \\
0 & 1 & 0 \\
1 & 0 & 1
\end{pmatrix},
\qquad
A_D^{(2)}
=
\begin{pmatrix}
0 & 1 & 0 \\
1 & 1 & 1 \\
0 & 1 & 0
\end{pmatrix}.
\end{align}

A unit entry in $P^{D(c)}$ specifies an allowed component of $D_{ijkl}^{(c)}$, whereas a vanishing entry denotes a forbidden component. The equality of the blocks at $(k,l)$ and $(l,k)$ shows that the support is invariant under $k\leftrightarrow l$. Since both $A_D^{(1)}$ and $A_D^{(2)}$ are symmetric matrices, the support is also invariant under $i\leftrightarrow j$. Thus,
\begin{align}
P_{ijkl}^{D(c)}
=P_{jikl}^{D(c)}
=P_{ijlk}^{D(c)}.
\end{align}

For $E_{ijkl}^{(c)}$, the support tensor is given by
\begin{align}
P^{E(c)}
=
\begin{pmatrix}
A_E^{(1)} & A_E^{(2)} & A_E^{(1)} \\
A_E^{(2)} & \mathbf{J} & A_E^{(2)} \\
A_E^{(1)} & A_E^{(2)} & A_E^{(1)}
\end{pmatrix}_{(k,l)},
\end{align}
with
\begin{align}
A_E^{(1)}
&=
\begin{pmatrix}
1 & 0 & 1 \\
0 & 1 & 0 \\
1 & 0 & 1
\end{pmatrix},
\qquad
A_E^{(2)}
=
\begin{pmatrix}
0 & 1 & 0 \\
1 & 1 & 1 \\
0 & 1 & 0
\end{pmatrix}.
\end{align}

As in the case of $D_{ijkl}^{(c)}$, the block structure is symmetric under $k\leftrightarrow l$, and each block is symmetric under $i\leftrightarrow j$. Therefore,
\begin{align}
P_{ijkl}^{E(c)}
=P_{jikl}^{E(c)}
=P_{ijlk}^{E(c)}.
\end{align}

The support of $E_{ijkl}^{(c)}$ admits a simple characterization. Let
\begin{align}
N_2
=
\delta_{i2}
+
\delta_{j2}
+
\delta_{k2}
+
\delta_{l2}
\end{align}
denote the number of indices equal to $2$. The only forbidden components are those for which exactly one of the four indices is equal to $2$. Hence,
\begin{align}
P_{ijkl}^{E(c)}
=
\begin{cases}
0, & N_2=1, \\
1, & N_2=0,2,3,4.
\end{cases}
\end{align}
Equivalently, the support tensor can be written as
\begin{align}
P_{ijkl}^{E(c)}
=1- \frac{1}{6}
N_2(2-N_2)(3-N_2)(4-N_2).
\end{align}
Since this expression depends only on $N_2$, the support is invariant under arbitrary permutations of the four index positions. In particular, it is invariant under $i\leftrightarrow j$, $k\leftrightarrow l$, and the pair exchange $(i,j)\leftrightarrow(k,l)$.
These relations refer to the support of the tensors. The full coefficient $D_{ijkl}^{(c)}$ possesses the corresponding exchange symmetries only when
\begin{align}
d_{ijkl}^{(c)}
=d_{jikl}^{(c)}
=d_{ijlk}^{(c)}.
\end{align}

Similarly, the full coefficient $E_{ijkl}^{(c)}$ possesses the full index-position permutation symmetry only when $e_{ijkl}^{(c)}$ is invariant under the corresponding permutations of $i$, $j$, $k$, and $l$.

    \item Assignment (d):

{\small
\setlength{\arraycolsep}{3pt}
\renewcommand{\arraystretch}{0.95}    
\begin{align}
D_{ijkl}^{(d)}&= 
\left( \begin{array}{ccc|ccc|ccc}
\checkmark & \checkmark & 0 & \checkmark & \checkmark & \checkmark & \checkmark & \checkmark & \checkmark \\
\checkmark & \checkmark & 0 & \checkmark & \checkmark & \checkmark & \checkmark & \checkmark & \checkmark \\
0 & 0 & \checkmark & \checkmark & \checkmark & 0 & \checkmark & \checkmark & 0 \\
\hline
\checkmark & \checkmark & \checkmark & \checkmark & \checkmark & \checkmark & \checkmark & \checkmark & \checkmark \\
\checkmark & \checkmark & \checkmark & \checkmark & \checkmark & \checkmark & \checkmark & \checkmark & \checkmark \\
\checkmark & \checkmark & 0 & \checkmark & \checkmark & \checkmark & \checkmark & \checkmark & \checkmark \\
\hline
\checkmark & \checkmark & \checkmark & \checkmark & \checkmark & \checkmark & \checkmark & \checkmark & \checkmark \\
\checkmark & \checkmark & \checkmark & \checkmark & \checkmark & \checkmark & \checkmark & \checkmark & \checkmark \\
\checkmark & \checkmark & 0 & \checkmark & \checkmark & \checkmark & \checkmark & \checkmark & \checkmark
\end{array} \right)
,\quad
E_{ijkl}^{(d)}&= 
\left( \begin{array}{ccc|ccc|ccc}
\checkmark & \checkmark & \checkmark
& \checkmark & \checkmark & 0
& \checkmark & \checkmark & \checkmark
\\
\checkmark & \checkmark & \checkmark
& \checkmark & \checkmark & 0
& \checkmark & \checkmark & \checkmark
\\
\checkmark & \checkmark & \checkmark
& 0 & 0 & 0
& \checkmark & \checkmark & 0
\\
\hline
\checkmark & \checkmark & 0
& \checkmark & \checkmark & 0
& 0 & 0 & \checkmark
\\
\checkmark & \checkmark & 0
& \checkmark & \checkmark & 0
& 0 & 0 & \checkmark
\\
0 & 0 & 0
& 0 & 0 & \checkmark
& \checkmark & \checkmark & 0
\\
\hline
\checkmark & \checkmark & \checkmark
& 0 & 0 & \checkmark
& \checkmark & \checkmark & 0
\\
\checkmark & \checkmark & \checkmark
& 0 & 0 & \checkmark
& \checkmark & \checkmark & 0
\\
\checkmark & \checkmark & 0
& \checkmark & \checkmark & 0
& 0 & 0 & \checkmark
\end{array} \right).
\end{align}
}

The textures of $D_{ijkl}^{(d)}$ and $E_{ijkl}^{(d)}$ can be organized as $3\times3$ block matrices. Let us introduce the support tensors $P_{ijkl}^{D(d)}$ and $P_{ijkl}^{E(d)}$ through
\begin{align}
D_{ijkl}^{(d)}
&=
d_{ijkl}^{(d)}P_{ijkl}^{D(d)},
&
E_{ijkl}^{(d)}
&=
e_{ijkl}^{(d)}P_{ijkl}^{E(d)},
\end{align}
where $d_{ijkl}^{(d)}$ and $e_{ijkl}^{(d)}$ denote independent coefficients for the allowed components.

For $D_{ijkl}^{(d)}$, the support tensor takes the form
\begin{align}
P^{D(d)}
=
\begin{pmatrix}
A_D^{(1)} & A_D^{(2)} & A_D^{(2)} \\
A_D^{(2)} & \mathbf{J} & \mathbf{J} \\
A_D^{(2)} & \mathbf{J} & \mathbf{J}
\end{pmatrix}_{(k,l)},
\end{align}
where the block rows and columns are labeled by $k$ and $l$, respectively, while each block is a $3\times3$ matrix in the indices $i$ and $j$. Here, we define
\begin{align}
A_D^{(1)}
&=
\begin{pmatrix}
1 & 1 & 0 \\
1 & 1 & 0 \\
0 & 0 & 1
\end{pmatrix},
\qquad
A_D^{(2)}
=
\begin{pmatrix}
1 & 1 & 1 \\
1 & 1 & 1 \\
1 & 1 & 0
\end{pmatrix}.
\end{align}

A unit entry in $P^{D(d)}$ specifies an allowed component of $D_{ijkl}^{(d)}$, whereas a vanishing entry denotes a forbidden component. The equality of the blocks at $(k,l)$ and $(l,k)$ shows that the support is invariant under $k\leftrightarrow l$. Since $A_D^{(1)}$ and $A_D^{(2)}$ are symmetric matrices, it is also invariant under $i\leftrightarrow j$. Therefore,
\begin{align}
P_{ijkl}^{D(d)}
=P_{jikl}^{D(d)}
=P_{ijlk}^{D(d)}.
\end{align}

For $E_{ijkl}^{(d)}$, the support tensor is given by
\begin{align}
P^{E(d)}
=
\begin{pmatrix}
\mathbf{J} & A_E^{(1)} & A_E^{(2)} \\
A_E^{(1)} & A_E^{(3)} & A_E^{(4)} \\
A_E^{(2)} & A_E^{(4)} & A_E^{(3)}
\end{pmatrix}_{(k,l)},
\end{align}
with
\begin{align}
A_E^{(1)}
&=
\begin{pmatrix}
1 & 1 & 0 \\
1 & 1 & 0 \\
0 & 0 & 0
\end{pmatrix},
\quad
A_E^{(2)}
=
\begin{pmatrix}
1 & 1 & 1 \\
1 & 1 & 1 \\
1 & 1 & 0
\end{pmatrix},
\quad
A_E^{(3)}
=
\begin{pmatrix}
1 & 1 & 0 \\
1 & 1 & 0 \\
0 & 0 & 1
\end{pmatrix},
\qquad
A_E^{(4)}
=
\begin{pmatrix}
0 & 0 & 1 \\
0 & 0 & 1 \\
1 & 1 & 0
\end{pmatrix}.
\end{align}
The block matrix is symmetric under $k\leftrightarrow l$, and each of the matrices $A_E^{(n)}$ is symmetric under $i\leftrightarrow j$. It follows that
\begin{align}
P_{ijkl}^{E(d)}
=P_{jikl}^{E(d)}
=P_{ijlk}^{E(d)}.
\end{align}
These relations concern the support of the tensors. The full coefficients possess the same exchange symmetries only when
\begin{align}
d_{ijkl}^{(d)}
=
d_{jikl}^{(d)}
=d_{ijlk}^{(d)},
\qquad
e_{ijkl}^{(d)}
=
e_{jikl}^{(d)}
=e_{ijlk}^{(d)}.
\end{align}

The support tensors $P^{E(b)}$ and $P^{E(d)}$ are related by an interchange of the first and third generations. Defining the permutation $\sigma$ by
\begin{align}
\sigma(1)=3,
\qquad
\sigma(2)=2,
\qquad
\sigma(3)=1,
\end{align}
one finds
\begin{align}
P_{ijkl}^{E(d)}
=P_{\sigma(i)\sigma(j)\sigma(k)\sigma(l)}^{E(b)}.
\end{align}
Thus, the two textures belong to the same equivalence class under a simultaneous relabeling of all four generation indices,
\begin{align}
E^{(b)}
\sim
E^{(d)}.
\end{align}
For the full coefficient tensors, the same relation holds when their nonzero coefficients are mapped according to
\begin{align}
e_{ijkl}^{(d)}
=e_{\sigma(i)\sigma(j)\sigma(k)\sigma(l)}^{(b)}.
\end{align}

    \item Assignment (e):

{\small
\setlength{\arraycolsep}{3pt}
\renewcommand{\arraystretch}{0.95}  
\begin{align}
D_{ijkl}^{(e)}&= 
\left( \begin{array}{ccc|ccc|ccc}
\checkmark & 0 & 0
& 0 & \checkmark & \checkmark
& \checkmark & 0 & 0
\\
0 & \checkmark & \checkmark
& \checkmark & \checkmark & \checkmark
& 0 & \checkmark & \checkmark
\\
0 & \checkmark & \checkmark
& \checkmark & \checkmark & \checkmark
& 0 & \checkmark & \checkmark
\\
\hline
0 & \checkmark & \checkmark
& \checkmark & \checkmark & \checkmark
& 0 & \checkmark & \checkmark
\\
\checkmark & \checkmark & \checkmark
& \checkmark & \checkmark & \checkmark
& \checkmark & \checkmark & \checkmark
\\
\checkmark & \checkmark & \checkmark
& \checkmark & \checkmark & \checkmark
& \checkmark & \checkmark & \checkmark
\\
\hline
\checkmark & 0 & 0
& 0 & \checkmark & \checkmark
& \checkmark & 0 & 0
\\
0 & \checkmark & \checkmark
& \checkmark & \checkmark & \checkmark
& 0 & \checkmark & \checkmark
\\
0 & \checkmark & \checkmark
& \checkmark & \checkmark & \checkmark
& 0 & \checkmark & \checkmark
\end{array} \right)
,\quad
E_{ijkl}^{(e)}&= 
\left( \begin{array}{ccc|ccc|ccc}
\checkmark & \checkmark & \checkmark & \checkmark & \checkmark & \checkmark & \checkmark & \checkmark & \checkmark \\
\checkmark & \checkmark & \checkmark & \checkmark & \checkmark & \checkmark & \checkmark & \checkmark & \checkmark \\
\checkmark & \checkmark & \checkmark & \checkmark & \checkmark & \checkmark & \checkmark & \checkmark & \checkmark \\
\hline
\checkmark & \checkmark & \checkmark & \checkmark & \checkmark & \checkmark & \checkmark & \checkmark & \checkmark \\
\checkmark & \checkmark & \checkmark & \checkmark & \checkmark & \checkmark & \checkmark & \checkmark & \checkmark \\
\checkmark & \checkmark & \checkmark & \checkmark & \checkmark & \checkmark & \checkmark & \checkmark & \checkmark \\
\hline
\checkmark & \checkmark & \checkmark & \checkmark & \checkmark & \checkmark & \checkmark & \checkmark & \checkmark \\
\checkmark & \checkmark & \checkmark & \checkmark & \checkmark & \checkmark & \checkmark & \checkmark & \checkmark \\
\checkmark & \checkmark & \checkmark & \checkmark & \checkmark & \checkmark & \checkmark & \checkmark & \checkmark
\end{array} \right).
\end{align}
}
The texture of $D_{ijkl}^{(e)}$ can be organized as a $3\times3$ block matrix. Let us introduce the support tensor $P_{ijkl}^{D(e)}$ through
\begin{align}
D_{ijkl}^{(e)}
=
d_{ijkl}^{(e)}P_{ijkl}^{D(e)},
\end{align}
where $d_{ijkl}^{(e)}$ denotes an independent coefficient for each allowed component. The support tensor takes the form
\begin{align}
P^{D(e)}
=
\begin{pmatrix}
A_D^{(1)} & A_D^{(2)} & A_D^{(1)} \\
A_D^{(2)} & \mathbf{J} & A_D^{(2)} \\
A_D^{(1)} & A_D^{(2)} & A_D^{(1)}
\end{pmatrix}_{(k,l)},
\end{align}
where the block rows and columns are labeled by $k$ and $l$, respectively, while each block is a $3\times3$ matrix in the indices $i$ and $j$. Here, we define
\begin{align}
A_D^{(1)}
&=
\begin{pmatrix}
1 & 0 & 0 \\
0 & 1 & 1 \\
0 & 1 & 1
\end{pmatrix},
\qquad
A_D^{(2)}
=
\begin{pmatrix}
0 & 1 & 1 \\
1 & 1 & 1 \\
1 & 1 & 1
\end{pmatrix}.
\end{align}

A unit entry in $P^{D(e)}$ specifies an allowed component of $D_{ijkl}^{(e)}$, whereas a vanishing entry denotes a forbidden component. The equality of the blocks at $(k,l)$ and $(l,k)$ shows that the support is invariant under $k\leftrightarrow l$. Since $A_D^{(1)}$ and $A_D^{(2)}$ are symmetric matrices, the support is also invariant under $i\leftrightarrow j$. Thus,
\begin{align}
P_{ijkl}^{D(e)}
=P_{jikl}^{D(e)}
=P_{ijlk}^{D(e)}.
\end{align}

These relations concern the support of the tensor. The full coefficient possesses the same exchange symmetries only when
\begin{align}
d_{ijkl}^{(e)}
=d_{jikl}^{(e)}
=d_{ijlk}^{(e)}.
\end{align}

\end{enumerate}

\subsubsection{$(\bar{R}R)(\bar{R}R)$}

In this section, we deal with the four-fermion operators:
\begin{align}
    F_{ijkl}Q_{ee} &= F_{ijkl}(\bar{e}_R^i\gamma_\mu e^j_R)(\bar{e}^k_R \gamma^\mu e^l_R),
    \nonumber\\
    F_{ijkl}Q_{dd} &= F_{ijkl}(\bar{d}_R^i\gamma_\mu d^j_R)(\bar{d}^k_R \gamma^\mu d^l_R),
    \nonumber\\
    F_{ijkl}Q_{ed} &= F_{ijkl}(\bar{e}_R^i\gamma_\mu e^j_R)(\bar{d}^k_R \gamma^\mu d^l_R),
    \nonumber\\
    G_{ijkl}Q_{eu} &= G_{ijkl}(\bar{e}_R^i\gamma_\mu e^j_R)(\bar{u}^k_R \gamma^\mu u^l_R),
    \nonumber\\
    H_{ijkl}Q_{uu} &= H_{ijkl}(\bar{u}_R^i\gamma_\mu u^j_R)(\bar{u}^k_R \gamma^\mu u^l_R),
    \nonumber\\
    G_{ijkl}Q_{ud}^{(1)} &= G_{ijkl}(\bar{u}_R^i\gamma_\mu u^j_R)(\bar{d}^k_R \gamma^\mu d^l_R),
    \nonumber\\
    G_{ijkl}Q_{ud}^{(8)} &= G_{ijkl}(\bar{u}_R^i\gamma_\mu T^A u^j_R)(\bar{d}^k_R \gamma^\mu T^A d^l_R).
\end{align}
Note that some operators exhibit the same textures due to the same charge assignment of $[d]=[e]$. We therefore denote some Wilson coefficients by the same symbols $F_{ijkl}$ and $G_{ijkl}$.

Let us focus on five representative assignments of matter fields listed in Table~\ref{tab:Z5Z5_tex2_1}.

\begin{enumerate}
    \item Assignment (a)

All components of $F_{ijkl}^{(a)}$ are non-vanishing, while $G_{ijkl}$ and $H_{ijkl}^{(a)}$ have the same textures as $D_{ijkl}^{(a)}$ and $E_{ijkl}^{(a)}$, respectively, i.e., 
\begin{align}
F_{ijkl}^{(a)}&\neq0
\quad
\text{for all }i,j,k,l,
\qquad
\operatorname{supp}(G_{ijkl}^{(a)})
=
\operatorname{supp}(D_{ijkl}^{(a)}),
\qquad
\operatorname{supp}(H_{ijkl}^{(a)})
=
\operatorname{supp}(C_{ijkl}^{(a)}).
\end{align}

    \item Assignment (b):

{\small
\setlength{\arraycolsep}{3pt}
\renewcommand{\arraystretch}{0.95}  
\begin{align}
\operatorname{supp}(F_{ijkl}^{(b)})
&=
\operatorname{supp}(C_{ijkl}^{(b)})
,\nonumber\\
G_{ijkl}^{(b)}&= 
\left( \begin{array}{ccc|ccc|ccc}
\checkmark & 0 & 0 & 0 & \checkmark & \checkmark & \checkmark & 0 & 0 \\ 0 & \checkmark & \checkmark & \checkmark & \checkmark & \checkmark & 0 & \checkmark & \checkmark \\ 0 & \checkmark & \checkmark & \checkmark & \checkmark & 0 & 0 & \checkmark & \checkmark \\ \hline 0 & \checkmark & \checkmark & \checkmark & \checkmark & 0 & 0 & 0 & \checkmark \\ \checkmark & \checkmark & \checkmark & \checkmark & \checkmark & 0 & 0 & 0 & \checkmark \\ \checkmark & \checkmark & 0 & 0 & 0 & \checkmark & \checkmark & \checkmark & \checkmark \\ \hline \checkmark & 0 & 0 & 0 & 0 & \checkmark & \checkmark & 0 & 0 \\ 0 & \checkmark & \checkmark & 0 & 0 & \checkmark & 0 & \checkmark & \checkmark \\ 0 & \checkmark & \checkmark & \checkmark & \checkmark & \checkmark & 0 & \checkmark & \checkmark
\end{array} \right)
,\quad
H_{ijkl}^{(b)}=
\left( \begin{array}{ccc|ccc|ccc}
\checkmark & 0 & \checkmark
& 0 & \checkmark & \checkmark
& \checkmark & \checkmark & \checkmark
\\
0 & \checkmark & \checkmark
& \checkmark & 0 & 0
& \checkmark & 0 & \checkmark
\\
\checkmark & \checkmark & \checkmark
& \checkmark & 0 & \checkmark
& \checkmark & \checkmark & \checkmark
\\
\hline
0 & \checkmark & \checkmark
& \checkmark & 0 & 0
& \checkmark & 0 & \checkmark
\\
\checkmark & 0 & 0
& 0 & \checkmark & 0
& 0 & 0 & \checkmark
\\
\checkmark & 0 & \checkmark
& 0 & 0 & \checkmark
& \checkmark & \checkmark & 0
\\
\hline
\checkmark & \checkmark & \checkmark
& \checkmark & 0 & \checkmark
& \checkmark & \checkmark & \checkmark
\\
\checkmark & 0 & \checkmark
& 0 & 0 & \checkmark
& \checkmark & \checkmark & 0
\\
\checkmark & \checkmark & \checkmark
& \checkmark & \checkmark & 0
& \checkmark & 0 & \checkmark
\end{array} \right)
\end{align}
}

The textures of $G_{ijkl}^{(b)}$ and $H_{ijkl}^{(b)}$ can be organized as $3\times3$ block matrices. Let us introduce the support tensors $P_{ijkl}^{G(b)}$ and $P_{ijkl}^{H(b)}$ through
\begin{align}
G_{ijkl}^{(b)}
&=
g_{ijkl}^{(b)}P_{ijkl}^{G(b)},
&
H_{ijkl}^{(b)}
&=
h_{ijkl}^{(b)}P_{ijkl}^{H(b)},
\end{align}
where $g_{ijkl}^{(b)}$ and $h_{ijkl}^{(b)}$ denote independent coefficients for the allowed components.

For $G_{ijkl}^{(b)}$, the support tensor takes the form
\begin{align}
P^{G(b)}
=
\begin{pmatrix}
A_G^{(1)} & A_G^{(2)} & A_G^{(1)} \\
A_G^{(2)} & A_G^{(3)} & A_G^{(4)} \\
A_G^{(1)} & A_G^{(4)} & A_G^{(1)}
\end{pmatrix}_{(k,l)},
\end{align}
where the block rows and columns are labeled by $k$ and $l$, respectively, while each block is a $3\times3$ matrix in the indices $i$ and $j$. The matrices $A_G^{(n)}$ are given by
\begin{align}
A_G^{(1)}
&=
\begin{pmatrix}
1 & 0 & 0 \\
0 & 1 & 1 \\
0 & 1 & 1
\end{pmatrix},
\quad
A_G^{(2)}
=
\begin{pmatrix}
0 & 1 & 1 \\
1 & 1 & 1 \\
1 & 1 & 0
\end{pmatrix},
\quad
A_G^{(3)}
=
\begin{pmatrix}
1 & 1 & 0 \\
1 & 1 & 0 \\
0 & 0 & 1
\end{pmatrix},
\quad
A_G^{(4)}
=
\begin{pmatrix}
0 & 0 & 1 \\
0 & 0 & 1 \\
1 & 1 & 1
\end{pmatrix}.
\end{align}
A unit entry in $P^{G(b)}$ specifies an allowed component of $G_{ijkl}^{(b)}$, whereas a vanishing entry denotes a forbidden component. The equality of the blocks at $(k,l)$ and $(l,k)$ shows that the support is invariant under $k\leftrightarrow l$. Since each of the matrices $A_G^{(n)}$ is symmetric, the support is also invariant under $i\leftrightarrow j$. Therefore,
\begin{align}
P_{ijkl}^{G(b)}
=
P_{jikl}^{G(b)}
=
P_{ijlk}^{G(b)}.
\end{align}

For $H_{ijkl}^{(b)}$, the support tensor is given by
\begin{align}
P^{H(b)}
=
\begin{pmatrix}
A_H^{(1)} & A_H^{(2)} & A_H^{(3)} \\
A_H^{(2)} & \mathbf{1} & A_H^{(4)} \\
A_H^{(3)} & A_H^{(4)} & A_H^{(5)}
\end{pmatrix}_{(k,l)},
\end{align}
with
\begin{align}
A_H^{(1)}
&=
\begin{pmatrix}
1 & 0 & 1 \\
0 & 1 & 1 \\
1 & 1 & 1
\end{pmatrix},
\quad
A_H^{(2)}
=
\begin{pmatrix}
0 & 1 & 1 \\
1 & 0 & 0 \\
1 & 0 & 1
\end{pmatrix},
\quad
A_H^{(3)}
=
\begin{pmatrix}
1 & 1 & 1 \\
1 & 0 & 1 \\
1 & 1 & 1
\end{pmatrix},
\quad
A_H^{(4)}
=
\begin{pmatrix}
1 & 0 & 1 \\
0 & 0 & 1 \\
1 & 1 & 0
\end{pmatrix},
\nonumber\\
A_H^{(5)}
&=
\begin{pmatrix}
1 & 1 & 1 \\
1 & 1 & 0 \\
1 & 0 & 1
\end{pmatrix},
\end{align}
where $\mathbf{1}$ denotes the $3\times3$ identity matrix. As in the case of $G_{ijkl}^{(b)}$, the block structure is symmetric under $k\leftrightarrow l$, and each block is symmetric under $i\leftrightarrow j$. Hence,
\begin{align}
P_{ijkl}^{H(b)}
=
P_{jikl}^{H(b)}
=
P_{ijlk}^{H(b)}.
\end{align}

These relations refer to the support of the tensors. The full coefficients possess the same exchange symmetries only when
\begin{align}
g_{ijkl}^{(b)}
&=
g_{jikl}^{(b)}
=
g_{ijlk}^{(b)},
&
h_{ijkl}^{(b)}
&=
h_{jikl}^{(b)}
=
h_{ijlk}^{(b)}.
\end{align}
    \item Assignment (c)

All components of $F_{ijkl}^{(c)}$, $G_{ijkl}^{(c)}$ and $H_{ijkl}^{(c)}$ are non-vanishing.

    \item Assignment (d)

All components of $F_{ijkl}^{(d)}$ are non-vanishing, and we find       
{\small
\setlength{\arraycolsep}{3pt}
\renewcommand{\arraystretch}{0.95}  
\begin{align}
G_{ijkl}^{(d)}&= 
\left( \begin{array}{ccc|ccc|ccc}
\checkmark & \checkmark & \checkmark & \checkmark & \checkmark & \checkmark & \checkmark & \checkmark & \checkmark \\ \checkmark & \checkmark & \checkmark & \checkmark & 0 & 0 & \checkmark & \checkmark & \checkmark \\ \checkmark & \checkmark & \checkmark & \checkmark & 0 & 0 & \checkmark & \checkmark & \checkmark \\ \hline \checkmark & \checkmark & \checkmark & \checkmark & 0 & 0 & 0 & \checkmark & \checkmark \\ \checkmark & 0 & 0 & 0 & \checkmark & \checkmark & \checkmark & \checkmark & \checkmark \\ \checkmark & 0 & 0 & 0 & \checkmark & \checkmark & \checkmark & \checkmark & \checkmark \\ \hline \checkmark & \checkmark & \checkmark & 0 & \checkmark & \checkmark & \checkmark & \checkmark & \checkmark \\ \checkmark & \checkmark & \checkmark & \checkmark & \checkmark & \checkmark & \checkmark & \checkmark & \checkmark \\ \checkmark & \checkmark & \checkmark & \checkmark & \checkmark & \checkmark & \checkmark & \checkmark & \checkmark
\end{array} \right)
,\qquad
\operatorname{supp}(H_{ijkl}^{(d)})
=
\operatorname{supp}(H_{ijkl}^{(b)}).
\end{align}
}
The texture of $G_{ijkl}^{(d)}$ can be organized as a $3\times3$ block matrix. Let us introduce the support tensor $P_{ijkl}^{G(d)}$ through
\begin{align}
G_{ijkl}^{(d)}
=
g_{ijkl}^{(d)}P_{ijkl}^{G(d)},
\end{align}
where $g_{ijkl}^{(d)}$ denotes an independent coefficient for each allowed component.

The support tensor takes the form
\begin{align}
P^{G(d)}
=
\begin{pmatrix}
\mathbf{J} & A_G^{(1)} & \mathbf{J} \\
A_G^{(1)} & A_G^{(2)} & A_G^{(3)} \\
\mathbf{J} & A_G^{(3)} & \mathbf{J}
\end{pmatrix}_{(k,l)},
\end{align}
where the block rows and columns are labeled by $k$ and $l$, respectively, while each block is a $3\times3$ matrix in the indices $i$ and $j$. Here, we define
\begin{align}
A_G^{(1)}
&=
\begin{pmatrix}
1 & 1 & 1 \\
1 & 0 & 0 \\
1 & 0 & 0
\end{pmatrix},
\qquad
A_G^{(2)}
=
\begin{pmatrix}
1 & 0 & 0 \\
0 & 1 & 1 \\
0 & 1 & 1
\end{pmatrix},
\qquad
A_G^{(3)}
=
\begin{pmatrix}
0 & 1 & 1 \\
1 & 1 & 1 \\
1 & 1 & 1
\end{pmatrix}.
\end{align}

A unit entry in $P^{G(d)}$ specifies an allowed component of $G_{ijkl}^{(d)}$, whereas a vanishing entry denotes a forbidden component. The equality of the blocks at $(k,l)$ and $(l,k)$ shows that the support is invariant under $k\leftrightarrow l$. Since each of the matrices $A_G^{(1)}$, $A_G^{(2)}$, and $A_G^{(3)}$ is symmetric, the support is also invariant under $i\leftrightarrow j$. Therefore,
\begin{align}
P_{ijkl}^{G(d)}
=
P_{jikl}^{G(d)}
=
P_{ijlk}^{G(d)}.
\end{align}

These relations refer to the support of the tensor. The full coefficient possesses the same exchange symmetries only when
\begin{align}
g_{ijkl}^{(d)}
=
g_{jikl}^{(d)}
=
g_{ijlk}^{(d)}.
\end{align}

    \item Assignment (e)

All components of $H_{ijkl}^{(e)}$ are non-vanishing, and we find       
\begin{align}
\operatorname{supp}(F_{ijkl}^{(e)})
=
\operatorname{supp}(C_{ijkl}^{(c)})
,\quad
\operatorname{supp}(G_{ijkl}^{(e)})
=
\operatorname{supp}(D_{ijkl}^{(c)})
.
\end{align}
    
\end{enumerate}

\subsubsection{$(\bar{L}R)(\bar{R}L)$ and $(\bar{L}R)(\bar{L}R)$}

Finally, we investigate the $(\bar{L}R)(\bar{R}L)$ and $(\bar{L}R)(\bar{L}R)$ four-fermion operators:
\begin{align}
    D'_{ijkl} Q_{\ell edq} &= D'_{ijkl}(\bar{\ell}_L^i e^j_R)(\bar{d}^k_R q^l_L),
    \nonumber\\
    I_{ijkl}Q_{quqd}^{(1)} &= I_{ijkl}(\bar{q}_L^{i,a} u^j_R)\epsilon_{ab}(\bar{q}^{k,b}_L d^l_R),
    \nonumber\\
    I_{ijkl}Q_{quqd}^{(8)} &= I_{ijkl}(\bar{q}_L^{i,a} T^A u^j_R)\epsilon_{ab}(\bar{q}^{k,b}_L T^A d^l_R),
    \nonumber\\
    J_{ijkl}Q_{\ell equ}^{(1)} &= J_{ijkl}(\bar{\ell}_L^{i,a}e^j_R)\epsilon_{ab}(\bar{q}^{k,b}_L u^l_R),
    \nonumber\\
    J_{ijkl}Q_{\ell equ}^{(3)} &= J_{ijkl}(\bar{\ell}_L^{i,a}\sigma_{\mu\nu} e^j_R)\epsilon_{ab}(\bar{q}^{k,b}_L \sigma^{\mu\nu} u^l_R).
    \label{operator-LRRL}
\end{align}
The textures are related by simple permutations of the flavor indices. For each assignment $x=a,b,c,d,e$, the texture of $D'^{(x)}_{ijkl}$ is obtained from that of $D^{(x)}_{ijkl}$ by interchanging the second and fourth indices:
\begin{align}
\operatorname{supp}\left(D'^{(x)}_{ijkl}\right)
=
\operatorname{supp}\left(D^{(x)}_{ilkj}\right).
\label{Dp-D}
\end{align}
Similarly, the textures of $I^{(x)}_{ijkl}$ and $J^{(x)}_{ijkl}$ are related by the same interchange:
\begin{align}
\operatorname{supp}\left(J^{(x)}_{ijkl}\right)
=
\operatorname{supp}\left(I^{(x)}_{ilkj}\right).
\label{I-J}
\end{align}
Hence, we present only the texture of $I_{ijkl}$ for five representative examples in the following.

\begin{enumerate}
\label{texture-LRRL}
    \item Assignment (a):

{\small
\setlength{\arraycolsep}{3pt}
\renewcommand{\arraystretch}{0.95}  
\begin{align}
I_{ijkl}^{(a)}= 
\begin{array}{cc}
& \begin{array}{ccc} \hspace{0.5em} l=1 \hspace{0.5em} & \hspace{0.5em} l=2 \hspace{0.5em} & \hspace{0.5em} l=3 \hspace{0.5em} \end{array} \\
\begin{array}{c} k=1 \\ \\ \\ k=2 \\ \\ \\ k=3 \end{array} & 
\left( \begin{array}{ccc|ccc|ccc}
0 & \checkmark & 0
& 0 & 0 & 0
& 0 & 0 & 0
\\
\checkmark & 0 & 0
& 0 & \checkmark & 0
& 0 & 0 & 0
\\
0 & 0 & 0
& 0 & 0 & 0
& 0 & 0 & 0
\\
\hline
\checkmark & 0 & 0
& 0 & \checkmark & 0
& 0 & 0 & 0
\\
0 & \checkmark & 0
& \checkmark & \checkmark & \checkmark
& 0 & \checkmark & \checkmark
\\
0 & 0 & \checkmark
& 0 & \checkmark & 0
& 0 & \checkmark & 0
\\
\hline
0 & 0 & 0
& 0 & 0 & 0
& 0 & 0 & 0
\\
0 & 0 & \checkmark
& 0 & \checkmark & 0
& 0 & \checkmark & 0
\\
0 & \checkmark & 0
& 0 & 0 & 0
& 0 & 0 & 0
\end{array} \right).
\end{array}
\label{I(a)1}
\end{align}
}
The texture of $I_{ijkl}^{(a)}$ can be organized as a $3\times3$ block matrix:
\begin{align}
I_{ijkl}^{(a)}
=f_{ijkl}^{(a)}P_{ijkl}^{I(a)},
\label{I(a)2}
\end{align}
where $f_{ijkl}^{(a)}$ denotes an independent coefficient for each allowed component. The support tensor $P_{ijkl}^{I(a)}$ is represented by
\begin{align}
P^{I(a)}
=\begin{pmatrix}
A_I^{(1)} & A_I^{(2)} & \mathbf{0} \\
\mathbf{1} & A_I^{(3)} & A_I^{(4)} \\
A_I^{(5)} & A_I^{(2)} & A_I^{(2)}
\end{pmatrix}_{(k,l)},
\label{I(a)3}
\end{align}
where the block rows and columns are labeled by $k$ and $l$, respectively, while each block is a $3\times3$ matrix in the indices $i$ and $j$, i.e.,
\begin{align}
A_I^{(1)}
&=
\begin{pmatrix}
0 & 1 & 0 \\
1 & 0 & 0 \\
0 & 0 & 0
\end{pmatrix},
\quad
A_I^{(2)}
=
\begin{pmatrix}
0 & 0 & 0 \\
0 & 1 & 0 \\
0 & 0 & 0
\end{pmatrix},
\quad
A_I^{(3)}
=
\begin{pmatrix}
0 & 1 & 0 \\
1 & 1 & 1 \\
0 & 1 & 0
\end{pmatrix},
\nonumber\\
A_I^{(4)}
&=
\begin{pmatrix}
0 & 0 & 0 \\
0 & 1 & 1 \\
0 & 1 & 0
\end{pmatrix},
\quad
A_I^{(5)}
=
\begin{pmatrix}
0 & 0 & 0 \\
0 & 0 & 1 \\
0 & 1 & 0
\end{pmatrix}.
\label{I(a)4}
\end{align}
Here, $\mathbf{1}$ denotes the $3\times3$ identity matrix. A unit entry in $P^{I(a)}$ specifies an allowed component of $I_{ijkl}^{(a)}$, whereas a vanishing entry denotes a forbidden component. Equivalently, the individual blocks are given by
\begin{align}
&P^{I(a);11}=A_I^{(1)},\qquad
P^{I(a);12}=A_I^{(2)},\qquad
P^{I(a);13}=0,
\nonumber\\
&P^{I(a);21}=\mathbf{1},\qquad
P^{I(a);22}=A_I^{(3)},\qquad
P^{I(a);23}=A_I^{(4)},
\nonumber\\
&P^{I(a);31}=A_I^{(5)},\qquad
P^{I(a);32}=A_I^{(2)},\qquad
P^{I(a);33}=A_I^{(2)},
\end{align}
where $P^{I(a);kl}$ denotes the $3\times3$ matrix in the indices $i$ and $j$ for fixed $(k,l)$.

    \item Assignment (b):

{\small
\setlength{\arraycolsep}{3pt}
\renewcommand{\arraystretch}{0.95}  
\begin{align}
I_{ijkl}^{(b)}= 
\begin{array}{cc}
& \begin{array}{ccc} \hspace{0.5em} l=1 \hspace{0.5em} & \hspace{0.5em} l=2 \hspace{0.5em} & \hspace{0.5em} l=3 \hspace{0.5em} \end{array} \\
\begin{array}{c} k=1 \\ \\ \\ k=2 \\ \\ \\ k=3 \end{array} & 
\left( \begin{array}{ccc|ccc|ccc}
0 & 0 & 0
& 0 & 0 & \checkmark
& \checkmark & 0 & 0
\\
\checkmark & 0 & \checkmark
& \checkmark & 0 & \checkmark
& 0 & \checkmark & \checkmark
\\
0 & 0 & 0
& \checkmark & 0 & \checkmark
& 0 & \checkmark & \checkmark
\\
\hline
\checkmark & 0 & \checkmark
& \checkmark & 0 & \checkmark
& 0 & \checkmark & \checkmark
\\
\checkmark & 0 & \checkmark
& \checkmark & \checkmark & \checkmark
& \checkmark & 0 & \checkmark
\\
\checkmark & \checkmark & \checkmark
& \checkmark & \checkmark & \checkmark
& \checkmark & 0 & \checkmark
\\
\hline
0 & 0 & 0
& \checkmark & 0 & \checkmark
& 0 & \checkmark & \checkmark
\\
\checkmark & \checkmark & \checkmark
& \checkmark & \checkmark & \checkmark
& \checkmark & 0 & \checkmark
\\
0 & \checkmark & 0
& \checkmark & \checkmark & \checkmark
& \checkmark & 0 & \checkmark
\end{array} \right)
\end{array}.
\end{align}
}
The texture of $I_{ijkl}^{(b)}$ can be organized as a $3\times3$ block matrix. We introduce the support tensor $P_{ijkl}^{I(b)}$ through
\begin{align}
I_{ijkl}^{(b)}
=f_{ijkl}^{(b)}P_{ijkl}^{I(b)},
\end{align}
where $f_{ijkl}^{(b)}$ denotes an independent coefficient for each allowed component. The support tensor takes the form
\begin{align}
P^{I(b)}
=
\begin{pmatrix}
A_I^{(1)} & A_I^{(2)} & A_I^{(3)} \\
A_I^{(4)} & A_I^{(5)} & A_I^{(6)} \\
A_I^{(7)} & A_I^{(5)} & A_I^{(6)}
\end{pmatrix}_{(k,l)},
\end{align}
where the block rows and columns are labeled by $k$ and $l$, respectively, while each block is a $3\times3$ matrix in the indices $i$ and $j$. The matrices $A_I^{(n)}$ are given by
\begin{align}
A_I^{(1)}
&=
\begin{pmatrix}
0 & 0 & 0 \\
1 & 0 & 1 \\
0 & 0 & 0
\end{pmatrix},
\qquad
A_I^{(2)}
=
\begin{pmatrix}
0 & 0 & 1 \\
1 & 0 & 1 \\
1 & 0 & 1
\end{pmatrix},
\qquad
A_I^{(3)}
=
\begin{pmatrix}
1 & 0 & 0 \\
0 & 1 & 1 \\
0 & 1 & 1
\end{pmatrix},
\qquad
A_I^{(4)}
=
\begin{pmatrix}
1 & 0 & 1 \\
1 & 0 & 1 \\
1 & 1 & 1
\end{pmatrix},
\nonumber\\
A_I^{(5)}
&=
\begin{pmatrix}
1 & 0 & 1 \\
1 & 1 & 1 \\
1 & 1 & 1
\end{pmatrix},
\quad
A_I^{(6)}
=
\begin{pmatrix}
0 & 1 & 1 \\
1 & 0 & 1 \\
1 & 0 & 1
\end{pmatrix},
\quad
A_I^{(7)}
=
\begin{pmatrix}
0 & 0 & 0 \\
1 & 1 & 1 \\
0 & 1 & 0
\end{pmatrix}.
\end{align}

A unit entry in $P^{I(b)}$ specifies an allowed component of $I_{ijkl}^{(b)}$, whereas a vanishing entry denotes a forbidden component. In contrast to the textures discussed above, the blocks at $(k,l)$ and $(l,k)$ are not generally equal. For example,
\begin{align}
P^{I(b);12}
=A_I^{(2)}
\neq
A_I^{(4)}
=P^{I(b);21},
\end{align}
and
\begin{align}
P^{I(b);13}
=
A_I^{(3)}
\neq
A_I^{(7)}
=
P^{I(b);31}.
\end{align}

The support is therefore not invariant under $k\leftrightarrow l$. Moreover, several of the individual blocks are not symmetric matrices, so the support is not invariant under $i\leftrightarrow j$ either. Thus,
\begin{align}
P_{ijkl}^{I(b)}
\neq
P_{jikl}^{I(b)},
\qquad
P_{ijkl}^{I(b)}
\neq
P_{ijlk}^{I(b)}
\end{align}
in general. The block representation is useful in this case because the texture does not reduce to a simple exchange-symmetric support tensor.

    \item Assignment (c):

{\small
\setlength{\arraycolsep}{3pt}
\renewcommand{\arraystretch}{0.95}  
\begin{align}
I_{ijkl}^{(c)}= 
\begin{array}{cc}
& \begin{array}{ccc} \hspace{0.5em} l=1 \hspace{0.5em} & \hspace{0.5em} l=2 \hspace{0.5em} & \hspace{0.5em} l=3 \hspace{0.5em} \end{array} \\
\begin{array}{c} k=1 \\ \\ \\ k=2 \\ \\ \\ k=3 \end{array} & 
\left( \begin{array}{ccc|ccc|ccc}
\checkmark & 0 & \checkmark
& 0 & \checkmark & 0
& 0 & \checkmark & 0
\\
0 & \checkmark & 0
& \checkmark & \checkmark & \checkmark
& \checkmark & \checkmark & \checkmark
\\
\checkmark & 0 & \checkmark
& 0 & \checkmark & 0
& 0 & \checkmark & 0
\\
\hline
0 & \checkmark & 0
& \checkmark & \checkmark & \checkmark
& \checkmark & \checkmark & \checkmark
\\
\checkmark & \checkmark & \checkmark
& \checkmark & \checkmark & \checkmark
& \checkmark & \checkmark & \checkmark
\\
0 & \checkmark & 0
& \checkmark & \checkmark & \checkmark
& \checkmark & \checkmark & \checkmark
\\
\hline
\checkmark & 0 & \checkmark
& 0 & \checkmark & 0
& 0 & \checkmark & 0
\\
0 & \checkmark & 0
& \checkmark & \checkmark & \checkmark
& \checkmark & \checkmark & \checkmark
\\
\checkmark & 0 & \checkmark
& 0 & \checkmark & 0
& 0 & \checkmark & 0
\end{array} \right).
\end{array}
\end{align}
}
The texture of $I_{ijkl}^{(c)}$ can be written as a $3\times3$ block matrix. We introduce the support tensor $P_{ijkl}^{I(c)}$ through
\begin{align}
I_{ijkl}^{(c)}=f_{ijkl}^{(c)}P_{ijkl}^{I(c)},
\end{align}
where $f_{ijkl}^{(c)}$ denotes an independent coefficient for each allowed component. The support tensor is given by
\begin{align}
P^{I(c)}=
\begin{pmatrix}
A_I^{(1)} & A_I^{(2)} & A_I^{(2)} \\
A_I^{(2)} & \mathbf{J} & \mathbf{J} \\
A_I^{(1)} & A_I^{(2)} & A_I^{(2)}
\end{pmatrix}_{(k,l)},
\end{align}
where the block rows and columns are labeled by $k$ and $l$, respectively, and each block is a $3\times3$ matrix in the indices $i$ and $j$. Here, we define
\begin{align}
A_I^{(1)}=
\begin{pmatrix}
1 & 0 & 1 \\
0 & 1 & 0 \\
1 & 0 & 1
\end{pmatrix},
\qquad
A_I^{(2)}=
\begin{pmatrix}
0 & 1 & 0 \\
1 & 1 & 1 \\
0 & 1 & 0
\end{pmatrix}.
\end{align}
A unit entry in $P^{I(c)}$ specifies an allowed component of $I_{ijkl}^{(c)}$, whereas a vanishing entry denotes a forbidden component. Since $A_I^{(1)}$, $A_I^{(2)}$, and $\mathbf{J}$ are symmetric matrices, the support is invariant under $i\leftrightarrow j$:
\begin{align}
P_{ijkl}^{I(c)}=P_{jikl}^{I(c)}.
\end{align}
Hence, the full coefficient is invariant under $i\leftrightarrow j$ only when $f_{ijkl}^{(c)}=f_{jikl}^{(c)}$.

    \item Assignment (d):

{\small
\setlength{\arraycolsep}{3pt}
\renewcommand{\arraystretch}{0.95}  
\begin{align}
I_{ijkl}^{(d)}= 
\begin{array}{cc}
& \begin{array}{ccc} \hspace{0.5em} l=1 \hspace{0.5em} & \hspace{0.5em} l=2 \hspace{0.5em} & \hspace{0.5em} l=3 \hspace{0.5em} \end{array} \\
\begin{array}{c} k=1 \\ \\ \\ k=2 \\ \\ \\ k=3 \end{array} & 
\left( \begin{array}{ccc|ccc|ccc}
\checkmark & 0 & \checkmark
& \checkmark & \checkmark & \checkmark
& \checkmark & \checkmark & \checkmark
\\
\checkmark & 0 & \checkmark
& \checkmark & \checkmark & \checkmark
& \checkmark & \checkmark & \checkmark
\\
\checkmark & \checkmark & 0
& \checkmark & 0 & \checkmark
& \checkmark & 0 & \checkmark
\\
\hline
\checkmark & 0 & \checkmark
& \checkmark & \checkmark & \checkmark
& \checkmark & \checkmark & \checkmark
\\
\checkmark & 0 & \checkmark
& \checkmark & \checkmark & \checkmark
& \checkmark & \checkmark & \checkmark
\\
\checkmark & \checkmark & 0
& \checkmark & 0 & \checkmark
& \checkmark & 0 & \checkmark
\\
\hline
\checkmark & \checkmark & 0
& \checkmark & 0 & \checkmark
& \checkmark & 0 & \checkmark
\\
\checkmark & \checkmark & 0
& \checkmark & 0 & \checkmark
& \checkmark & 0 & \checkmark
\\
0 & 0 & \checkmark
& \checkmark & 0 & 0
& \checkmark & 0 & 0
\end{array} \right).
\end{array}
\end{align}
}

The texture of $I_{ijkl}^{(d)}$ can be written as a $3\times3$ block matrix. We define the support tensor $P_{ijkl}^{I(d)}$ by
\begin{align}
I_{ijkl}^{(d)}=f_{ijkl}^{(d)}P_{ijkl}^{I(d)},
\end{align}
where $f_{ijkl}^{(d)}$ is an independent coefficient assigned to each allowed component. The support tensor is
\begin{align}
P^{I(d)}=
\begin{pmatrix} 
A_I^{(1)} & A_I^{(2)} & A_I^{(2)} \\ 
A_I^{(1)} & A_I^{(2)} & A_I^{(2)} \\ 
A_I^{(3)} & A_I^{(4)} & A_I^{(4)} 
\end{pmatrix}_{(k,l)},
\end{align}
where the block rows and columns are labeled by $k$ and $l$, respectively, and each block is a $3\times3$ matrix in the indices $i$ and $j$. The four blocks are
\begin{align}
A_I^{(1)}=
\begin{pmatrix} 
1 & 0 & 1 \\ 
1 & 0 & 1 \\ 
1 & 1 & 0 
\end{pmatrix},
\qquad 
A_I^{(2)}=
\begin{pmatrix} 
1 & 1 & 1 \\ 
1 & 1 & 1 \\ 
1 & 0 & 1 
\end{pmatrix},
\qquad 
A_I^{(3)}=
\begin{pmatrix} 
1 & 1 & 0 \\ 
1 & 1 & 0 \\ 
0 & 0 & 1 
\end{pmatrix},
\qquad 
A_I^{(4)}
=\begin{pmatrix} 
1 & 0 & 1 \\ 
1 & 0 & 1 \\ 
1 & 0 & 0 
\end{pmatrix}.
\end{align}
A simplified expression can be obtained by introducing three mutually exclusive row sectors,
\begin{align}
R_1(k,i)=(1-\delta_{k3})(1-\delta_{i3}),\quad R_2(k,i)=\delta_{k3}+\delta_{i3}-2\delta_{k3}\delta_{i3},\quad R_3(k,i)=\delta_{k3}\delta_{i3},
\end{align}
which satisfy $R_1(k,i)+R_2(k,i)+R_3(k,i)=1$. We also define
\begin{align}
Q_1(l,j)&=1-\delta_{l1}\delta_{j2},
\qquad 
Q_2(l,j)=1-\delta_{l1}\delta_{j3}-(1-\delta_{l1})\delta_{j2},
\nonumber\\
Q_3(l,j)&=\delta_{l1}\delta_{j3}+(1-\delta_{l1})\delta_{j1}.
\end{align}
The support tensor is then expressed as
\begin{align}
P_{ijkl}^{I(d)}=R_1(k,i)Q_1(l,j)+R_2(k,i)Q_2(l,j)+R_3(k,i)Q_3(l,j).
\end{align}
Here, $R_1=1$ for $k,i\in\{1,2\}$, $R_2=1$ when exactly one of $k$ and $i$ is equal to $3$, and $R_3=1$ for $(k,i)=(3,3)$. The support is not invariant under either $i\leftrightarrow j$ or $k\leftrightarrow l$. 
On the other hand, the support tensor is invariant under the independent relabelings $i:1\leftrightarrow2$, $k:1\leftrightarrow2$, and $l:2\leftrightarrow3$. Explicitly, it satisfies
\begin{align}
P_{1jkl}^{I(d)}=P_{2jkl}^{I(d)},\qquad P_{ij1l}^{I(d)}=P_{ij2l}^{I(d)},\qquad P_{ijk2}^{I(d)}=P_{ijk3}^{I(d)}.
\end{align}
These relations characterize only the zero and nonzero structure. Since the coefficients $f_{ijkl}^{(d)}$ are independent, the full tensor $I_{ijkl}^{(d)}$ does not in general satisfy the same relations.

    \item Assignment (e):

{\small
\setlength{\arraycolsep}{3pt}
\renewcommand{\arraystretch}{0.95}  
\begin{align}
I_{ijkl}^{(e)}= 
\begin{array}{cc}
& \begin{array}{ccc} \hspace{0.5em} l=1 \hspace{0.5em} & \hspace{0.5em} l=2 \hspace{0.5em} & \hspace{0.5em} l=3 \hspace{0.5em} \end{array} \\
\begin{array}{c} k=1 \\ \\ \\ k=2 \\ \\ \\ k=3 \end{array} & 
\left( \begin{array}{ccc|ccc|ccc}
0 & \checkmark & \checkmark
& \checkmark & \checkmark & \checkmark
& 0 & \checkmark & \checkmark
\\
\checkmark & \checkmark & \checkmark
& \checkmark & \checkmark & \checkmark
& \checkmark & \checkmark & \checkmark
\\
\checkmark & \checkmark & \checkmark
& \checkmark & \checkmark & \checkmark
& \checkmark & \checkmark & \checkmark
\\
\hline
\checkmark & \checkmark & \checkmark
& \checkmark & \checkmark & \checkmark
& \checkmark & \checkmark & \checkmark
\\
\checkmark & 0 & 0
& \checkmark & \checkmark & \checkmark
& \checkmark & 0 & 0
\\
\checkmark & 0 & 0
& \checkmark & \checkmark & \checkmark
& \checkmark & 0 & 0
\\
\hline
\checkmark & \checkmark & \checkmark
& \checkmark & \checkmark & \checkmark
& \checkmark & \checkmark & \checkmark
\\
\checkmark & 0 & 0
& \checkmark & \checkmark & \checkmark
& \checkmark & 0 & 0
\\
\checkmark & 0 & 0
& \checkmark & \checkmark & \checkmark
& \checkmark & 0 & 0
\end{array} \right).
\end{array}
\end{align}
}
The texture of $I_{ijkl}^{(e)}$ can be written as a $3\times3$ block matrix. We define the support tensor $P_{ijkl}^{I(e)}$ by
\begin{align}
I_{ijkl}^{(e)}=f_{ijkl}^{(e)}P_{ijkl}^{I(e)},
\end{align}
where $f_{ijkl}^{(e)}$ is an independent coefficient assigned to each allowed component. The support tensor is given by
\begin{align}
P^{I(e)}=
\begin{pmatrix}
A_I^{(1)} & \mathbf{J} & A_I^{(1)} \\
A_I^{(2)} & \mathbf{J} & A_I^{(2)} \\
A_I^{(2)} & \mathbf{J} & A_I^{(2)}
\end{pmatrix}_{(k,l)},
\end{align}
where the block rows and columns are labeled by $k$ and $l$, respectively, and each block is a $3\times3$ matrix in the indices $i$ and $j$. Here, we define
\begin{align}
A_I^{(1)}&=
\begin{pmatrix}
0 & 1 & 1 \\
1 & 1 & 1 \\
1 & 1 & 1
\end{pmatrix},
\qquad
A_I^{(2)}=
\begin{pmatrix}
1 & 1 & 1 \\
1 & 0 & 0 \\
1 & 0 & 0
\end{pmatrix}.
\end{align}
The same support can be expressed directly in terms of Kronecker deltas as
\begin{align}
P_{ijkl}^{I(e)}
=1-(1-\delta_{l2})
\left[
\delta_{i1}\delta_{j1}\delta_{k1}
+
(1-\delta_{i1})(1-\delta_{j1})(1-\delta_{k1})
\right].
\end{align}
For $l=2$, the factor $1-\delta_{l2}$ vanishes, and hence all components are allowed. For $l=1$ or $l=3$, a component is forbidden only when either $i=j=k=1$ or $i,j,k\in\{2,3\}$. All remaining components are allowed.

The support is invariant under arbitrary permutations of the first three indices $i$, $j$, and $k$. Denoting $(a_1,a_2,a_3)=(i,j,k)$, this property can be written as
\begin{align}
P_{ijkl}^{I(e)}
=
P_{a_{\sigma(1)}a_{\sigma(2)}a_{\sigma(3)}l}^{I(e)},
\qquad
\sigma\in S_3.
\end{align}
In particular,
\begin{align}
P_{ijkl}^{I(e)}
=P_{jikl}^{I(e)}
=P_{kjil}^{I(e)}.
\end{align}
The support is also invariant under the relabeling $2\leftrightarrow3$ for each of the indices $i$, $j$, and $k$, as well as under $l:1\leftrightarrow3$. These relations characterize only the zero and nonzero structure. Since the coefficients $f_{ijkl}^{(e)}$ are independent, the full tensor $I_{ijkl}^{(e)}$ possesses the same permutation properties only when $f_{ijkl}^{(e)}$ satisfies the corresponding relations.

\end{enumerate}

\subsection{Two-fermion operators}
\label{sec:2fermi}

In this section, we classify the textures of dimension-six operators including Higgs and fermions under the charge assignments in Table~\ref{tab:Z5Z5_tex2_1}. 

In the two-Higgs doublet model discussed in this paper, the SM Higgs is defined in terms of a mixing parameter $\beta$:
\begin{align}
\label{eq:def_Higgs}
    H&= s_\beta\,\Phi_u + c_\beta\,\Phi_d,
\end{align}
with $c_\beta\equiv \cos(\beta)$ and $s_\beta \equiv \sin(\beta)$, 
indicating that the charge of the SM Higgs can be inferred from those of $\Phi_u$ and $\Phi_d$.

Here, we focus on the flavor structures induced by the non-invertible
selection rules after projecting the two Higgs doublets onto the SM-like
Higgs direction. If the Higgs doublet orthogonal to $H$ is heavy, a
complete matching onto the SMEFT requires integrating out the heavy
doublet, which may generate additional contributions to the SMEFT Wilson
coefficients. Such matching contributions depend on the details of the
Higgs potential and the UV parameters and are beyond the scope of the
present analysis. We therefore restrict ourselves to the flavor textures
directly inherited from the non-invertible selection rules.

\subsubsection{$\psi^2 X H$}

In this section, we investigate the dimension-six operators of the form $\psi^2 X H$:
\begin{align}
C_{\ell_ie_j h}Q_{eW} &=C_{\ell_ie_j h}
(\bar{\ell}_i\sigma^{\mu\nu}e_j)\tau^I H W_{\mu\nu}^I ,
\nonumber\\
C_{\ell_ie_j h}Q_{eB} &=C_{\ell_ie_j h}
(\bar{\ell}_i\sigma^{\mu\nu}e_j)H B_{\mu\nu} ,
\nonumber\\
C_{q_iu_jh}Q_{uG} &=C_{q_iu_jh}
(\bar{q}_i\sigma^{\mu\nu}T^A u_j)\widetilde{H}\,G_{\mu\nu}^A ,
\nonumber\\
C_{q_iu_jh}Q_{uW} &=C_{q_iu_jh}
(\bar{q}_i\sigma^{\mu\nu}u_j)\tau^I\widetilde{H}\,W_{\mu\nu}^I ,
\nonumber\\
C_{q_iu_jh}Q_{uB} &=C_{q_iu_jh}
(\bar{q}_i\sigma^{\mu\nu}u_j)\widetilde{H}\,B_{\mu\nu} ,
\nonumber\\
C_{\ell_i e_jh}Q_{dG} &=C_{\ell_i e_jh}
(\bar{q}_i\sigma^{\mu\nu}T^A d_j)H\,G_{\mu\nu}^A ,
\nonumber\\
C_{\ell_i e_jh}Q_{dW} &=C_{\ell_i e_jh}
(\bar{q}_i\sigma^{\mu\nu}d_j)\tau^I H\,W_{\mu\nu}^I ,
\nonumber\\
C_{\ell_i e_jh}Q_{dB} &=C_{\ell_i e_jh}
(\bar{q}_i\sigma^{\mu\nu}d_j)H\,B_{\mu\nu} .
\end{align}
Note that some of the Wilson coefficients are described by the same ones due to the identification of classes \eqref{eq:charge}, and the textures of $\psi^2 X H$ type operators can be understood from the following two operators:
\begin{align}
C_{\ell_ie_j h}    Q_{eW} &= C_{\ell_ie_j h}(\bar{\ell}_L^i \sigma^{\mu\nu}e_R^j)\tau^I H W_{\mu\nu}^I
    \nonumber\\
    &\equiv C_{\ell_i e_j \Phi_u}s_\beta(\bar{\ell}_L^i \sigma^{\mu\nu}e_R^j)\tau^I \Phi_u W_{\mu\nu}^I
    + C_{\ell_i e_j \Phi_d}c_\beta(\bar{\ell}_L^i \sigma^{\mu\nu}e_R^j)\tau^I \Phi_d W_{\mu\nu}^I,
    \nonumber\\
C_{q_iu_jh}Q_{uG} &=C_{q_iu_jh}
(\bar{q}_i\sigma^{\mu\nu}T^A u_j)\widetilde{H}\,G_{\mu\nu}^A 
\nonumber\\
&\equiv C_{q_iu_j\Phi_u} s_\beta
(\bar{q}_i\sigma^{\mu\nu}T^A u_j)\widetilde{\Phi}_u\,G_{\mu\nu}^A 
+ C_{q_iu_j\Phi_d} c_\beta
(\bar{q}_i\sigma^{\mu\nu}T^A u_j)\widetilde{\Phi}_d\,G_{\mu\nu}^A.
\label{QeWQuG}
\end{align}

For the five charge assignments in Table~\ref{tab:Z5Z5_tex2_1}, we present the explicit textures of the above Wilson coefficients:
\begin{enumerate}
    \item $Q_{eW}$, $Q_{eB}$, $Q_{dG}$, $Q_{dW}$, and $Q_{dB}$
\begin{align}
C_{\ell_i e_j \Phi_u}^{(a)}&=
\begin{pmatrix}
0 & 0 & 0 \\
\checkmark & 0 & 0 \\
0 & 0 & 0
\end{pmatrix}
,\qquad
C_{\ell_i e_j \Phi_u}^{(b)}=
\begin{pmatrix}
0 & 0 & \checkmark \\
0 & 0 & 0 \\
0 & \checkmark & 0
\end{pmatrix}
,\qquad
C_{\ell_i e_j \Phi_u}^{(c)}=
\begin{pmatrix}
\checkmark & 0 & 0 \\
0 & \checkmark & 0 \\
0 & 0 & 0
\end{pmatrix}
,\nonumber\\
C_{\ell_i e_j \Phi_u}^{(d)}&=
\begin{pmatrix}
0 & 0 & \checkmark \\
0 & \checkmark & 0 \\
0 & 0 & 0
\end{pmatrix}
,\qquad
C_{\ell_i e_j \Phi_u}^{(e)}=
\begin{pmatrix}
0 & \checkmark & 0 \\
0 & 0 & 0 \\
0 & 0 & 0
\end{pmatrix},
\nonumber\\[1ex]
C_{\ell_i e_j \Phi_d}^{(a)}&=
C_{\ell_i e_j \Phi_d}^{(b)}=
C_{\ell_i e_j \Phi_d}^{(c)}=
C_{\ell_i e_j \Phi_d}^{(d)}=
C_{\ell_i e_j \Phi_d}^{(e)}
=
\begin{pmatrix}
0 & \checkmark & 0 \\
\checkmark & \checkmark & \checkmark \\
0 & \checkmark & \checkmark
\end{pmatrix}.
\label{eq:lepton-dipole}
\end{align}
Hence, the Wilson coefficients $C_{\ell_ie_j h}$ have the following textures:
\begin{align}
C_{\ell_i e_j h}^{(a)}&=
C_{\ell_i e_j h}^{(e)}=
\begin{pmatrix}
0 & \checkmark & 0 \\
\checkmark & \checkmark & \checkmark \\
0 & \checkmark & \checkmark
\end{pmatrix}
,\quad
C_{\ell_i e_j h}^{(b)}=
C_{\ell_i e_j h}^{(d)}=
\begin{pmatrix}
0 & \checkmark & \checkmark \\
\checkmark & \checkmark & \checkmark \\
0 & \checkmark & \checkmark
\end{pmatrix}
,\quad
C_{\ell_i e_j h}^{(c)}=
\begin{pmatrix}
\checkmark & \checkmark & 0 \\
\checkmark & \checkmark & \checkmark \\
0 & \checkmark & \checkmark
\end{pmatrix}.
\end{align}

    \item $Q_{uG}$, $Q_{uW}$ and $Q_{uB}$

\begin{align}
C_{q_i u_j \Phi_d}^{(a)}&=
\begin{pmatrix}
0 & 0 & 0 \\
0 & \checkmark & 0 \\
0 & 0 & 0
\end{pmatrix}
,\qquad
C_{q_i u_j \Phi_d}^{(b)}=
\begin{pmatrix}
0 & 0 & \checkmark \\
\checkmark & 0 & \checkmark \\
\checkmark & 0 & \checkmark
\end{pmatrix}
,\qquad
C_{q_i u_j \Phi_d}^{(c)}=
\begin{pmatrix}
0 & \checkmark & 0 \\
\checkmark & \checkmark & \checkmark \\
0 & \checkmark & 0
\end{pmatrix}
,\nonumber\\
C_{q_i u_j \Phi_d}^{(d)}&=
\begin{pmatrix}
0 & 0 & \checkmark \\ 
\checkmark & 0 & \checkmark \\ 
\checkmark & 0 & 0
\end{pmatrix}
,\qquad
C_{q_i u_j \Phi_d}^{(e)}=
\begin{pmatrix}
\checkmark & \checkmark & \checkmark \\
\checkmark & 0 & 0 \\
\checkmark & 0 & 0
\end{pmatrix},
\nonumber\\[1ex]
C_{q_i u_j \Phi_u}^{(a)}&=
C_{q_i u_j \Phi_u}^{(b)}=
C_{q_i u_j \Phi_u}^{(c)}=
C_{q_i u_j \Phi_u}^{(d)}=
C_{q_i u_j \Phi_u}^{(e)}
=
\begin{pmatrix}
\checkmark & 0 & 0 \\
0 & \checkmark & 0 \\
0 & 0 & \checkmark
\end{pmatrix}.
\end{align}
Hence, the Wilson coefficients $C_{q_iu_j h}$ can be described by
\begin{align}
C_{q_i u_j h}^{(a)}&=
\begin{pmatrix}
\checkmark & 0 & 0 \\
0 & \checkmark & 0 \\
0 & 0 & \checkmark
\end{pmatrix}
,\quad
C_{q_i u_j h}^{(b)}=
C_{q_i u_j h}^{(d)}=
\begin{pmatrix}
\checkmark & 0 & \checkmark \\
\checkmark & \checkmark & \checkmark \\
\checkmark & 0 & \checkmark
\end{pmatrix}
,\quad
C_{q_i u_j h}^{(c)}=
\begin{pmatrix}
\checkmark & \checkmark & 0 \\
\checkmark & \checkmark & \checkmark \\
0 & \checkmark & \checkmark
\end{pmatrix}
,\nonumber\\
C_{q_i u_j h}^{(e)}&=
\begin{pmatrix}
\checkmark & \checkmark & \checkmark \\
\checkmark & \checkmark & 0 \\
\checkmark & 0 & \checkmark
\end{pmatrix}.
\end{align}

\end{enumerate}

\subsubsection{$\psi^2 H^2 D$}

Next, we investigate the $\psi^2 H^2 D$ operators:
\begin{align}
C_{\ell_i \ell_j hh}Q_{Hl}^{(1)} &= C_{\ell_i \ell_j hh}
\left(H^\dagger i\overleftrightarrow{D}_{\mu}H\right)
(\bar{\ell}_i\gamma^\mu \ell_j) ,
\nonumber\\
C_{\ell_i \ell_j hh}Q_{Hl}^{(3)} &= C_{\ell_i \ell_j hh}
\left(H^\dagger i\overleftrightarrow{D}_{\mu}^{\,I}H\right)
(\bar{\ell}_i\tau^I\gamma^\mu \ell_j) ,
\nonumber\\
C_{e_i e_j hh}Q_{He} &= C_{e_i e_j hh}
\left(H^\dagger i\overleftrightarrow{D}_{\mu}H\right)
(\bar{e}_i\gamma^\mu e_j) ,
\nonumber\\
C_{\ell_i \ell_j hh}Q_{Hq}^{(1)} &=C_{\ell_i \ell_j hh}
\left(H^\dagger i\overleftrightarrow{D}_{\mu}H\right)
(\bar{q}_i\gamma^\mu q_j) ,
\nonumber\\
C_{\ell_i \ell_j hh}Q_{Hq}^{(3)} &=C_{\ell_i \ell_j hh}
\left(H^\dagger i\overleftrightarrow{D}_{\mu}^{\,I}H\right)
(\bar{q}_i\tau^I\gamma^\mu q_j) ,
\nonumber\\
C_{u_i u_j hh}Q_{Hu} &= C_{u_i u_j hh}
\left(H^\dagger i\overleftrightarrow{D}_{\mu}H\right)
(\bar{u}_i\gamma^\mu u_j) ,
\nonumber\\
C_{e_i e_j hh}Q_{Hd} &= C_{e_i e_j hh}
\left(H^\dagger i\overleftrightarrow{D}_{\mu}H\right)
(\bar{d}_i\gamma^\mu d_j) ,
\nonumber\\
C_{u_i d_j hh}Q_{Hud} &= C_{u_i d_j hh}
\left(i\widetilde{H}^{\dagger}D_\mu H\right)
(\bar{u}_i\gamma^\mu d_j) ,
\end{align}
where some of the Wilson coefficients are described by the same ones due to the identification of classes \eqref{eq:charge}, and the textures of $\psi^2 H^2 D$ type operator can be understood by the following four operators:
\begin{align}
C_{\ell_i \ell_j hh}Q_{Hl}^{(1)} &= C_{\ell_i \ell_j hh}
\left(H^\dagger i\overleftrightarrow{D}_{\mu}H\right)
(\bar{\ell}_i\gamma^\mu \ell_j) 
\nonumber\\
&\equiv \Big[ C_{\ell_i \ell_j \Phi_d \Phi_d}\, c_\beta^2 \left(\Phi_d^\dagger i\overleftrightarrow{D}_{\mu}\Phi_d\right) + C_{\ell_i \ell_j \Phi_u \Phi_u}\, s_\beta^2 \left(\Phi_u^\dagger i\overleftrightarrow{D}_{\mu}\Phi_u\right) \nonumber\\ &\hspace{3.5em} + C_{\ell_i \ell_j \Phi_u \Phi_d}\, c_\beta s_\beta \left(\Phi_d^\dagger i\overleftrightarrow{D}_{\mu}\Phi_u\right) + C_{\ell_i \ell_j \Phi_u \Phi_d}\, s_\beta c_\beta \left(\Phi_u^\dagger i\overleftrightarrow{D}_{\mu}\Phi_d\right) \Big] (\bar{\ell}_i\gamma^\mu \ell_j) .
\nonumber\\
C_{e_i e_j hh}Q_{He} &= C_{e_i e_j hh}
\left(H^\dagger i\overleftrightarrow{D}_{\mu}H\right)
(\bar{e}_i\gamma^\mu e_j) 
\nonumber\\
&\equiv \Big[ C_{e_i e_j \Phi_d \Phi_d}\, c_\beta^2 \left(\Phi_d^\dagger i\overleftrightarrow{D}_{\mu}\Phi_d\right) + C_{e_i e_j \Phi_u \Phi_u}\, s_\beta^2 \left(\Phi_u^\dagger i\overleftrightarrow{D}_{\mu}\Phi_u\right) \nonumber\\ &\hspace{3.5em} + C_{e_i e_j \Phi_u \Phi_d}\, c_\beta s_\beta \left(\Phi_d^\dagger i\overleftrightarrow{D}_{\mu}\Phi_u\right) + C_{e_i e_j \Phi_u \Phi_d}\, s_\beta c_\beta \left(\Phi_u^\dagger i\overleftrightarrow{D}_{\mu}\Phi_d\right) \Big]
(\bar{e}_i\gamma^\mu e_j) ,
\nonumber\\
C_{u_i u_j hh}Q_{Hu} &= C_{u_i u_j hh}
\left(H^\dagger i\overleftrightarrow{D}_{\mu}H\right)
(\bar{u}_i\gamma^\mu u_j) 
\nonumber\\
&\equiv \Big[ C_{u_i u_j \Phi_d \Phi_d}\, c_\beta^2 \left(\Phi_d^\dagger i\overleftrightarrow{D}_{\mu}\Phi_d\right) + C_{u_i u_j \Phi_u \Phi_u}\, s_\beta^2 \left(\Phi_u^\dagger i\overleftrightarrow{D}_{\mu}\Phi_u\right) \nonumber\\ &\hspace{3.5em} + C_{u_i u_j \Phi_u \Phi_d}\, c_\beta s_\beta \left(\Phi_d^\dagger i\overleftrightarrow{D}_{\mu}\Phi_u\right) + C_{u_i u_j \Phi_u \Phi_d}\, s_\beta c_\beta \left(\Phi_u^\dagger i\overleftrightarrow{D}_{\mu}\Phi_d\right) \Big] (\bar{u}_i\gamma^\mu u_j),
\nonumber\\
 C_{u_i d_j hh}Q_{Hud} &= C_{u_i d_j hh}
\left(i\widetilde{H}^{\dagger}D_\mu H\right)
(\bar{u}_i\gamma^\mu d_j)
\nonumber\\
&\equiv \Big[ C_{u_i d_j \Phi_d \Phi_d}\, c_\beta^2 \left(i\widetilde{\Phi}_d^{\dagger}D_\mu \Phi_d\right) + C_{u_i d_j \Phi_u \Phi_u}\, s_\beta^2 \left(i\widetilde{\Phi}_u^{\dagger}D_\mu \Phi_u\right) \nonumber\\ &\hspace{3.5em} + C_{u_i d_j \Phi_u \Phi_d}\, c_\beta s_\beta \left(i\widetilde{\Phi}_d^{\dagger}D_\mu \Phi_u\right) + C_{u_i d_j \Phi_u \Phi_d}\, s_\beta c_\beta \left(i\widetilde{\Phi}_u^{\dagger}D_\mu \Phi_d\right) \Big] (\bar{u}_i\gamma^\mu d_j) .
\end{align}

For the five charge assignments in Table~\ref{tab:Z5Z5_tex2_1}, we find that all entries of the Wilson
coefficients $C_{\ell_i\ell_jhh}$, $C_{e_i e_jhh}$, and $C_{u_i u_jhh}$ are allowed.
For $C_{u_i d_jhh}$, associated with $Q_{Hud}$, all nine entries are allowed in
cases~(b)--(d), whereas one entry is forbidden in cases~(a) and~(e).
Specifically, the $(i,j)=(1,3)$ entry is forbidden in case~(a), while the
$(i,j)=(3,1)$ entry is forbidden in case~(e).
Thus, $Q_{Hud}$ contains eight allowed flavor components in cases~(a) and~(e),
and nine in cases~(b)--(d).

\subsubsection{$\psi^2 H^3$}

In this section, we list the Wilson coefficients for the $\psi^2 H^3$ operators: 
\begin{align}
C_{\ell_i e_j hhh}Q_{eH} &= C_{\ell_i e_j hhh}
(H^\dagger H)(\bar{l}_i e_j H),
\nonumber\\
C_{q_i u_j hhh}Q_{uH} &= C_{q_i u_j hhh}
(H^\dagger H)(\bar{q}_i u_j\widetilde{H}) ,
\nonumber\\
C_{\ell_i e_j hhh}Q_{dH} &= C_{\ell_i e_j hhh}
(H^\dagger H)(\bar{q}_i d_j H) ,
\end{align}
where some of the Wilson coefficients are described by the same ones due to the identification of classes \eqref{eq:charge}, and the textures of $\psi^2 H^3$ type operator can be understood by the following two operators:
\begin{align}
C_{\ell_i e_j hhh}Q_{eH} &= C_{\ell_i e_j hhh}
(H^\dagger H)(\bar{l}_i e_j H)
\nonumber\\
&\equiv 
C_{\ell_i e_j \Phi_d \Phi_d \Phi_d}\, c_\beta^3 (\Phi_d^\dagger \Phi_d)(\bar{l}_i e_j \Phi_d) 
+ C_{\ell_i e_j \Phi_d \Phi_d \Phi_u}\, c_\beta^2 s_\beta (\Phi_d^\dagger \Phi_d)(\bar{l}_i e_j \Phi_u) \nonumber\\ &\hspace{1em} 
+ C_{\ell_i e_j \Phi_d \Phi_u \Phi_u}\, s_\beta^2 c_\beta (\Phi_u^\dagger \Phi_u)(\bar{l}_i e_j \Phi_d) 
+ C_{\ell_i e_j \Phi_u \Phi_u \Phi_u}\, s_\beta^3 (\Phi_u^\dagger \Phi_u)(\bar{l}_i e_j \Phi_u) \nonumber\\ &\hspace{3.5em} 
+ C_{\ell_i e_j \Phi_d \Phi_d \Phi_u}\, s_\beta c_\beta^2 (\Phi_d^\dagger \Phi_u)(\bar{l}_i e_j \Phi_d) 
+ C_{\ell_i e_j \Phi_d \Phi_u \Phi_u}\, s_\beta^2c_\beta  (\Phi_d^\dagger \Phi_u)(\bar{l}_i e_j \Phi_u) \nonumber\\ &\hspace{1em} 
+ C_{\ell_i e_j \Phi_d \Phi_d \Phi_u}\, s_\beta c_\beta^2 (\Phi_u^\dagger \Phi_d)(\bar{l}_i e_j \Phi_d) 
+ C_{\ell_i e_j \Phi_d \Phi_u \Phi_u}\, s_\beta^2 c_\beta(\Phi_u^\dagger \Phi_d)(\bar{l}_i e_j \Phi_u) ,
\nonumber\\
 C_{q_i u_j hhh}Q_{uH} &= C_{q_i u_j hhh}
(H^\dagger H)(\bar{q}_i u_j\widetilde{H})
\nonumber\\
&\equiv 
C_{q_i u_j \Phi_d \Phi_d \Phi_d}\, c_\beta^3 (\Phi_d^\dagger \Phi_d)(\bar{q}_i u_j\widetilde{\Phi}_d) 
+ C_{q_i u_j \Phi_d \Phi_d \Phi_u}\, s_\beta c_\beta^2 (\Phi_d^\dagger \Phi_d)(\bar{q}_i u_j\widetilde{\Phi}_u) \nonumber\\ &\hspace{1em} 
+ C_{q_i u_j \Phi_d \Phi_u \Phi_u}\, s_\beta^2 c_\beta (\Phi_u^\dagger \Phi_u)(\bar{q}_i u_j\widetilde{\Phi}_d) 
+ C_{q_i u_j \Phi_u \Phi_u \Phi_u}\, s_\beta^3 (\Phi_u^\dagger \Phi_u)(\bar{q}_i u_j\widetilde{\Phi}_u) \nonumber\\ &\hspace{1em} 
+ C_{q_i u_j \Phi_d \Phi_d \Phi_u}\, s_\beta c_\beta^2 (\Phi_d^\dagger \Phi_u)(\bar{q}_i u_j\widetilde{\Phi}_d) 
+ C_{q_i u_j \Phi_d \Phi_u \Phi_u}\, s_\beta^2 c_\beta (\Phi_d^\dagger \Phi_u)(\bar{q}_i u_j\widetilde{\Phi}_u) \nonumber\\ &\hspace{1em} 
+ C_{q_i u_j \Phi_d \Phi_d \Phi_u}\, s_\beta c_\beta^2 (\Phi_u^\dagger \Phi_d)(\bar{q}_i u_j\widetilde{\Phi}_d) 
+ C_{q_i u_j \Phi_d \Phi_u \Phi_u}\, s_\beta^2 c_\beta (\Phi_u^\dagger \Phi_d)(\bar{q}_i u_j\widetilde{\Phi}_u).
\end{align}
For the five charge assignments in Table~\ref{tab:Z5Z5_tex2_1}, we find that all entries of Wilson coefficients $C_{\ell_i e_j hhh}$ and $C_{q_i u_j hhh}$ are allowed.

\subsection{Counting the number of dimension-six operators}

In this section, we count the number of physically distinct parameters in the Wilson coefficients. 

First, let us consider $(\bar{L}L)(\bar{L}L)$ operators for charge assignment (a). For the texture considered in this case, each Wilson-coefficient matrix contains 21 non-vanishing entries. If all non-zero entries are treated as independent complex parameters, this corresponds to 21 complex, or equivalently 42 real, degrees of freedom for each operator.

We then impose Hermiticity of the effective Lagrangian:
\begin{align}
C_{ijkl}=C_{jilk}^{\ast}.
\end{align}
This condition identifies each off-diagonal entry with its complex conjugate, while diagonal entries are required to be real. In the present texture, there are nine diagonal entries and six off-diagonal complex-conjugate pairs. Therefore, the number of independent real parameters is
\begin{align}
9+2\times 6=21.
\end{align}
Equivalently, these parameters can be decomposed into CP-even and CP-odd parts. The nine diagonal real coefficients and the real parts of the six complex pairs are CP-even, while the imaginary parts of the six complex pairs are CP-odd. Thus, for operators subject only to Hermiticity, one obtains
\begin{align}
N_{\rm CP\text{-}even}=9+6=15,
\qquad
N_{\rm CP\text{-}odd}=6.
\end{align}

For operators built from two identical left-handed currents, such as $Q_{\ell\ell}$ and $Q_{qq}^{(1,3)}$, one must additionally impose the symmetry under interchange of the two current bilinears:
\begin{align}
C_{ijkl}=C_{klij}.
\end{align}
This further identifies several independent entries. For the texture under consideration, the 21 non-zero entries are reduced to 15 independent ones. Combining this current-exchange symmetry with Hermiticity leaves nine real coefficients and three complex-conjugate pairs, corresponding to $9+2\times 3=15$ independent real parameters. In this case, the CP-even parameters consist of the nine real coefficients and the real parts of the three complex pairs, while the CP-odd parameters are given by the imaginary parts of the three complex pairs. Hence,
\begin{align}
N_{\rm CP\text{-}even}=9+3=12,
\qquad
N_{\rm CP\text{-}odd}=3.
\end{align}

In contrast, for semileptonic operators such as $Q_{\ell q}^{(1,3)}$, the two fermion currents are of different species, and the interchange symmetry does not apply. Consequently, only the Hermiticity condition is imposed, yielding 21 independent real parameters for each semileptonic operator, decomposed into 15 CP-even and 6 CP-odd parameters.
The results including the other charge assignments are summarized in Table~\ref{tab:LLbarLL_CP_count_operator_all}.
\begin{table}[H]
\centering
\scriptsize
\renewcommand{\arraystretch}{1.25}
\setlength{\tabcolsep}{3pt}
\caption{Numbers of CP-even and CP-odd degrees of freedom for the LL$\overline{\mathrm{LL}}$ operators for assignments (a)--(e) after imposing Hermiticity and identical-current exchange symmetries.}
\label{tab:LLbarLL_CP_count_operator_all}
\resizebox{\textwidth}{!}{%
\begin{tabular}{c|c|c|cc|cc|cc|cc|cc}
\hline
Operator
& Coefficient
& Conditions
& \multicolumn{2}{c|}{(a)}
& \multicolumn{2}{c|}{(b)}
& \multicolumn{2}{c|}{(c)}
& \multicolumn{2}{c|}{(d)}
& \multicolumn{2}{c}{(e)}
\\
\cline{4-13}
&
&
&
CP-even & CP-odd
& CP-even & CP-odd
& CP-even & CP-odd
& CP-even & CP-odd
& CP-even & CP-odd
\\
\hline

$Q_{\ell\ell}$
& $C_{ijkl}$
& Hermiticity $+$ current exchange
& $12$ & $3$
& $17$ & $8$
& $17$ & $8$
& $17$ & $8$
& $27$ & $18$
\\

\hline

$Q_{\ell q}^{(1)}$
& $C_{ijkl}$
& Hermiticity
& $15$ & $6$
& $25$ & $16$
& $25$ & $16$
& $25$ & $16$
& $45$ & $36$
\\

$Q_{\ell q}^{(3)}$
& $C_{ijkl}$
& Hermiticity
& $15$ & $6$
& $25$ & $16$
& $25$ & $16$
& $25$ & $16$
& $45$ & $36$
\\

\hline

$Q_{qq}^{(1)}$
& $C_{ijkl}$
& Hermiticity $+$ current exchange
& $12$ & $3$
& $17$ & $8$
& $17$ & $8$
& $17$ & $8$
& $27$ & $18$
\\

$Q_{qq}^{(3)}$
& $C_{ijkl}$
& Hermiticity $+$ current exchange
& $12$ & $3$
& $17$ & $8$
& $17$ & $8$
& $17$ & $8$
& $27$ & $18$
\\

\hline

\multicolumn{3}{c|}{Total}
& $\mathbf{66}$ & $\mathbf{21}$
& $\mathbf{101}$ & $\mathbf{56}$
& $\mathbf{101}$ & $\mathbf{56}$
& $\mathbf{101}$ & $\mathbf{56}$
& $\mathbf{171}$ & $\mathbf{126}$
\\

\hline
\end{tabular}%
}
\end{table}

Similarly, one can discuss $(\bar{L}L)(\bar{R}R)$ operators. 
In this case, we only impose Hermiticity of the effective Lagrangian. 
This condition identifies each off-diagonal entry of the Wilson-coefficient matrix with its complex conjugate, while diagonal entries are required to be real. Since the two currents in the $(\bar{L}L)(\bar{R}R)$ operators are composed of different chiral fermions, no additional identical-current exchange symmetry is imposed. 
From the textures of Wilson coefficients in the $(\bar{L}L)(\bar{R}R)$ operators, we count the CP-even and -odd degrees of freedom as shown in Table~\ref{tab:LLRR_CP_count_operator_all}. 
\begin{table}[H]
\centering
\scriptsize
\renewcommand{\arraystretch}{1.25}
\setlength{\tabcolsep}{3pt}
\caption{Numbers of CP-even and CP-odd degrees of freedom for the LLRR operators for assignments (a)--(e) after imposing Hermiticity.}
\label{tab:LLRR_CP_count_operator_all}
\resizebox{\textwidth}{!}{%
\begin{tabular}{c|c|cc|cc|cc|cc|cc}
\hline
Operator
& Coefficient
& \multicolumn{2}{c|}{(a)}
& \multicolumn{2}{c|}{(b)}
& \multicolumn{2}{c|}{(c)}
& \multicolumn{2}{c|}{(d)}
& \multicolumn{2}{c}{(e)}
\\
\cline{3-12}
&
&
CP-even & CP-odd
& CP-even & CP-odd
& CP-even & CP-odd
& CP-even & CP-odd
& CP-even & CP-odd
\\
\hline

$Q_{\ell e}$
& $D_{ijkl}$
& $29$ & $20$
& $32$ & $23$
& $35$ & $26$
& $41$ & $32$
& $35$ & $26$
\\

$Q_{\ell d}$
& $D_{ijkl}$
& $29$ & $20$
& $32$ & $23$
& $35$ & $26$
& $41$ & $32$
& $35$ & $26$
\\

$Q_{qe}$
& $D_{ijkl}$
& $29$ & $20$
& $32$ & $23$
& $35$ & $26$
& $41$ & $32$
& $35$ & $26$
\\

\hline

$Q_{\ell u}$
& $E_{ijkl}$
& $15$ & $6$
& $30$ & $21$
& $29$ & $20$
& $30$ & $21$
& $45$ & $36$
\\

$Q_{qu}^{(1)}$
& $E_{ijkl}$
& $15$ & $6$
& $30$ & $21$
& $29$ & $20$
& $30$ & $21$
& $45$ & $36$
\\

$Q_{qu}^{(8)}$
& $E_{ijkl}$
& $15$ & $6$
& $30$ & $21$
& $29$ & $20$
& $30$ & $21$
& $45$ & $36$
\\

\hline

$Q_{qd}^{(1)}$
& $D_{ijkl}$
& $29$ & $20$
& $32$ & $23$
& $35$ & $26$
& $41$ & $32$
& $35$ & $26$
\\

$Q_{qd}^{(8)}$
& $D_{ijkl}$
& $29$ & $20$
& $32$ & $23$
& $35$ & $26$
& $41$ & $32$
& $35$ & $26$
\\

\hline

\multicolumn{2}{c|}{Total}
& $\mathbf{190}$ & $\mathbf{118}$
& $\mathbf{250}$ & $\mathbf{178}$
& $\mathbf{262}$ & $\mathbf{190}$
& $\mathbf{295}$ & $\mathbf{223}$
& $\mathbf{310}$ & $\mathbf{238}$
\\

\hline
\end{tabular}%
}
\end{table}

The number of Wilson coefficients for the $(\bar{R}R
)(\bar{R}R)$ operators can be calculated in the same way as those for the $(\bar{L}L)(\bar{L}L)$ case, which is summarized in Table~\ref{tab:RRRR_CP_count_operator_all}. 
In addition to Hermiticity and the symmetry under
the interchange of the two fermion bilinears, 
the operator $Q_{ee}$ is subject to a
further identification following from the Fierz identity for right-handed vector currents:
\begin{align}
(\bar e_i\gamma_\mu e_j)(\bar e_k\gamma^\mu e_l)
=
(\bar e_k\gamma_\mu e_j)(\bar e_i\gamma^\mu e_l).
\end{align}
At the level of flavor indices, this relation identifies
\begin{align}
C_{ijkl}= C_{kjil},
\end{align}
and therefore reduces the number of independent flavor contractions beyond the reduction implied by Hermiticity and current exchange alone.

\begin{table}[H]
\centering
\scriptsize
\renewcommand{\arraystretch}{1.25}
\setlength{\tabcolsep}{3pt}
\caption{Numbers of CP-even and CP-odd degrees of freedom for the RRRR operators for assignments (a)--(e). The Fierz identity is included for $Q_{ee}$.}
\label{tab:RRRR_CP_count_operator_all}
\resizebox{\textwidth}{!}{%
\begin{tabular}{c|c|c|cc|cc|cc|cc|cc}
\hline
Operator
& Coefficient
& Conditions
& \multicolumn{2}{c|}{(a)}
& \multicolumn{2}{c|}{(b)}
& \multicolumn{2}{c|}{(c)}
& \multicolumn{2}{c|}{(d)}
& \multicolumn{2}{c}{(e)}
\\
\cline{4-13}
&
&
&
CP-even & CP-odd
& CP-even & CP-odd
& CP-even & CP-odd
& CP-even & CP-odd
& CP-even & CP-odd
\\
\hline

$Q_{ee}$
& $F_{ijkl}$
& Hermiticity $+$ current exchange $+$ Fierz
& $21$ & $15$
& $13$ & $7$
& $21$ & $15$
& $21$ & $15$
& $13$ & $7$
\\

$Q_{dd}$
& $F_{ijkl}$
& Hermiticity $+$ current exchange
& $27$ & $18$
& $17$ & $8$
& $27$ & $18$
& $27$ & $18$
& $17$ & $8$
\\

$Q_{ed}$
& $F_{ijkl}$
& Hermiticity
& $45$ & $36$
& $25$ & $16$
& $45$ & $36$
& $45$ & $36$
& $25$ & $16$
\\

\hline

$Q_{eu}$
& $G_{ijkl}$
& Hermiticity
& $29$ & $20$
& $29$ & $20$
& $45$ & $36$
& $38$ & $29$
& $35$ & $26$
\\

$Q_{uu}$
& $H_{ijkl}$
& Hermiticity $+$ current exchange
& $12$ & $3$
& $20$ & $11$
& $27$ & $18$
& $20$ & $11$
& $27$ & $18$
\\

$Q_{ud}^{(1)}$
& $G_{ijkl}$
& Hermiticity
& $29$ & $20$
& $29$ & $20$
& $45$ & $36$
& $38$ & $29$
& $35$ & $26$
\\

$Q_{ud}^{(8)}$
& $G_{ijkl}$
& Hermiticity
& $29$ & $20$
& $29$ & $20$
& $45$ & $36$
& $38$ & $29$
& $35$ & $26$
\\

\hline

\multicolumn{3}{c|}{Total}
& $\mathbf{192}$ & $\mathbf{132}$
& $\mathbf{162}$ & $\mathbf{102}$
& $\mathbf{255}$ & $\mathbf{195}$
& $\mathbf{227}$ & $\mathbf{167}$
& $\mathbf{187}$ & $\mathbf{127}$
\\

\hline
\end{tabular}%
}
\end{table}

We move to the $(\bar{L}R)(\bar{R}L)$ and $(\bar{L}R)(\bar{L}R)$ operators. 
In this case, the scalar and tensor four-fermion operators are generally non-Hermitian. Therefore, in the effective Lagrangian they should be accompanied by their Hermitian conjugates as
\begin{align}
\mathcal{L}_{\rm eff} \supset C\mathcal{O}+C^\ast\mathcal{O}^\dagger ,
\end{align}
meaning that we do not impose the Hermiticity condition on the Wilson coefficients of them unlike the current-current operators. 
Consequently, the Wilson coefficients are counted as independent complex parameters. 
The results are summarized in Table~\ref{tab:LRRL_CP_count_operator_all}.

\begin{table}[H]
\centering
\scriptsize
\renewcommand{\arraystretch}{1.25}
\setlength{\tabcolsep}{3pt}
\caption{Numbers of CP-even and CP-odd degrees of freedom for scalar and tensor four-fermion operators for assignments (a)--(e).}
\label{tab:LRRL_CP_count_operator_all}
\resizebox{\textwidth}{!}{%
\begin{tabular}{c|c|cc|cc|cc|cc|cc}
\hline
Operator
& Coefficient
& \multicolumn{2}{c|}{(a)}
& \multicolumn{2}{c|}{(b)}
& \multicolumn{2}{c|}{(c)}
& \multicolumn{2}{c|}{(d)}
& \multicolumn{2}{c}{(e)}
\\
\cline{3-12}
&
&
CP-even & CP-odd
& CP-even & CP-odd
& CP-even & CP-odd
& CP-even & CP-odd
& CP-even & CP-odd
\\
\hline

$Q_{\ell edq}$
& $D_{ijkl}^{\prime}$
& $49$ & $49$
& $55$ & $55$
& $61$ & $61$
& $73$ & $73$
& $61$ & $61$
\\

\hline

$Q_{quqd}^{(1)}$
& $I_{ijkl}$
& $18$ & $18$
& $51$ & $51$
& $53$ & $53$
& $59$ & $59$
& $63$ & $63$
\\

$Q_{quqd}^{(8)}$
& $I_{ijkl}$
& $18$ & $18$
& $51$ & $51$
& $53$ & $53$
& $59$ & $59$
& $63$ & $63$
\\

\hline

$Q_{\ell equ}^{(1)}$
& $J_{ijkl}$
& $18$ & $18$
& $51$ & $51$
& $53$ & $53$
& $59$ & $59$
& $63$ & $63$
\\

$Q_{\ell equ}^{(3)}$
& $J_{ijkl}$
& $18$ & $18$
& $51$ & $51$
& $53$ & $53$
& $59$ & $59$
& $63$ & $63$
\\

\hline

\multicolumn{2}{c|}{Total}
& $\mathbf{121}$ & $\mathbf{121}$
& $\mathbf{259}$ & $\mathbf{259}$
& $\mathbf{273}$ & $\mathbf{273}$
& $\mathbf{309}$ & $\mathbf{309}$
& $\mathbf{313}$ & $\mathbf{313}$
\\

\hline
\end{tabular}%
}
\end{table}

Finally, we show the number of CP-even and -odd degrees of freedom for the dipole operators in Table~\ref{tab:dipole_CP_count_operator_all}, and the other operators such as $\psi^2 H^2 D$ in Table~\ref{tab:psi2H2D_CP_count_operator_all} and $\psi^2 H^3$ in Table~\ref{tab:psi2H3_CP_count_operator_all}. 

Since the SM-like Higgs doublet is expressed as in Eq.~\eqref{eq:def_Higgs}, each dipole operator is decomposed into two independent operator structures,
one involving $\Phi_u$ and the other involving $\Phi_d$. 
For instance,
\begin{align}
C_{\ell_i e_j h}
(\bar{\ell}_L^i \sigma^{\mu\nu}e_R^j)\tau^I H W_{\mu\nu}^I
\rightarrow
&
\;
C_{\ell_i e_j\Phi_u}\,s_\beta
(\bar{\ell}_L^i \sigma^{\mu\nu}e_R^j)\tau^I\Phi_u W_{\mu\nu}^I
\nonumber\\
&
+
C_{\ell_i e_j\Phi_d}\,c_\beta
(\bar{\ell}_L^i \sigma^{\mu\nu}e_R^j)\tau^I\Phi_d W_{\mu\nu}^I .
\label{eq:higgs_decomposition}
\end{align}
In the context of the two-Higgs doublet model, the Wilson coefficients multiplying the $\Phi_u$ and $\Phi_d$ structures are counted independently, 
since they correspond to distinct operators in the two-Higgs-doublet description. 
In the context of the SMEFT, however, both terms originate from the same SMEFT operator after identifying the SM-like Higgs direction. 
Hence, to compare the number of independent operators with the SMEFT subject to flavor symmetries, we count the independent Wilson coefficients $C_{\ell_i e_j h}$ in the case of Eq.~\eqref{eq:higgs_decomposition}, i.e., focusing on the Wilson coefficients in the SMEFT. 
The same counting is applied to the cases with $\psi^2 H^2 D$ and $\psi^2 H^3$. 
Note that the dipole operators and $\psi^2 H^3$ types are non-Hermitian and appear in the
effective Lagrangian together with their Hermitian conjugates:
\begin{align}
\mathcal{L}_{\rm eff}
\supset
C\,Q+C^\ast Q^\dagger .
\end{align}
Among the $\psi^2 H^2 D$ operators, $Q_{Hud}$ is the only non-Hermitian operator. 

\begin{table}[H]
\centering
\scriptsize
\renewcommand{\arraystretch}{1.25}
\setlength{\tabcolsep}{3pt}
\caption{Numbers of CP-even and CP-odd degrees of freedom for dipole operators for assignments (a)--(e).}
\label{tab:dipole_CP_count_operator_all}
\resizebox{\textwidth}{!}{%
\begin{tabular}{c|c|cc|cc|cc|cc|cc}
\hline
Operator
& Coefficient
& \multicolumn{2}{c|}{(a)}
& \multicolumn{2}{c|}{(b)}
& \multicolumn{2}{c|}{(c)}
& \multicolumn{2}{c|}{(d)}
& \multicolumn{2}{c}{(e)}
\\
\cline{3-12}
&
&
CP-even & CP-odd
& CP-even & CP-odd
& CP-even & CP-odd
& CP-even & CP-odd
& CP-even & CP-odd
\\
\hline

$Q_{eW}$
& $C_{\ell_i e_j h}$
& $6$ & $6$
& $7$ & $7$
& $7$ & $7$
& $7$ & $7$
& $6$ & $6$
\\

$Q_{eB}$
& $C_{\ell_i e_j h}$
& $6$ & $6$
& $7$ & $7$
& $7$ & $7$
& $7$ & $7$
& $6$ & $6$
\\

\hline

$Q_{uG}$
& $C_{q_i u_j h}$
& $3$ & $3$
& $7$ & $7$
& $7$ & $7$
& $7$ & $7$
& $7$ & $7$
\\

$Q_{uW}$
& $C_{q_i u_j h}$
& $3$ & $3$
& $7$ & $7$
& $7$ & $7$
& $7$ & $7$
& $7$ & $7$
\\

$Q_{uB}$
& $C_{q_i u_j h}$
& $3$ & $3$
& $7$ & $7$
& $7$ & $7$
& $7$ & $7$
& $7$ & $7$
\\

\hline

$Q_{dG}$
& $C_{\ell_i e_j h}$
& $6$ & $6$
& $7$ & $7$
& $7$ & $7$
& $7$ & $7$
& $6$ & $6$
\\

$Q_{dW}$
& $C_{\ell_i e_j h}$
& $6$ & $6$
& $7$ & $7$
& $7$ & $7$
& $7$ & $7$
& $6$ & $6$
\\

$Q_{dB}$
& $C_{\ell_i e_j h}$
& $6$ & $6$
& $7$ & $7$
& $7$ & $7$
& $7$ & $7$
& $6$ & $6$
\\

\hline

\multicolumn{2}{c|}{Total}
& $\mathbf{39}$ & $\mathbf{39}$
& $\mathbf{56}$ & $\mathbf{56}$
& $\mathbf{56}$ & $\mathbf{56}$
& $\mathbf{56}$ & $\mathbf{56}$
& $\mathbf{51}$ & $\mathbf{51}$
\\

\hline
\end{tabular}%
}
\end{table}

\begin{table}[H]
\centering
\scriptsize
\renewcommand{\arraystretch}{1.25}
\setlength{\tabcolsep}{3pt}
\caption{Numbers of CP-even and CP-odd degrees of freedom for
$\psi^2H^2D$ operators for assignments (a)--(e).}
\label{tab:psi2H2D_CP_count_operator_all}
\resizebox{\textwidth}{!}{%
\begin{tabular}{c|c|cc|cc|cc|cc|cc}
\hline
Operator
& Coefficient
& \multicolumn{2}{c|}{(a)}
& \multicolumn{2}{c|}{(b)}
& \multicolumn{2}{c|}{(c)}
& \multicolumn{2}{c|}{(d)}
& \multicolumn{2}{c}{(e)}
\\
\cline{3-12}
&
&
CP-even & CP-odd
& CP-even & CP-odd
& CP-even & CP-odd
& CP-even & CP-odd
& CP-even & CP-odd
\\
\hline

$Q_{Hl}^{(1)}$
& $C_{\ell_i\ell_jhh}$
& $6$ & $3$
& $6$ & $3$
& $6$ & $3$
& $6$ & $3$
& $6$ & $3$
\\

$Q_{Hl}^{(3)}$
& $C_{\ell_i\ell_jhh}$
& $6$ & $3$
& $6$ & $3$
& $6$ & $3$
& $6$ & $3$
& $6$ & $3$
\\

\hline

$Q_{He}$
& $C_{e_ie_jhh}$
& $6$ & $3$
& $6$ & $3$
& $6$ & $3$
& $6$ & $3$
& $6$ & $3$
\\

\hline

$Q_{Hq}^{(1)}$
& $C_{\ell_i\ell_jhh}$
& $6$ & $3$
& $6$ & $3$
& $6$ & $3$
& $6$ & $3$
& $6$ & $3$
\\

$Q_{Hq}^{(3)}$
& $C_{\ell_i\ell_jhh}$
& $6$ & $3$
& $6$ & $3$
& $6$ & $3$
& $6$ & $3$
& $6$ & $3$
\\

\hline

$Q_{Hu}$
& $C_{u_iu_jhh}$
& $6$ & $3$
& $6$ & $3$
& $6$ & $3$
& $6$ & $3$
& $6$ & $3$
\\

$Q_{Hd}$
& $C_{e_ie_jhh}$
& $6$ & $3$
& $6$ & $3$
& $6$ & $3$
& $6$ & $3$
& $6$ & $3$
\\

\hline

$Q_{Hud}$
& $C_{u_id_jhh}$
& $8$ & $8$
& $9$ & $9$
& $9$ & $9$
& $9$ & $9$
& $8$ & $8$
\\

\hline

\multicolumn{2}{c|}{Total}
& $\mathbf{50}$ & $\mathbf{29}$
& $\mathbf{51}$ & $\mathbf{30}$
& $\mathbf{51}$ & $\mathbf{30}$
& $\mathbf{51}$ & $\mathbf{30}$
& $\mathbf{50}$ & $\mathbf{29}$
\\

\hline
\end{tabular}%
}
\end{table}

\begin{table}[H]
\centering
\scriptsize
\renewcommand{\arraystretch}{1.25}
\setlength{\tabcolsep}{3pt}
\caption{Numbers of CP-even and CP-odd degrees of freedom for
$\psi^2H^3$ operators for assignments (a)--(e).}
\label{tab:psi2H3_CP_count_operator_all}
\resizebox{\textwidth}{!}{%
\begin{tabular}{c|c|cc|cc|cc|cc|cc}
\hline
Operator
& Coefficient
& \multicolumn{2}{c|}{(a)}
& \multicolumn{2}{c|}{(b)}
& \multicolumn{2}{c|}{(c)}
& \multicolumn{2}{c|}{(d)}
& \multicolumn{2}{c}{(e)}
\\
\cline{3-12}
&
&
CP-even & CP-odd
& CP-even & CP-odd
& CP-even & CP-odd
& CP-even & CP-odd
& CP-even & CP-odd
\\
\hline

$Q_{eH}$
& $C_{\ell_i e_j hhh}$
& $9$ & $9$
& $9$ & $9$
& $9$ & $9$
& $9$ & $9$
& $9$ & $9$
\\

$Q_{uH}$
& $C_{q_i u_j hhh}$
& $9$ & $9$
& $9$ & $9$
& $9$ & $9$
& $9$ & $9$
& $9$ & $9$
\\

$Q_{dH}$
& $C_{\ell_i e_j hhh}$
& $9$ & $9$
& $9$ & $9$
& $9$ & $9$
& $9$ & $9$
& $9$ & $9$
\\

\hline

\multicolumn{2}{c|}{Total}
& $\mathbf{27}$ & $\mathbf{27}$
& $\mathbf{27}$ & $\mathbf{27}$
& $\mathbf{27}$ & $\mathbf{27}$
& $\mathbf{27}$ & $\mathbf{27}$
& $\mathbf{27}$ & $\mathbf{27}$
\\

\hline
\end{tabular}%
}
\end{table}

\subsection{Summary}

Tables~\ref{tab:flavour_symmetry_comparison} and~\ref{tab:NISR_cases_comparison} summarize the numbers of independent CP-even and CP-odd Wilson coefficients. 
In the absence of flavor symmetry, there are 1350 CP-even and 1149 CP-odd coefficients, giving 2499 independent coefficients in total. These numbers are reduced to 52 CP-even and 17 CP-odd coefficients for $U(3)^5$ at $\mathcal{O}(Y_{e,d,u})$, and to 212 CP-even and 111 CP-odd coefficients for $U(2)^5$ at $\mathcal{O}(V,\Delta)$.

The non-invertible selection rules also reduce the number of independent coefficients, 
although the restrictions are weaker than those obtained for $U(3)^5$ and $U(2)^5$ at the spurion orders. 
Among Cases (a)--(e), Case (a) gives the smallest number of coefficients, with 694 CP-even and 493 CP-odd coefficients. 
The purely bosonic operator classes are independent of the matter-field assignments and contain the same numbers of coefficients in all five cases. 
The numbers for the $\psi^2 H^3$ class are common to Cases~(a)--(e), while the
$\psi^2 H^2 D$ class differs slightly due to the flavor texture of $Q_{Hud}$ in
Cases~(a) and~(e). 
By contrast, the $\psi^2XH$ and four-fermion sectors depend on the assignments. 
Most of the variation in the total number of coefficients comes from the four-fermion operators. 
In particular, Case (a) is strongly restricted in the $(\bar L L)(\bar L L)$, $(\bar L L)(\bar R R)$, and scalar four-fermion sectors. 

Although the reduction in the number of coefficients is less pronounced than in the $U(3)^5$ and $U(2)^5$ cases, the flavor structures obtained here are not simply determined by the Yukawa textures. 
In particular, the Wilson coefficients enjoy flavor structures different from those of the Yukawa couplings, 
in contrast to the MFV hypothesis. Their phenomenological consequences are discussed in the next section. 
Note that in the present analysis, we have focused on the charge assignments
reproducing the Yukawa textures in Eq.~\eqref{eq:Yftexture}. 
A systematic classification
of dimension-six operators for other Yukawa textures is beyond the scope
of this work and will be investigated elsewhere.

\begin{table}[H]
\centering
\scriptsize
\renewcommand{\arraystretch}{1.25}
\setlength{\tabcolsep}{2.5pt}
\caption{Number of CP-even and CP-odd independent Wilson coefficients for different flavor structures, including $U(2)^5$, $U(3)^5$ (MFV) and without symmetry.
The numbers shown here are taken from Ref.\cite{Faroughy:2020ina}, where operators are counted according to the minimal number of spurion insertions required to reproduce the Yukawa structure.
The spurions $Y_{e,d,u}$, $V$ and $\Delta$ are those required to reproduce the Yukawa structure in $U(3)^5$ and $U(2)^5$ cases. 
}
\label{tab:flavour_symmetry_comparison}
\resizebox{\textwidth}{!}{%
\begin{tabular}{c|cc|cc|cc}
\hline
\multirow{2}{*}{Operator class}
& \multicolumn{2}{c|}{No symmetry, 3 Gen.}
& \multicolumn{2}{c|}{$U(3)^5$, $\mathcal{O}(Y_{e,d,u}^1)$}
& \multicolumn{2}{c}{$U(2)^5$, $\mathcal{O}(V^1,\Delta^1)$} \\
\cline{2-7}
&
CP-even
& CP-odd
&
CP-even
& CP-odd
&
CP-even
& CP-odd \\
\hline
$X^3$, $H^6$, $H^4D^2$, $X^2H^2$
& $9$
& $6$
& $9$
& $6$
& $9$
& $6$ \\
\hline
$\psi^2H^3$
& $27$
& $27$
& $3$
& $3$
& $9$
& $9$ \\
$\psi^2XH$
& $72$
& $72$
& $8$
& $8$
& $24$
& $24$ \\
$\psi^2H^2D$
& $51$
& $30$
& $7$
& --
& $19$
& $5$ \\
\hline
$(\bar L L)(\bar L L)$
& $171$
& $126$
& $8$
& --
& $40$
& $17$ \\
$(\bar L L)(\bar R R)$
& $360$
& $288$
& $8$
& --
& $53$
& $21$ \\
$(\bar R R)(\bar R R)$
& $255$
& $195$
& $9$
& --
& $29$
& -- \\
$(\bar L R)(\bar R L)$ and $(\bar L R)(\bar L R)$
& $405$
& $405$
& --
& --
& $29$
& $29$ \\
\hline
Total
& $1350$
& $1149$
& $52$
& $17$
& $212$
& $111$ \\
\hline
\end{tabular}%
}
\end{table}

\begin{table}[H]
\centering
\scriptsize
\renewcommand{\arraystretch}{1.25}
\setlength{\tabcolsep}{2.5pt}
\caption{Number of CP-even and CP-odd independent Wilson coefficients for Cases (a)--(e).}
\label{tab:NISR_cases_comparison}
\resizebox{\textwidth}{!}{%
\begin{tabular}{c|cc|cc|cc|cc|cc}
\hline

\multirow{2}{*}{Operator class}
&
\multicolumn{2}{c|}{Case (a)}
&
\multicolumn{2}{c|}{Case (b)}
&
\multicolumn{2}{c|}{Case (c)}
&
\multicolumn{2}{c|}{Case (d)}
&
\multicolumn{2}{c}{Case (e)}
\\

\cline{2-11}

&
CP-even & CP-odd
&
CP-even & CP-odd
&
CP-even & CP-odd
&
CP-even & CP-odd
&
CP-even & CP-odd
\\

\hline

$X^3$, $H^6$, $H^4D^2$, $X^2H^2$
&
$9$ & $6$
&
$9$ & $6$
&
$9$ & $6$
&
$9$ & $6$
&
$9$ & $6$
\\

\hline

$\psi^2H^3$
&
$27$ & $27$
&
$27$ & $27$
&
$27$ & $27$
&
$27$ & $27$
&
$27$ & $27$
\\

$\psi^2XH$
&
$39$ & $39$
&
$56$ & $56$
&
$56$ & $56$
&
$56$ & $56$
&
$51$ & $51$
\\

$\psi^2H^2D$
&
$50$ & $29$
&
$51$ & $30$
&
$51$ & $30$
&
$51$ & $30$
&
$50$ & $29$
\\

\hline

$(\bar L L)(\bar L L)$
&
$66$ & $21$
&
$101$ & $56$
&
$101$ & $56$
&
$101$ & $56$
&
$171$ & $126$
\\

$(\bar L L)(\bar R R)$
&
$190$ & $118$
&
$250$ & $178$
&
$262$ & $190$
&
$295$ & $223$
&
$310$ & $238$
\\

$(\bar R R)(\bar R R)$
&
$192$ & $132$
&
$162$ & $102$
&
$255$ & $195$
&
$227$ & $167$
&
$187$ & $127$
\\

$(\bar L R)(\bar R L)$ and $(\bar L R)(\bar L R)$
&
$121$ & $121$
&
$259$ & $259$
&
$273$ & $273$
&
$309$ & $309$
&
$313$ & $313$
\\

\hline

Total
&
$\mathbf{694}$ & $\mathbf{493}$
&
$\mathbf{915}$ & $\mathbf{714}$
&
$\mathbf{1034}$ & $\mathbf{833}$
&
$\mathbf{1075}$ & $\mathbf{874}$
&
$\mathbf{1118}$ & $\mathbf{917}$
\\

\hline
\end{tabular}%
}
\end{table}

\section{Phenomenological studies}
\label{sec:pheno}

In this section, we discuss several phenomenological consequences of the flavor textures obtained in the previous sections. We first examine the right-handed quark and charged-lepton mixing matrices for case (a) to move to the mass basis. 
We then investigate the phenomenological implications of the four-fermion operators for semileptonic $B$-meson decays, focusing on the $R(D^{(*)})$ anomaly. We also study the flavor structure of the leptonic dipole operators and their implications for charged-lepton flavor-violating decays.

The $R(D^{(*)})$ anomaly refers to the deviation of the lepton flavor universality ratios in semileptonic $B$-meson decays.
Although recent measurements have somewhat reduced the tension, a non-negligible discrepancy with the SM remains~\cite{HeavyFlavorAveragingGroupHFLAV:2024ctg}.
The anomaly may indicate a violation of lepton flavor universality in the $b\to c\ell\bar{\nu}$ transition.
A possible NP contribution is expected to enhance $b\to c\tau\bar{\nu}$, while being suppressed for $b\to c\mu\bar \nu$ and $b\to c e\bar \nu$.
This suggests a hierarchical flavor structure with NP predominantly coupled to the 3rd generation.
In the SMEFT framework, $b\to c\tau\bar \nu$ can be generated by the semileptonic operators $Q_{lq}^{(3)}$, $Q_{ledq}$, $Q_{lequ}^{(1)}$, and $Q_{lequ}^{(3)}$~\cite{Jenkins:2017jig}.
We discuss the flavor structure of the Wilson coefficients relevant to the $R(D^{(*)})$ anomaly in our model.

Recently, a deviation from the SM prediction has also been reported in $B^+ \to K^+ \nu\bar{\nu}$, with the measured branching fraction lying above the SM expectation~\cite{Belle-II:2023esi}.
Since $Q_{lq}^{(3)}$, which contributes to $b\to c\tau\bar{\nu}$, also induces $b\to s\nu\bar{\nu}$, we therefore investigate the impact of $Q_{lq}^{(3)}$ on the $b\to s\nu\bar{\nu}$ process.

Finally, we investigate the flavor structure of the leptonic dipole operators $Q_{eW}$ and $Q_{eB}$, which can induce charged-lepton flavor-violating decays through their flavor-off-diagonal components.
We discuss how these operators are generated in our model and examine the resulting implications for processes such as $\mu\to e\gamma$ and $\tau\to\ell\gamma$.

\subsection{Right-handed mixing matrices of quarks and leptons}

To transform the Wilson coefficients in the flavor basis to the ones in the mass basis, we discuss the right-handed mixing matrices for quarks and charged leptons as well as the left handed ones. 
Since the up-type quark mass matrix is diagonal, we discuss the down-type mass matrix, which is given in terms of powers of the Cabibbo angle $\lambda\simeq 0.2$ 
and is chosen to reproduce masses and CKM matrix elements:
 \begin{align}
 	Y_d \sim\begin{pmatrix}
 		0  &  \lambda^3 & 0\\
 		 \lambda^3 &  e^{i \phi_d}\lambda^2 & \lambda^2\\
 		0 & 1 &1
 	\end{pmatrix}_{LR}\,,
 \end{align}
 where the CP phase $\phi_d$ is put in the (2,2) entry without
 loss of generality.

Let us introduce the notation for the quark and lepton fields in the flavor and the mass bases:
\begin{align}
&d_L =  L_d^\dagger \, d_{Lm}\,, \quad
d_R =  R_{d}^\dagger \, d_{Rm}\,,\quad  
u_L =  \, u_{Lm}\,, \quad  
u_R =  \, u_{Rm}\,, \nonumber\\
&e_L=  L_e\, e_{Lm}\,, \ \quad
e_R=  R_{e}\, e_{Rm}\,,\ \quad
\nu_L= \nu _{Lm}\,,
\label{basis-def}
\end{align}
where the subscripts $L$ and $R$ denote the left-handed and right-handed fields, respectively, and the index $m$ denotes a mass eigenstate. 
The unitary matrices $L_d$, $R_d$, $L_e$ and $R_e$
satisfy
\begin{align}
L_d Y_d R_{d}^\dagger={\rm diag}\,(y_d,\, y_s, \,y_b)\,,\qquad
L_e^\dagger Y_e R_{e}= {\rm diag}\,(y_e,\, y_\mu, \,y_\tau)\,.
\end{align}
The right-handed mixing matrix satisfies
  \begin{align}
 R_{d} Y_d^\dagger  Y_d R_{d}^\dagger= 
 {\rm diag}\,(y_d^2,\, y_s^2, \,y_b^2)\,.
 \end{align}
We obtain explicitly:
 \begin{align}
Y_d^\dagger  Y_d\sim
\begin{pmatrix}
\lambda^6  &   e^{i \phi_d}\lambda^5&  \lambda^5\\
 e^{-i \phi_d}\lambda^5 & 1 & 1\\
 \lambda^5 & 1 &1
\end{pmatrix}
= P_d^*
\begin{pmatrix}
\lambda^6  &  \lambda^5&   e^{-i \phi_d}\lambda^5\\
\lambda^5 & 1 & 1\\
 e^{i \phi_d} \lambda^5 & 1 &1
\end{pmatrix}P_d\,, \qquad  
{P_d=
\begin{pmatrix}
e^{-i \phi_d} &  0&   0\\
0 & 1 & 0\\
0 & 0 &1
\end{pmatrix}\,.
}
\end{align}
  The phase matrix $P_d$ can be absorbed into a redefinition of the right-handed down-type quark fields. The effect of the CP-violating phase on the mixing matrix obtained from $Y_d^\dagger Y_d$ is expected to be small. 
Therefore, we neglect the CP phase to obtain the right-handed mixing angles.
We then obtain the following approximate form of $R_d^\dagger$ matrix:
 \begin{align}
  R_d^\dagger
\simeq\frac{1}{\sqrt{2}}
\begin{pmatrix} 
\sqrt{2} &  \sqrt{2}\lambda&   \lambda^5\\
-\lambda & 1 & 1\\
 \lambda & -1 &1
\end{pmatrix}\,.
\label{Rd}
\end{align}
Next, we discuss the right-handed mixing matrix of the charged lepton mass matrix, given by 
\begin{align}
 	Y_{e}\sim 
 	\begin{pmatrix}
 		0  & \lambda^{3} & 0\\
 		 \lambda^{3}  &    e^{i \phi_e} \kappa\lambda^2&1\\
 	0  &  \lambda^2 & 1
 	\end{pmatrix}_{LR}\, ,
 \end{align}
  where the CP phase $\phi_e$ is put in the (2,2) entry without
 loss of generality, and $\kappa$ is put to reproduce the proper mass ratios. 
 The right-handed mixing matrix satisfies
 \begin{align}
 R_{e}^\dagger Y_e^\dagger  Y_e R_{e}=
 {\rm diag}\,(y_e^2,\, y_\mu^2, \,y_\tau^2)\,.
 \end{align}
 We obtain explicitly:
   \begin{align}
 Y_e^\dagger  Y_e\sim
 \begin{pmatrix}
 \lambda^6  &   e^{i \phi_e}\kappa\lambda^5&  \lambda^3\\
 e^{-i \phi_e}\kappa\lambda^5 & \kappa^2\lambda^4 & 
 e^{-i \phi_e}\kappa\lambda^2+\lambda^2\\
 \lambda^3 &  e^{i \phi_e}\kappa\lambda^2+\lambda^2 &1
 \end{pmatrix}\,,
 \end{align}
 where $\kappa\simeq 3$ is chosen to reproduce the observed charged lepton mass ratios.
 Defining $\kappa=\zeta/\lambda$, where $\zeta\simeq 1/2$,
 we obtain 
  \begin{align}
& Y_e^\dagger  Y_e\sim
 \begin{pmatrix}
 \lambda^6  &   e^{i \phi_e}\zeta\lambda^4&  \lambda^3\\
 {e^{-i \phi_e}\zeta\lambda^{4}}
 & \zeta^2\lambda^2 & 
 e^{-i \phi_e}\zeta\lambda\\
 \lambda^3 &  e^{i \phi_e}\zeta\lambda &1
 \end{pmatrix}
 = P_e^*
 \begin{pmatrix}
  \lambda^6  &   \zeta\lambda^4&   \lambda^3\\
\zeta\lambda^4 & \zeta^2\lambda^2 & 
\zeta\lambda\\
  \lambda^3 &  \zeta\lambda &1
 \end{pmatrix}P_e\,, \nonumber \\ 
 \nonumber \\ 
 &{P_e
 =\begin{pmatrix}
 1 &  0&   0\\
 0 &  e^{i \phi_e}  & 0\\
 0 & 0 &1
 \end{pmatrix}\,.
 }
 \end{align}
The phase matrix $P_e$ can be absorbed into a redefinition of the right-handed charged-lepton fields.
The effect of the CP-violating phase on the right-handed mixing angles obtained from $Y_e^\dagger Y_e$ is expected to be small. 
Therefore, we also neglect the CP phase to obtain the right-handed mixing angles.
We then obtain the $R_e$ matrix as follows:
 \begin{align}
 R_e
 \simeq 
 \begin{pmatrix} 
 1 &  \lambda^2/\zeta&    \zeta\lambda^3\\
 - \lambda^2/\zeta & 1 & \zeta\lambda\\
\zeta \lambda^3 & -\zeta\lambda &1
 \end{pmatrix}\,,
 \label{Re}
 \end{align}
 where $\zeta\simeq 1/2$.

On the other hand, the flavor basis is in agreement with the mass basis for the up-type quarks and the neutrinos as seen in Eq.~\eqref{basis-def}.
Therefore, by using the convention of the Particle Data Group \cite{ParticleDataGroup:2024cfk}, we have 
\begin{align}
L_{d}^\dagger= V_{CKM}
\,,\qquad
L_{e}^\dagger=U_{PMNS}\,,
\end{align}
 for the left handed mixing matrices.
Hereafter, we use the abbreviations $V$ for $V_{CKM}$
and  $U$ for $U_{PMNS}$. 
The right-handed rotation matrices $R_d$ and $R_e$ are given in 
Eqs.~\eqref{Rd} and \eqref{Re}, respectively.

\subsection{$Q^{(1)}_{\ell q}$ and $Q^{(3)}_{\ell q}$ in case (a)}

We discuss the phenomenology of the following two operators for case (a):
\begin{align}
 Q_{\ell q}^{(1)} = (\bar{\ell}_L^i\gamma_\mu \ell^j_L)(\bar{q}^k_L \gamma^\mu q^l_L), \qquad
    Q_{\ell q}^{(3)} =(\bar{\ell}_L^i\gamma_\mu \tau^I \ell^j_L)(\bar{q}^k_L \gamma^\mu \tau^I q^l_L)\,,
\end{align}
where the corresponding Wilson coefficients are both given by $C_{ijkl}$.
While only $Q_{\ell q}^{(3)}$ contributes to the charged-current process $b\to c\ell\bar{\nu_\ell}$, we also discuss $Q_{\ell q}^{(1)}$ in the same manner since it has the same flavor structure. Both $Q_{\ell q}^{(3)}$ and $Q_{\ell q}^{(1)}$ contribute to the neutral-current process $b\to s\nu\bar{\nu}$.

\subsubsection{$b\to c \ell\bar\nu_\ell$}

Let us discuss  $C_{ijkl}^{(a)}$ for case (a), which is given in the flavor basis as
seen in Eq.~\eqref{CLLL(a)}
\begin{align}
C_{ijkl}^{(a)}= f_{ijkl}\left(\delta_{ij}\delta_{kl} +\delta_{ik}\delta_{jl} + \delta_{il}\delta_{jk}\right),
\nonumber
\end{align}
where $f_{ijkl}$ is an arbitrary tensor parametrizing the non-vanishing entries.
First, we discuss the NP contribution through $C_{ijkl}$ to the charged-current process 
$b\to c \ell\bar\nu_\ell$ where $\ell$ denotes $\tau$, $\mu$ or $e$.
In the mass basis, the coefficient is generally written as:
\begin{align}
\tilde C_{pqrs}^{(a)}
&=\sum_{i,\ell=1}^3 U_{pi}  V_{s\ell}^* C_{ijk\ell}^{(a)} \nonumber \\
&=\sum_{i,\ell=1}^3 U_{pi}  V_{s\ell}^* f_{ijkl}\left(\delta_{ij}\delta_{kl} +\delta_{ik}\delta_{jl} + \delta_{il}\delta_{jk}\right)\,,
\label{mass4L(a)3}
\end{align}
where 
$\tilde C$ denotes the Wilson coefficient in the mass basis, and
the subscripts  $p,\,q$ stand for  mass eigenstates of leptons
as well as $r,\,s$ of quarks.
It is noted that  the flavor basis is in agreement with the mass basis for 
the up-type quarks and neutrinos, that is, $q=j$ and $k=r$ in Eq.~\eqref{mass4L(a)3}.
Therefore, we have
 \begin{align}
\tilde C_{pqrs}^{(a)}
&=
f _{qqrr }\,U_{pq}V^{*}_{sr}
+
f _{rqrq }\,U_{pr}V^{*}_{sq}
+
\sum_{\ell=1}^{3}f_{\ell r r \ell}  U_{p\ell}V^{*}_{s\ell}\,\delta_{qr}\,.
\label{Cpqrs}
\end{align}
For $b\to c \ell\bar\nu_\ell$, which corresponds to $q=p$, $r=2$ and $s=3$,
we have
 \begin{align}
\tilde C_{pp23}^{(a)}
&=
f _{qq22 }\,U_{qq}V^{*}_{32}
+
f _{2q2q }\,U_{q2}V^{*}_{3q}
+
\sum_{\ell=1}^{3}f_{\ell 22 \ell}  U_{q\ell}V^{*}_{3\ell}\,\delta_{q2}\,.
\end{align}
Therefore, for $b\to c \tau\bar\nu_\tau$ ($p=q=3$), we obtain
  \begin{align}
\tilde C_{3323}^{(a)}
=
f _{3322}\,U_{33}V^{*}_{32}
+
f _{2323}\,U_{32}V^{*}_{33}\,,
\end{align}
 for $b\to c \mu\bar\nu_\mu$ ($p=q=2$),
 \begin{align}
\tilde C_{2223}^{(a)}
=
f _{2222 }\,U_{22}V^{*}_{32}
+
f _{2222 }\,U_{22}V^{*}_{32}
+
\sum_{\ell=1}^{3}f_{\ell 22 \ell}  U_{2\ell}V^{*}_{3\ell}\,,
\end{align}
 for $b\to c e\bar\nu_e$ ($p=q=1$),
 \begin{align}
\tilde C_{1123}^{(a)}
=
f _{1122 }\,U_{11}V^{*}_{32}
+
f _{2121 }\,U_{12}V^{*}_{31}\,.
\end{align}
The CKM matrix $V$ is well measured  and the magnitudes of the elements of the PMNS matrix $U$
have also been observed precisely. Therefore, 
we obtain the leading terms of the coefficient:
 \begin{align}
|\tilde C_{3323}^{(a)}|
\simeq  |\tilde C_{2223}^{(a)}| \gg |\tilde C_{1123}^{(a)}|\,,
\end{align}
where the universal coupling $f_{ijkl}=f$ is taken, and  we use $V_{33}\simeq 1$,  $V_{32}\simeq -\lambda^2$,
$V_{31}\simeq \lambda^3$,  $U_{11}\simeq 2/\sqrt{6}$, 
$U_{12}\simeq U_{22}\simeq U_{32} \simeq 1/\sqrt{3}$ and $U_{33}\simeq 1/\sqrt{2}$.
That is,
the NP contribution on  $b\to c \tau\bar\nu_\tau$ is
comparable to that of 
 $b\to c \mu\bar\nu_\mu$, but is suppressed in $b\to c e\bar\nu_e$.

Putting $q=p=3$, $r=1$ and $s=3$ in Eq.~\eqref{Cpqrs},
 we can also estimate the magnitude of the new physics in the  
 $b\to u \tau\bar\nu_\tau$ decay.
 It is  comparable to the one in the $b\to c \tau\bar\nu_\tau$ decay.

\subsubsection{$b\to s \nu \bar\nu$}

Next, we concentrate on the case that quarks are down-type ones and leptons are
 the neutrinos, for example, the  $b\to s \nu \bar\nu$ process.
In the mass eigenstates of fermions, the coefficient is rewritten 
by the unitary transformation of the left-handed down-type quark as:
\begin{align}
\tilde C_{ijrs}^{(a)} 
&=\sum_{k,\ell=1}^3 V_{rk} V_{s\ell}^* f_{ijkl}\left(\delta_{ij}\delta_{kl} +\delta_{ik}\delta_{jl} + \delta_{il}\delta_{jk}\right),
\label{mass4L(a)1}
\end{align}
where the subscripts  $r,\,s$ denote mass eigenstates of down-type quarks,
while the flavor basis is in agreement with the mass basis for neutrinos
in our lepton texture.
In Eq.~\eqref{mass4L(a)1},  $V_{rk}$ denotes the  CKM matrix element.
We consider a specific case $b\to s \nu_i \bar\nu_i$, that is,
$(i,i,r,s)=(i,i,2,3)$. 
By using Eq.~\eqref{mass4L(a)1},  the sum of $i=1-3$ of the Wilson coefficient
is given explicitly
\begin{align}
\sum_{i=1}^3 \tilde  C_{ii23}^{(a)}=\sum_{i,k=1}^3 f_{iikk} V_{2k} V_{3k}^*
+ \sum_{k=1}^3  f_{k k kk } V_{2k} V_{3k}^*
+ \sum_{\ell=1}^3 f_{\ell \ell \ell \ell}V_{2\ell} V_{3\ell}^*\,.
\label{mass4L(a)2}
\end{align}
If parameters $f_{iikk}$ are common for all non-vanishing ones,
this coefficient vanishes due to the unitarity of the CKM matrix.
That is to say,
 $\sum_{i=1}^3 C_{ii23}^{(a)}=0$  in the flavor basis.
The assumption of $f_{iikk}=f$ may be reasonable since the left-handed four-fermion coupling can be derived by the flavor independent gauge-like interactions in the UV theory. 
It may be interesting to comment on  the case of $i\not= j$, that is, the $b\to s \nu_i \bar\nu_j$ process. In this case, the second and third terms in r.h.s of Eq.~\eqref{mass4L(a)1} give the Cabibbo unsuppressed coefficients $f_{2323}$ and $f_{3223}$ for $b\to s \nu_2 \bar\nu_3$ and $b\to s \nu_3 \bar\nu_2$, respectively, while other coefficients  of $i\not = j$ is suppressed by the Cabibbo angle. 
Therefore, the NP contribution of the operator $ Q_{\ell q}^{(1),(3)}$ to the neutral-current $b \to s \nu \bar{\nu}$ is expected to be suppressed.

\subsection{
$Q_{\ell edq}$ and  $Q^{(1),(3)}_{\ell equ}$  in case (a)
}

We consider the operators  $Q_{\ell edq}$ and  $Q^{(1)}_{\ell equ}$ 
in Eq.~\eqref{operator-LRRL},
\begin{align}
    Q_{\ell edq} &= (\bar{\ell}_L^i e^j_R)(\bar{d}^k_R q^l_L),
   \nonumber\\
   Q_{\ell equ}^{(1)} &= (\bar{\ell}_L^{i,a}e^j_R)\epsilon_{ab}(\bar{q}^{k,b}_L u^l_R),
   \ \ \    
   Q_{\ell equ}^{(3)} = (\bar{\ell}_L^{i,a} \sigma_{\mu \nu} e^j_R)\epsilon_{ab}(\bar{q}^{k,b}_L \sigma^{\mu \nu} u^l_R) ,
\end{align}
where the corresponding Wilson coefficients are $D'_{ijkl}$ and $J_{ijkl}$, respectively.
The semileptonic process
$b_R\to c_L e^j_R \bar \nu^i$ corresponds to the operator 
$(\bar{\nu}_L^i e^j_R)(\bar{d}^k_R c^l_L)$.
In the mass basis of quarks and  charged leptons,
the coefficient is given as:
\begin{align}
 \tilde  D'_{pqrs}= \sum_{j,\,k=1}^3 (R_e)_{qj} D'_{pjks} (R_d)_{kr}=  
 \sum_{j,\,k=1}^3   (R_e)_{qj} D_{pskj} (R_d)_{kr},
\end{align}
where $D'_{ijkl}$ is given in Eqs.\eqref{Dijkl} and \eqref{Dp-D}.
By taking  $(r,\,s)=(3,\,2)$,
\begin{align}
 \tilde  D'_{pq32}= \sum_{j,\,k=1}^3 (R_e)_{qj} D'_{pjk2} (R_d)_{k3}=  
\sum_{j,\,k=1}^3   (R_e)_{qj} D_{p2kj} (R_d)_{k3}.
\end{align}
Therefore, 
for $b_R\to c_L \tau_R \bar \nu_\tau$, we obtain 
\begin{align}
  \tilde D'_{3332}= 
  \sum_{j,\,k=1}^3 (R_e)_{3j} D'_{3jk2} (R_d)_{k3}
  =\sum_{j,\,k=1}^3(R_e)_{3j} D_{32kj} (R_d)_{k3}
   \simeq D_{3233} + D_{3223} .
\end{align}
We also obtain for $b_R\to c_L \mu_R \bar \nu_\mu$,
\begin{align}
  \tilde D'_{2232}= 
  \sum_{j,\,k=1}^3 (R_e)_{2j} D'_{2jk2} (R_d)_{k3}
  =\sum_{j,\,k=1}^3(R_e)_{2j} D_{22kj} (R_d)_{k3}
   \simeq D_{2232},
\end{align}
and for $b_R\to c_L e_R \bar \nu_e$,
\begin{align}
  \tilde D'_{1132}= 
   \sum_{j,\,k=1}^3 (R_e)_{1j} D'_{1jk2} (R_d)_{k3}
   =\sum_{j,\,k=1}^3 (R_e)_{1j} D_{12kj} (R_d)_{k3}
   \simeq D_{1231}-D_{1221}=- D_{1221},
\end{align}
since $ D_{1231}$ vanish. 
Thus, the magnitude of the NP contribution of $b_R\to c_L \tau_R \bar \nu_\tau$ is comparable to $b_R\to c_L \mu_R \bar \nu_\mu$ and   $b_R\to c_L e_R \bar \nu_e$ if  magnitudes of all non-vanishing elements of $D_{ijkl}$ are in same order.

On the other hand, the  $b_L\to c_R \tau_R \bar \nu_\tau$ process corresponds to
$(\bar{\ell}_L^{i,a} \Gamma e^j_R)\epsilon_{ab}(\bar{q}^{k,b}_L \Gamma  u^l_R)$,
where $\Gamma$ denotes the Lorentz structure of the bilinear, including scalar and tensor structures.
Let us consider the case (a) where the coefficient  $J^{(a)}_{ijkl}$
is given in Eq.~\eqref{I-J}. 
In the mass basis, we have 
\begin{align}
  \tilde J_{pqrs}=\sum_{j,\,k=1}^3 (R_e)_{qj} J_{pjks} (L_d)_{kr}=  
  \sum_{j,\,k=1}^3 (R_e)_{qj} J_{ijk\ell} (V^\dagger)_{kr}=\sum_{j,\,k=1}^3(R_e)_{qj} J_{ijk\ell} V^*_{rk}.
\end{align}
We set $(r,\,s)=(3,\,2)$ 
for $b_L\to c_R e^j_R \bar \nu_i$, where $i=p,\,\ell=s$. Then we obtain 
\begin{align}
  \tilde J_{pq32}= \sum_{j,\,k=1}^3 (R_e)_{qj} J_{qjk2} (V^\dagger)_{k3}=
 \sum_{j,\,k=1}^3  (R_e)_{qj} J_{qjk2} V^*_{3k},
\end{align}
where $J_{ijk\ell}$ is given in Eqs.~\eqref{I-J},
\eqref{I(a)1}, \eqref{I(a)2}, \eqref{I(a)3} and \eqref{I(a)4}
explicitly.
We obtain
\begin{align}
  \tilde J_{3332}= \sum_{j,\,k=1}^3 (R_e)_{3j} J_{3jk2} (V^*)_{3k}=
 \sum_{j,\,k=1}^3 (R_e)_{3j} I_{32kj} V^*_{3k}\simeq I_{3233}-\lambda I_{3232}-\lambda^2 I_{3223} =-\lambda^2 I_{3223}\,, 
\end{align}
for the process  $b_L\to c_R \tau_R \bar \nu_\tau$, where  $I_{3233}=I_{3232}=0$, 
and 
\begin{align}
 \tilde  J_{2232}= \sum_{j,\,k=1}^3 (R_e)_{2j} J_{2jk2} (V^*)_{3k}=
\sum_{j,\,k=1}^3 (R_e)_{2j} I_{22kj} V^*_{3k}\simeq 
  I_{2232}\,,
\end{align}
for the process  $b_L\to c_R \mu_R \bar \nu_\mu$.
For the process  $b_L\to c_R e_R \bar \nu_e$, we have 
\begin{align}
  \tilde J_{1132}= &\sum_{j,\,k=1}^3 (R_e)_{1j} J_{1jk2} (V^*)_{3k}=
\sum_{j,\,k=1}^3  (R_e)_{1j} I_{12kj} V^*_{3k} \nonumber\\
&\simeq I_{1231}+\lambda^2 I_{1232}-\lambda^2 I_{1221}+\lambda^4 I_{1222} = \lambda^4 I_{1222}\,,
\end{align}
where $I_{1231}= I_{1232}=I_{1221}=0$. 
Thus, the NP contribution of $b_L\to c_R \mu_R \bar \nu_\mu$ is dominant
while   $b_L\to c_R \tau_R \bar \nu_\tau$ and
$b_L\to c_R e_R \bar \nu_e$
are Cabibbo suppressed ones, where  the magnitudes of all elements of non-vanishing $J_{ijkl}$ are supposed to be in same order. 
We find that, although the $B$-meson anomalies favor an NP structure with a strong coupling to the third generation, which is compatible with a $U(2)$ flavor structure (see, e.g., Ref. \cite{Fuentes-Martin:2019mun}), our model realizes a distinct flavor structure.

\subsection{Leptonic dipole operators in case (a)}

For the leptonic dipole operators, 
the relevant dimension-six operators are:
\begin{align}
Q_{eW} &=
(\bar{l}_i\sigma^{\mu\nu}e_j)\tau^I H W_{\mu\nu}^I \,,\qquad
Q_{eB} =
(\bar{l}_i\sigma^{\mu\nu}e_j)H B_{\mu\nu}\, .
\end{align}
where the corresponding Wilson coefficients are both given by $C_{\ell_ie_j h}$.
Since we consider the two-Higgs doublet model  in this paper, the SM Higgs is defined in terms of a mixing parameter $\beta$
as given in Eq.~\eqref{eq:def_Higgs}, i.e.,
\begin{align}
    H&= s_\beta\,\Phi_u + c_\beta\,\Phi_d\,, \nonumber
\end{align}
with $c_\beta\equiv \cos(\beta)$ and $s_\beta \equiv \sin(\beta)$, 
indicating that the charge of SM Higgs can be understood by those of $\Phi_u$ and $\Phi_d$. 
The textures of these operators can be given by
the sum of the  two operators
as seen in Eq.\eqref{QeWQuG}:
\begin{align}
    C_{\ell_ie_j h}Q_{eW(eB)} &= C_{\ell_ie_j h}(\bar{\ell}_L^i \sigma^{\mu\nu}e_R^j)\tau^I H W_{\mu\nu}^I(B_{\mu\nu}^I)
    \nonumber\\
    &\equiv C_{\ell_i e_j \Phi_u}s_\beta(\bar{\ell}_L^i \sigma^{\mu\nu}e_R^j)\tau^I \Phi_u W_{\mu\nu}^I(B_{\mu\nu}^I)
    + C_{\ell_i e_j \Phi_d}c_\beta(\bar{\ell}_L^i \sigma^{\mu\nu}e_R^j)\tau^I \Phi_d W_{\mu\nu}^I(B_{\mu\nu}^I)\,.
    \nonumber
\end{align}
In the mass basis of quarks and the charged leptons, the coefficients
of the case (a) are given as:
\begin{align}
&\tilde C^{(a)}_{\ell_p e_q \Phi_d}=\sum_{i,\,j=1}^3(L_e^\dagger)_{pi}C^{(a)}_{\ell_i e_j  \Phi_d} ( R_e)_{jq} 
=\sum_{i,\,j=1}^3 U_{pi}C^{(a)}_{\ell_i e_j  \Phi_d} ( R_e)_{jq}\,, \nonumber\\
&\tilde C^{(a)}_{\ell_p e_q \Phi_u}=
\sum_{i,\,j=1}^3(L_e^\dagger)_{pi}C^{(a)}_{\ell_i e_j  \Phi_u} ( R_e)_{jq} 
=\sum_{i,\,j=1}^3 U_{pi}C^{(a)}_{\ell_i e_j  \Phi_u} ( R_e)_{jq}\,,
\end{align}
where $C^{(a)}_{\ell_i e_j \Phi_d}$ and $C^{(a)}_{\ell_i e_j \Phi_u}$ 
are given in Eq.~\eqref{eq:lepton-dipole}.
Since   $C^{(a)}_{\ell_i e_j \Phi_d}$ is the same texture as  the Yukawa
matrix of the charged lepton, we suppose then 
$\tilde C^{(a)}_{\ell_p e_q \Phi_d}$
is almost diagonal in the mass basis.  On the other hand, 
$ C^{(a)}_{\ell_i e_j \Phi_u}$ has a texture different from that of $C^{(a)}_{\ell_i e_j \Phi_d}$ as seen 
in Eq.~\eqref{eq:lepton-dipole},
$C^{(a)}_{\ell_i e_j \Phi_u}= f \delta_{i2}\delta_{j1} $.
Then, $C^{(a)}_{\ell_i e_j \Phi_u}$ is transformed into the mass basis as: 
\begin{align}
\tilde C^{(a)}_{\ell_p e_q \Phi_u}=
\sum_{i,\,j=1}^3 U_{pi} f \delta_{i2}\delta_{j1}( R_e)_{jq}
= f U_{p2}  ( R_e)_{1q}\,.
\label{dipole-massbasis}
\end{align}
Therefore, the decay amplitude of  $\mu_L\to e_R \gamma$ is proportional to
$ U_{22}  ( R_e)_{11}$ ($p=2$,\,$q=1$),
while $\tau_L\to e_R \gamma$ 
is proportional to $ U_{23}  ( R_e)_{11}$ ($p=3,\,q=1$).
Putting the numerical values given in Eq.~\eqref{Re}, together with $U_{22} \simeq U_{32} \simeq   1/\sqrt{3},\,U_{23} \simeq 1/\sqrt{2}$, we find that the amplitudes for $\mu_L\to e_R \gamma$ and $\tau_L\to e_R \gamma$ are comparable in magnitude. 
On the other hand, the magnitude of $\tau_L\to \mu_R\gamma$ is proportional to
$ U_{32}  ( R_e)_{12}$ ($p=3$,\,$q=2$), which is suppressed by 
$(R_e)_{12}\simeq \lambda^2$.

We also investigate  the $\mu_R\to e_L \gamma$ process.
By taking $p=1$ and $q=2$ in Eq.~\eqref{dipole-massbasis}, 
its magnitude is given by $f\,U_{12}  (R_e)_{12}$ ($p=1$,\,$q=2$), which is suppressed by ${\cal O}(\lambda^2)$ relative to the magnitude for $\mu_L\to e_R \gamma$, as seen in Eq.~\eqref{Re}.
The magnitude of $\tau_R\to e_L(\mu_L)\gamma$ is also suppressed
compared with $\tau_L\to e_R(\mu_R)\gamma$ due to the factor $ ( R_e)_{13}\simeq \lambda^3$.
This result is in contrast to those obtained in other models with the conventional flavor symmetries, for example, models based on $A_4$ modular symmetry \cite{Kobayashi:2021pav,Kobayashi:2022jvy} and $U(2)$ symmetry~\cite{Isidori:2021gqe,Tanimoto:2023hse}, in which $\mu_R\to e_L\gamma$ dominates the $\mu\to e \gamma$ decay.
These predictions could be tested in future experiments.

We note that above result depends significantly on the zero structures of  $C_{\ell_i e_j \Phi_u}$ in Eq.~\eqref{eq:lepton-dipole}, while $C_{\ell_i e_j\Phi_d}$ has the same structure in all cases (a)--(e). 
Thus, the leptonic dipole operators provide a crucial probe for distinguishing among the cases (a)--(e).

\section{Conclusions}
\label{sec:con}

In this paper, we have studied the flavor structures of dimension-six operators in the Standard Model Effective Field Theory using non-invertible selection rules. We considered a four-dimensional effective field theory whose interactions are restricted by fusion rules arising from the $\mathbb{Z}_2$ gauging of $\mathbb{Z}_5$ symmetries. These fusion rules reproduce realistic Yukawa textures and allow textures that cannot be obtained from conventional group-theoretical symmetries.

We classified the flavor textures of all baryon-number-conserving dimension-six SMEFT operators for the matter-field assignments considered in this work. The fusion rules determine both the texture zeros and the allowed tensor structures of the Wilson coefficients. 
Furthermore, the coefficients can be analytically expressed in terms of a reduced number of independent parameters, giving rise to correlations among different flavor violating processes. 
Remarkably, the flavor structures of the higher-dimensional operators are not necessarily aligned with those of the Yukawa couplings. This is different from the Minimal Flavor Violation hypothesis, in which the Yukawa couplings control the flavor dependence of higher-dimensional operators. Non-invertible selection rules thus allow the Yukawa couplings and SMEFT Wilson coefficients to have different flavor structures while being governed by the same fusion rules.

We also studied several phenomenological consequences of the resulting textures in the case (a). 
For the semileptonic four-fermion operators, we examined their contributions to neutral- and charged-current processes. 
In particular, we quantitatively analyzed the $b\to c \tau\bar\nu_\tau$ and $b\to s\nu \bar\nu$ processes, finding that our model exhibits a flavor structure distinct from that of $U(2)$ symmetry. 
For the leptonic dipole operators, the texture zeros lead to characteristic chirality-dependent suppressions. In the case (a), the amplitude for $\mu_R\to e_L\gamma$ is suppressed by $\mathcal{O}(\lambda^2)={\mathcal{O}(0.04)}$ relative to that for $\mu_L\to e_R\gamma$. 
The amplitudes for $\tau_R\to e_L\gamma$ and $\tau_R\to\mu_L\gamma$ are also suppressed by the small right-handed charged-lepton mixing. 
Consequently, the dominant chirality in $\mu\to e\gamma$ differs from that predicted in models based on $A_4$ modular symmetry and $U(2)$ symmetry. 
These results may provide a way to distinguish between these flavor structures in future measurements.

Further phenomenological studies of other processes may also provide useful tests of the relations among Wilson coefficients implied by the fusion rules. 
It would also be interesting to extend the present
analysis to other Yukawa textures realized by non-invertible selection
rules and to systematically investigate the resulting flavor structures
of higher-dimensional operators. We leave these directions for future work.

\acknowledgments

This work was supported by JSPS KAKENHI Grant Numbers JP23K03375 (T.K.), JP25H01539 (H.O.),  JP26K07087 (H.O.) and  JP26K07097(K.Y.).

\appendix

\bibliography{references}{}

\providecommand{\href}[2]{#2}\begingroup\raggedright\begin{thebibliography}{10}

\bibitem{Weinberg:1977hb}
S.~Weinberg, \emph{{The Problem of Mass}}, \href{https://doi.org/10.1111/j.2164-0947.1977.tb02958.x}{\emph{Trans. New York Acad. Sci.} {\bfseries 38} (1977) 185}.

\bibitem{Fritzsch:1977vd}
H.~Fritzsch, \emph{{Weak Interaction Mixing in the Six - Quark Theory}}, \href{https://doi.org/10.1016/0370-2693(78)90524-5}{\emph{Phys. Lett. B} {\bfseries 73} (1978) 317}.

\bibitem{Kobayashi:2024yqq}
T.~Kobayashi and H.~Otsuka, \emph{{Non-invertible flavor symmetries in magnetized extra dimensions}}, \href{https://doi.org/10.1007/JHEP11(2024)120}{\emph{JHEP} {\bfseries 11} (2024) 120} [\href{https://arxiv.org/abs/2408.13984}{{\ttfamily 2408.13984}}].

\bibitem{Kobayashi:2024cvp}
T.~Kobayashi, H.~Otsuka and M.~Tanimoto, \emph{{Yukawa textures from non-invertible symmetries}}, \href{https://doi.org/10.1007/JHEP12(2024)117}{\emph{JHEP} {\bfseries 12} (2024) 117} [\href{https://arxiv.org/abs/2409.05270}{{\ttfamily 2409.05270}}].

\bibitem{Kobayashi:2025znw}
T.~Kobayashi, Y.~Nishioka, H.~Otsuka and M.~Tanimoto, \emph{{More about quark Yukawa textures from selection rules without group actions}}, \href{https://doi.org/10.1007/JHEP05(2025)177}{\emph{JHEP} {\bfseries 05} (2025) 177} [\href{https://arxiv.org/abs/2503.09966}{{\ttfamily 2503.09966}}].

\bibitem{Liang:2025dkm}
Q.~Liang and T.T.~Yanagida, \emph{{Non-invertible symmetry as an axion-less solution to the strong CP problem}}, \href{https://doi.org/10.1016/j.physletb.2025.139706}{\emph{Phys. Lett. B} {\bfseries 868} (2025) 139706} [\href{https://arxiv.org/abs/2505.05142}{{\ttfamily 2505.05142}}].

\bibitem{Kobayashi:2025ldi}
T.~Kobayashi, H.~Otsuka, M.~Tanimoto and H.~Uchida, \emph{{Lepton mass textures from non-invertible multiplication rules}}, \href{https://doi.org/10.1007/JHEP08(2025)189}{\emph{JHEP} {\bfseries 08} (2025) 189} [\href{https://arxiv.org/abs/2505.07262}{{\ttfamily 2505.07262}}].

\bibitem{Jiang:2025psz}
Z.~Jiang, B.-Y.~Qu and G.-J.~Ding, \emph{{Texture-zeros in minimal seesaw from noninvertible symmetry fusion rules}}, \href{https://doi.org/10.1103/d29s-cw34}{\emph{Phys. Rev. D} {\bfseries 112} (2025) 115029} [\href{https://arxiv.org/abs/2510.07236}{{\ttfamily 2510.07236}}].

\bibitem{Qu:2026omn}
B.-Y.~Qu, Z.~Jiang and G.-J.~Ding, \emph{{Two-zero textures of the Majorana neutrino mass matrix from $\mathbb{Z}_3$ gauging of $\mathbb{Z}_N$ non-invertible symmetry}},  \href{https://arxiv.org/abs/2602.24214}{{\ttfamily 2602.24214}}.

\bibitem{Kitagawa:2026eck}
H.~Kitagawa, C.~Miyao, S.~Nishimura and H.~Otsuka, \emph{{Revisiting One-Zero and Two-Zero Neutrino Mass Textures in Light of Recent Oscillation and Cosmological Data}},  \href{https://arxiv.org/abs/2607.08384}{{\ttfamily 2607.08384}}.

\bibitem{Kobayashi:2025thd}
T.~Kobayashi, H.~Otsuka and T.T.~Yanagida, \emph{{Noninvertible symmetry as a solution to the strong CP problem in a GUT-inspired standard model}}, \href{https://doi.org/10.1103/vbd5-5cp3}{\emph{Phys. Rev. D} {\bfseries 113} (2026) 055016} [\href{https://arxiv.org/abs/2508.12287}{{\ttfamily 2508.12287}}].

\bibitem{Kobayashi:2025rpx}
T.~Kobayashi, H.~Otsuka, M.~Tanimoto and T.T.~Yanagida, \emph{{GUT-motivated noninvertible symmetry as a solution to the strong CP problem and the neutrino CP-violating phase}}, \href{https://doi.org/10.1103/f7bw-qlgs}{\emph{Phys. Rev. D} {\bfseries 113} (2026) 095034} [\href{https://arxiv.org/abs/2510.01680}{{\ttfamily 2510.01680}}].

\bibitem{Chen:2026mvi}
Z.-Q.~Chen, W.-H.~Jiang and Y.-L.~Zhou, \emph{{Universal two-zero texture in SO(10): implications of JUNO and realization from non-invertible symmetries}},  \href{https://arxiv.org/abs/2606.24571}{{\ttfamily 2606.24571}}.

\bibitem{HeavyFlavorAveragingGroupHFLAV:2024ctg}
{\scshape Heavy Flavor Averaging Group (HFLAV)} collaboration, \emph{{Averages of b-hadron, c-hadron, and {\ensuremath{\tau}}-lepton properties as of 2023}}, \href{https://doi.org/10.1103/x87q-tld5}{\emph{Phys. Rev. D} {\bfseries 113} (2026) 012008} [\href{https://arxiv.org/abs/2411.18639}{{\ttfamily 2411.18639}}].

\bibitem{Grzadkowski:2010es}
B.~Grzadkowski, M.~Iskrzynski, M.~Misiak and J.~Rosiek, \emph{{Dimension-Six Terms in the Standard Model Lagrangian}}, \href{https://doi.org/10.1007/JHEP10(2010)085}{\emph{JHEP} {\bfseries 10} (2010) 085} [\href{https://arxiv.org/abs/1008.4884}{{\ttfamily 1008.4884}}].

\bibitem{Alonso:2013hga}
R.~Alonso, E.E.~Jenkins, A.V.~Manohar and M.~Trott, \emph{{Renormalization Group Evolution of the Standard Model Dimension Six Operators III: Gauge Coupling Dependence and Phenomenology}}, \href{https://doi.org/10.1007/JHEP04(2014)159}{\emph{JHEP} {\bfseries 04} (2014) 159} [\href{https://arxiv.org/abs/1312.2014}{{\ttfamily 1312.2014}}].

\bibitem{Chivukula:1987py}
R.S.~Chivukula and H.~Georgi, \emph{{Composite Technicolor Standard Model}}, \href{https://doi.org/10.1016/0370-2693(87)90713-1}{\emph{Phys. Lett. B} {\bfseries 188} (1987) 99}.

\bibitem{DAmbrosio:2002vsn}
G.~D'Ambrosio, G.F.~Giudice, G.~Isidori and A.~Strumia, \emph{{Minimal flavor violation: An Effective field theory approach}}, \href{https://doi.org/10.1016/S0550-3213(02)00836-2}{\emph{Nucl. Phys. B} {\bfseries 645} (2002) 155} [\href{https://arxiv.org/abs/hep-ph/0207036}{{\ttfamily hep-ph/0207036}}].

\bibitem{Barbieri:2011ci}
R.~Barbieri, G.~Isidori, J.~Jones-Perez, P.~Lodone and D.M.~Straub, \emph{{$U(2)$ and Minimal Flavour Violation in Supersymmetry}}, \href{https://doi.org/10.1140/epjc/s10052-011-1725-z}{\emph{Eur. Phys. J. C} {\bfseries 71} (2011) 1725} [\href{https://arxiv.org/abs/1105.2296}{{\ttfamily 1105.2296}}].

\bibitem{Barbieri:2012uh}
R.~Barbieri, D.~Buttazzo, F.~Sala and D.M.~Straub, \emph{{Flavour physics from an approximate $U(2)^3$ symmetry}}, \href{https://doi.org/10.1007/JHEP07(2012)181}{\emph{JHEP} {\bfseries 07} (2012) 181} [\href{https://arxiv.org/abs/1203.4218}{{\ttfamily 1203.4218}}].

\bibitem{Faroughy:2020ina}
D.A.~Faroughy, G.~Isidori, F.~Wilsch and K.~Yamamoto, \emph{{Flavour symmetries in the SMEFT}}, \href{https://doi.org/10.1007/JHEP08(2020)166}{\emph{JHEP} {\bfseries 08} (2020) 166} [\href{https://arxiv.org/abs/2005.05366}{{\ttfamily 2005.05366}}].

\bibitem{Brivio:2017vri}
I.~Brivio and M.~Trott, \emph{{The Standard Model as an Effective Field Theory}}, \href{https://doi.org/10.1016/j.physrep.2018.11.002}{\emph{Phys. Rept.} {\bfseries 793} (2019) 1} [\href{https://arxiv.org/abs/1706.08945}{{\ttfamily 1706.08945}}].

\bibitem{Isidori:2023pyp}
G.~Isidori, F.~Wilsch and D.~Wyler, \emph{{The standard model effective field theory at work}}, \href{https://doi.org/10.1103/RevModPhys.96.015006}{\emph{Rev. Mod. Phys.} {\bfseries 96} (2024) 015006} [\href{https://arxiv.org/abs/2303.16922}{{\ttfamily 2303.16922}}].

\bibitem{Greljo:2022cah}
A.~Greljo, A.~Palavri{\'c} and A.E.~Thomsen, \emph{{Adding Flavor to the SMEFT}}, \href{https://doi.org/10.1007/JHEP10(2022)005}{\emph{JHEP} {\bfseries 10} (2022) 010} [\href{https://arxiv.org/abs/2203.09561}{{\ttfamily 2203.09561}}].

\bibitem{Loisa:2024xuk}
E.~Loisa and J.~Talbert, \emph{{Froggatt-Nielsen meets the SMEFT}}, \href{https://doi.org/10.1007/JHEP10(2024)017}{\emph{JHEP} {\bfseries 10} (2024) 017} [\href{https://arxiv.org/abs/2402.16940}{{\ttfamily 2402.16940}}].

\bibitem{Kobayashi:2021pav}
T.~Kobayashi, H.~Otsuka, M.~Tanimoto and K.~Yamamoto, \emph{{Modular symmetry in the SMEFT}}, \href{https://doi.org/10.1103/PhysRevD.105.055022}{\emph{Phys. Rev. D} {\bfseries 105} (2022) 055022} [\href{https://arxiv.org/abs/2112.00493}{{\ttfamily 2112.00493}}].

\bibitem{Kobayashi:2022jvy}
T.~Kobayashi, H.~Otsuka, M.~Tanimoto and K.~Yamamoto, \emph{{Lepton flavor violation, lepton (g {\ensuremath{-}} 2)$_{\mu, e}$ and electron EDM in the modular symmetry}}, \href{https://doi.org/10.1007/JHEP08(2022)013}{\emph{JHEP} {\bfseries 08} (2022) 013} [\href{https://arxiv.org/abs/2204.12325}{{\ttfamily 2204.12325}}].

\bibitem{Moreno-Sanchez:2025bzz}
A.~Moreno-S{\'a}nchez and A.~Palavri{\'c}, \emph{{Leptonic flavor from a modular A4 symmetry: UV mediators and SMEFT realizations}}, \href{https://doi.org/10.1103/hdgd-cq45}{\emph{Phys. Rev. D} {\bfseries 112} (2025) 075002} [\href{https://arxiv.org/abs/2505.01535}{{\ttfamily 2505.01535}}].

\bibitem{Kang:2026qgi}
L.-J.~Kang, H.~Sun and J.-H.~Yu, \emph{{Complete operator basis for the modular invariant SMEFT}}, \href{https://doi.org/10.1007/JHEP07(2026)130}{\emph{JHEP} {\bfseries 07} (2026) 130} [\href{https://arxiv.org/abs/2601.23060}{{\ttfamily 2601.23060}}].

\bibitem{Kobayashi:2021uam}
T.~Kobayashi and H.~Otsuka, \emph{{On stringy origin of minimal flavor violation}}, \href{https://doi.org/10.1140/epjc/s10052-022-09986-4}{\emph{Eur. Phys. J. C} {\bfseries 82} (2022) 25} [\href{https://arxiv.org/abs/2108.02700}{{\ttfamily 2108.02700}}].

\bibitem{Dong:2025pah}
J.~Dong, T.~Kobayashi, R.~Nishida, S.~Nishimura and H.~Otsuka, \emph{{Coupling selection rules in heterotic Calabi-Yau compactifications}}, \href{https://doi.org/10.1007/JHEP09(2025)012}{\emph{JHEP} {\bfseries 09} (2025) 012} [\href{https://arxiv.org/abs/2504.09773}{{\ttfamily 2504.09773}}].

\bibitem{Kobayashi:2025ocp}
T.~Kobayashi, R.~Nishida and H.~Otsuka, \emph{{Non-invertible selection rules on heterotic non-Abelian orbifolds}}, \href{https://doi.org/10.1007/JHEP03(2026)158}{\emph{JHEP} {\bfseries 03} (2026) 158} [\href{https://arxiv.org/abs/2509.10019}{{\ttfamily 2509.10019}}].

\bibitem{Dong:2025jra}
J.~Dong, T.~Jeric, T.~Kobayashi, R.~Nishida and H.~Otsuka, \emph{{Discrete gauging and noninvertible selection rules}}, \href{https://doi.org/10.1103/nsvv-l2dy}{\emph{Phys. Rev. D} {\bfseries 113} (2026) 056028} [\href{https://arxiv.org/abs/2507.02375}{{\ttfamily 2507.02375}}].

\bibitem{Jenkins:2017jig}
E.E.~Jenkins, A.V.~Manohar and P.~Stoffer, \emph{{Low-Energy Effective Field Theory below the Electroweak Scale: Operators and Matching}}, \href{https://doi.org/10.1007/JHEP03(2018)016}{\emph{JHEP} {\bfseries 03} (2018) 016} [\href{https://arxiv.org/abs/1709.04486}{{\ttfamily 1709.04486}}].

\bibitem{Belle-II:2023esi}
{\scshape Belle-II} collaboration, \emph{{Evidence for B+{\textrightarrow}K+{\ensuremath{\nu}}{\ensuremath{\nu}}{\textasciimacron} decays}}, \href{https://doi.org/10.1103/PhysRevD.109.112006}{\emph{Phys. Rev. D} {\bfseries 109} (2024) 112006} [\href{https://arxiv.org/abs/2311.14647}{{\ttfamily 2311.14647}}].

\bibitem{ParticleDataGroup:2024cfk}
{\scshape Particle Data Group} collaboration, \emph{{Review of particle physics}}, \href{https://doi.org/10.1103/PhysRevD.110.030001}{\emph{Phys. Rev. D} {\bfseries 110} (2024) 030001}.

\bibitem{Fuentes-Martin:2019mun}
J.~Fuentes-Mart{\'\i}n, G.~Isidori, J.~Pag{\`e}s and K.~Yamamoto, \emph{{With or without U(2)? Probing non-standard flavor and helicity structures in semileptonic B decays}}, \href{https://doi.org/10.1016/j.physletb.2019.135080}{\emph{Phys. Lett. B} {\bfseries 800} (2020) 135080} [\href{https://arxiv.org/abs/1909.02519}{{\ttfamily 1909.02519}}].

\bibitem{Isidori:2021gqe}
G.~Isidori, J.~Pag{\`e}s and F.~Wilsch, \emph{{Flavour alignment of New Physics in light of the (g {\ensuremath{-}} 2)$_{\mu}$ anomaly}}, \href{https://doi.org/10.1007/JHEP03(2022)011}{\emph{JHEP} {\bfseries 03} (2022) 011} [\href{https://arxiv.org/abs/2111.13724}{{\ttfamily 2111.13724}}].

\bibitem{Tanimoto:2023hse}
M.~Tanimoto and K.~Yamamoto, \emph{{Electron EDM and LFV decays in the light of Muon $(g-2)_\mu $ with U(2) flavor symmetry}}, \href{https://doi.org/10.1140/epjc/s10052-024-12623-x}{\emph{Eur. Phys. J. C} {\bfseries 84} (2024) 252} [\href{https://arxiv.org/abs/2310.16325}{{\ttfamily 2310.16325}}].

\end{thebibliography}\endgroup
\bibliographystyle{JHEP}

\end{document}